\documentclass[aps,reprint,groupedaddress]{revtex4-1}
\usepackage[latin1]{inputenc}
\usepackage{graphicx}
\usepackage{url}
\usepackage{rotating}
\usepackage{longtable}
\usepackage{amsfonts}
\usepackage{xspace}
\usepackage{feynmf}
\usepackage{subfigure}
\usepackage{amsfonts}
\usepackage{amsmath,amssymb}
\usepackage{cleveref}
\usepackage{appendix}
\usepackage{epstopdf}
\begin{document}
\title{\uppercase{The beam exchange of a circulator cooler ring with a ultrafast harmonic kicker}}
\author{Gunn Tae Park, Jiquan Guo, Shaoheng Wang, Robert A. Rimmer, and Haipeng Wang\\
Accelerator Division, Jefferson Lab, Newport News, VA 23606, USA}

\begin{abstract}
In this paper, we describe a harmonic kicker system used in the beam exchange scheme for the Circulator Cooling Ring (CCR) of the Jefferson Lab Electron Ion Collider(JLEIC). By delivering an ultra-fast deflecting kick, a kicker directs electron bunches selectively in/out of the CCR without degrading the beam dynamics of the CCR optimized for ion beam cooling. We will discuss the design principle of the kicker system and demonstrate its performance with various numerical simulations. In particular, the degrading effects of realistic harmonic kicks on the beam dynamics, such as 3D kick field profiles interacting with the magnetized beam, is studied in detail with a scheme that keeps the cooling efficiency within allowable limits. 
\end{abstract}
\maketitle

\section{Introduction}
\par
\begin{figure}[hbt]
  \centering
    \subfigure[ Overview of the CCR. The grey colored box is the beam exchange region.]{\label{fig:CCR_view}\includegraphics[scale=0.28]{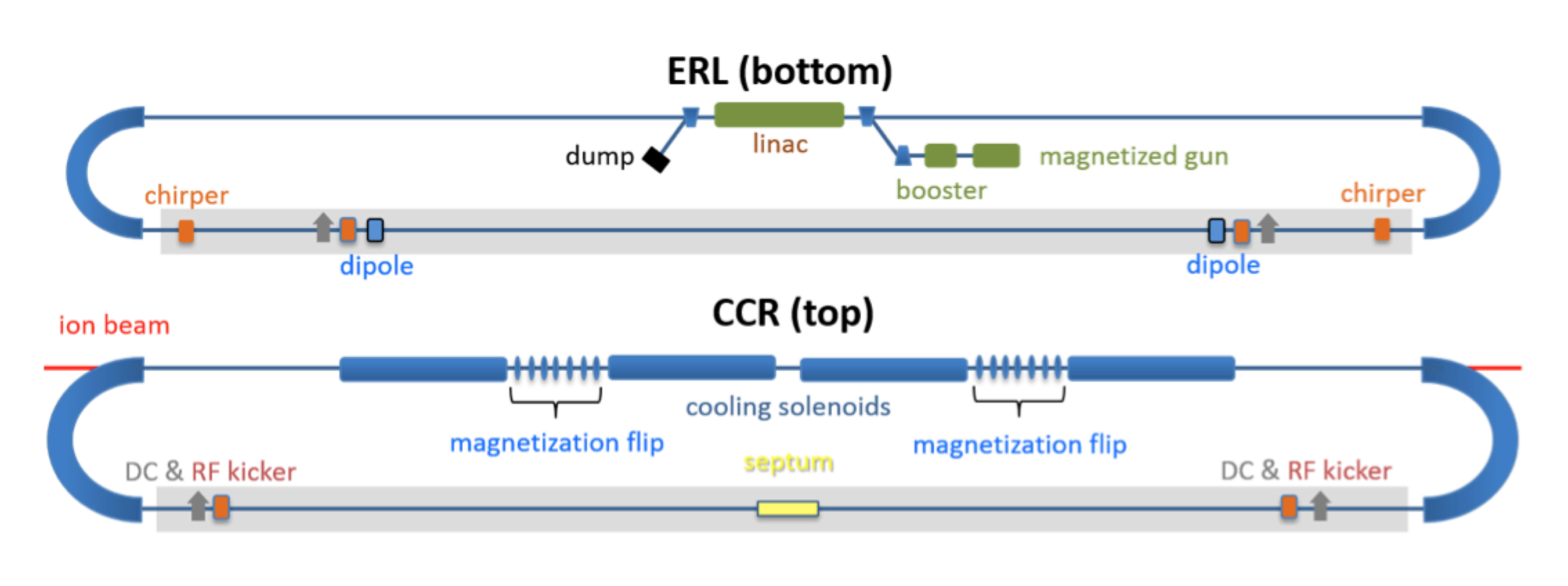}}\\
    \subfigure[ The expanded view of the exchange region.]{\label{fig:exchange}\includegraphics[scale=0.35]{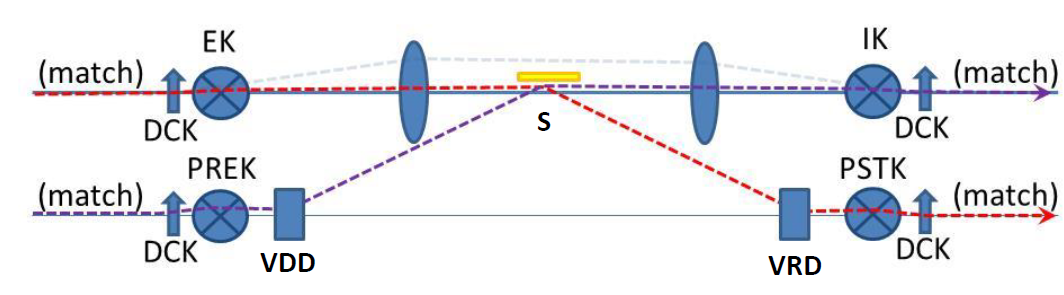}}
  \caption{The schematic view of the CCR.}
  \label{fig:CCR}
\end{figure}

 A proposal~\cite{pCDR} for the Jefferson Laboratory Electron-Ion Collider (JLEIC) includes the Circulator Cooling Ring (CCR), which can dramatically increase luminosity of the electron-ion collision at a 45\,GeV center-of-mass energy by cooling the ion beam in a storage ring at an energy of  up to 100\,GeV/nucleon. The cooling is done by passing the ion beam through a series of cooling solenoid channels\textemdash located in an overlapping segment of the CCR and the ion storage ring\textemdash along with a co-moving electron beam (see Fig.\,\ref{fig:CCR_view}), whose beam parameters are dictated mostly by the ion beam parameters and listed in Table\,\ref{table:bp}~\cite{benson}. In order to deliver an electron beam current of 0.76\,A at a bunch repetition rate of $f_b=476.3$\,MHz to the cooling channels while taking into account the technological limitation on the injection current from the gun (whose state-of-art limit is $\sim75$\,mA), the CCR is designed to increase the current in the cooler by causing the electron bunches to re-circulate in the ring for 11\,turns. The schematic layout of the CCR complex and the exchange region with a kicker system is shown in Fig.\,\ref{fig:CCR}. In Fig.\,\ref{fig:CCR_view}, the electron beam from a magnetized RF gun at 43.3\.MHz gets accelerated by the energy recovery linac (ERL) to a nominal 55\,MeV energy and enters the exchange region (grey strip) to be transferred to the upper level. On the upper level, the electron beam joins the CCR populated with the re-circulating beam at 476.3\,MHz and circulates in the ring for 11 turns before exiting through the exchange region and going back to the ERL, and eventually to the beam dump. In a closer look at the exchange region illustrated in Fig.\,\ref{fig:exchange}, an injected bunch follows the purple dashed line: goes through a pre-kicker cavity(PREK), gets bent towards the upper level by a large-angle deflecting magnet (VDD) and bent back level via a septum (S) magnet. It is then kicked down by an injection kicker (IK) to merge with the CCR beam (the grey dashed line), and then kicked up (together with the re-circulating bunches) by a DC kicker magnet (DCK) to start circulation. During the re-circulation, the bunches in the exchange region avoid the influence of the septum by following the path of the grey dashed line created by a series of magnets\textemdash a DCK, a pair of focusing magnets, and another DCK. During re-circulation, the bunches do not experience kicks due to phase mismatch of the harmonic modes of the kick. After 11\,passes, an extracted bunch follows the red dashed line: after the DCK magnet, it gets kicked down by an extraction kicker (EK), gets transferred down to the ERL ring via the septum magnet and another large-angle deflecting magnet (VRD), and goes through a post kicker (PSTK) cavity before returning to the ERL. 
\par According to the proposed beam exchange scheme~\cite{pCDR}, every 11th bunch at the injection/extraction points in the CCR must be injected/extracted into/out of the ring at a kick frequency of $f_k=43.3$\,MHz\textemdash The ion beam bunch frequency will double to $952.6$\,MHz in a future upgrade, leading to the doubled exchange/CCR bunch frequencies of the electron beam. The design of the kicker system was based on $f_k=86.6$\,MHz as a preparation for this upgrade. Such an exchange scheme would need an ultra-fast kicker system that selectively delivers a deflecting kick of 2.5\,mrad angle to the exchanged bunches only. This implies the rise-fall time of the kick must be much smaller than the bunch spacing of $1/f_b=2.1$\,ns. The most promising candidate for a fast kicker is a harmonic kicker based on a quarter wave resonator (QWR) whose kick profile is made of a linear combination of RF harmonic modes so that it has a sharply peaked temporal profile around an exchanged bunch. 
 \par In a CCR, a magnetized beam has some advantages over a non-magnetized beam, including a strong suppression of the CSR microbunching/energy spread growth~\cite{tsai} and stronger cooling~\cite{burov0}. For a high energy electron beam, the transverse velocity spread in the beam frame is enlarged by the Lorentz factor and hence the transverse temperature is usually much higher than the longitudinal velocity. For a strongly magnetized electron beam, the Larmor radius is much smaller than the impact factor. The ions interact with the Larmor circles instead of the free electron. The cooling time is mainly determined by the longitudinal temperature of the electrons and a stronger cooling, i.e. a shorter cooling time, can be achieved~\cite{budker},\cite{derbenev},\cite{parkhomchun}. To maintain a constant electron bunch size in the solenoid, the beta function is determined by the momentum of the electron and the magnetic field. In the JLEIC configuration, the beta function of the electron beam is much smaller than the beta function of the ion beam. To make sure the electron beam size matches the ion beam size, the emittance of an unmagnetized electron beam will have to be very large, which leads to very high temperature and lowers the cooling rate. But for a magnetized electron beam, the beam size is determined by the drift emittance, while the cooling rate is determined by the Larmor emittance. We could simultaneously achieve a large drift emittance to match the beam size and a small Larmor emittance to obtain a good cooling rate~\cite{zhang}. A round magnetized beam within the cooling solenoid can be achieved by generating a round magnetized beam at a photocathode gun immersed in a solenoid and propagating it through a rotationally invariant and ``decoupled" beamline to the cooler, as first conceived in~\cite{burov1} and later adopted in the CCR design with an extension of the scheme to the entire CCR (i.e. to come back to the cooler after circulation)~\cite{pCDR}. At the start and end point of a globally invariant beamline, the beam has the same canonical angular momentum (CAM) and consequently roundness of the beam is preserved. If the beamline is decoupled as well, then the Larmor (rotational) motion of the beam is decoupled from the drift (Larmor center) motion, implying the Larmor emittance as a measure of the Larmor motion of the beam is conserved. The lattice design of an optimized beamline in a CCR without kickers can be found in~\cite{tennant}. 

  \begin{table}[hbt]
   \centering
   \caption{Beam parameters of the CCR at the kicker and at the photocathode gun. The Twiss parameters are at the kickers. The value in the bracket for the Lamor emittance refers to the tolerance limit.}
  \vspace{5pt}
   \begin{tabular}{lcc}
       \toprule
       Parameters & Unit & Magnetized beam
       \vspace{3pt}\\
       \hline
           Beam energy $E$     &  MeV  &  $20-55$  \\
           Bunch frequency $f_b$ & MHz & 476.3  \\
           Bunch charge $Q_b$ & nC& 1.6 \\
           Kick frequency $f_k$ & MHz & 86.6 \\
           Kick angle $\theta$ & mrad  &  2.5 \\
           Bunch distr.$_\perp$ & - & Uniform-ellipse \\
           Bunch distr.$_\parallel$ & - & Top-hat \\
           Bunch length $l$ &   cm &  3 \\
           $rms$ Energy spread $\delta E/E$ & $\times 10^{-4}$ & 3\\
           Effective emittance (hor.) $\varepsilon_{eff,x}$  &  mm\,mrad & 36  \\
           Effective emittance (vert.) $\varepsilon_{eff,y}$  &  mm\,mrad & 36  \\
           Drift emittance $\varepsilon_{drift}$  &  mm\,mrad & 36 \\
           Larmor emittance $\varepsilon_{Larmor}$ & mm\,mrad & 1(19) \\
           Twiss parameter (hor.) $\beta_x$ & m & 10 \\
           Twiss parameter (hor.) $\alpha_x$ & - & 0 \\
           Twiss parameter (vert.) $\beta_y$ & m & 120  \\
           Twiss parameter (vert.) $\alpha_y$ & - & 0 \\
           Magnetization $M$ & mm\,mrad & $35$ \\
         \hline
           Gun frequency $f_g$ & MHz & 43.3 \\
           Gun voltage $V_g$ & kV & 400  \\
           Cathode spot radius $\sigma_{cath}$ & mm & 2.2 \\
           Cathode magnetic field $B_{cath}$ & T & 0.1 \\
           Cooler Solenoid field $B_{cool}$ & T & 1 \\
           Beam spot radius $\sigma_{cool}$ & mm & 0.7 \\
           Electron beta at the cooler & cm & 36 \\
        \botrule
   \end{tabular}
   \label{table:bp}
\end{table}

\par In this note, we describe design of a harmonic kicker system and demonstrate by numerical simulations that the optimized beam dynamics of the CCR for the maximum cooling efficiency can be maintained (within tolerance limits) after a harmonic kicker system is implemented. The basic ideas in the design of a harmonic kicker presented here are not new. The first prototype of a harmonic kicker was developed in \cite{huang1},\cite{huang2} for different beam dynamics of the CCR. In \cite{huang1}, a linear combination of 10 harmonic modes, distributed over three different cavities, was designed  as a kick profile, the idea of using two kickers, injection (IK) and extraction (EK), with an intervening betatron phase advance of $\pi$ to cancel out the residual fields of the kick for the re-circulating bunches was conceived, and pre/post kickers were introduced to flatten the RF curvature of harmonic kick on the exchanged bunches, preventing longitudinal profiles of angular distribution from bending into a ``banana" shape. Subsequently these ideas were demonstrated by the numerical simulation studies using the particle tracking code ELEGANT~\cite{ele} based on a simple model of the kick fields, where the only non-trivial component of the kick is in the kick direction with the spatial profile being transversely uniform and longitudinally $\delta$ function-like or ``impulsive", i.e., $\vec{F}_L=eV_k(t)\delta(z)\hat{x}$ ( $\vec{F}_L$ is the Lorentz force as a kick acting on an electron, $e$ is an electron charge, $V_k$ is a kick voltage, $\hat{x}$ is the kick direction-\textemdash regardless of any physical direction and might be vertical\textemdash and $z$ is a longitudinal coordinate with the origin at the cavity center). The analysis of the beam propagation in \cite{huang1} was limited to a non-magnetized beam. The current work is an adaptation of the aforementioned basic approach to a beam dynamics with updated parameters (See Table\,\ref{table:bp}) of the CCR, with improvement on the number of modes for the kick profile to the 5 modes within a single cavity~\cite{gpark1}. Furthermore, a kick model is now generalized to resemble the realistic kick field of the kicker cavity. The 3D field map of the QWR can be obtained from the RF simulations~\cite{gpark2} using the 3D FEA code CST-MWS~\cite{cst}, although its direct application to analysis is impractical\textemdash not applicable in the ELEGANT\textemdash and therefore the kick model that appropriately approximates the profile needs to be introduced. In contract to the aforementioned simple kick model, the realistic kick fields are not transversely uniform, nor are they ``impulsive". Through the transversely non-uniform fields, the electrons at different offsets see different kick voltages, leading to serious degradation of beam dynamics since the cancellation scheme is not effective anymore. As long as the beam trajectory remains flat, i.e., at zero-slope, throughout the effective range of the kick fields, a phase space transform with non-zero offsets can be systematically described by a multipole expansion of the RF fields near the beam axis analogous to expansions of static magnet fields, as was first done in~\cite{abell},\cite{barranco},\cite{garcia}. Then the multipole fields can be implemented into ELEGANT as beamline elements. We will present a modified configuration of a kicker system that cancels the multipole contributions in the residual kicks, as demonstrated by the ELEGANT simulations. On the other hand, the longitudinally extended profile of the kick fields, which closely resembles the pseudo-Gaussian profiles, allows the offsets of the beam to evolve over the effective field range. This evolution becomes more evident with the magnetized beam whose canonical angular momentum (CAM) defines non-trivial transverse slopes. Moreover, this evolution can not be cancelled by any cancellation scheme. We will show that the accumulated offset evolution reduces the magnetization and increases the Larmor emittance of the beam, leading to decreased the cooling efficiency. In the ELEGANT simulation, the extended kick profile is modeled as a series of impulsive kicks over the effective field range and the resulting Larmor emittance increase is shown to be smaller than 19\,mm\,mrad, the tolerance limit for the efficient cooling. 

\section{The baseline design of a harmonic kicker system \label{section:baseline}}
\par In this section, the baseline design of a harmonic kicker system is presented based on a simplified model of the CCR beam dynamics: the non-magnetized electron beam with a simple kick model. First we present the design principle of the kicker that leads to the implementation of the aforementioned ideas in a single cavity with 5 harmonic modes. Then a few important beam parameters are analytically computed and finally a proof of principle using the ELEGANT simulations is demonstrated. The beam dynamics based on a more realistic kick model will be discussed in the following sections.

\subsection{Harmonic kick design \label{subsec:baseline_p}} 
\par
\begin{figure}[htb]
   \centering
   \includegraphics*[width=90mm]{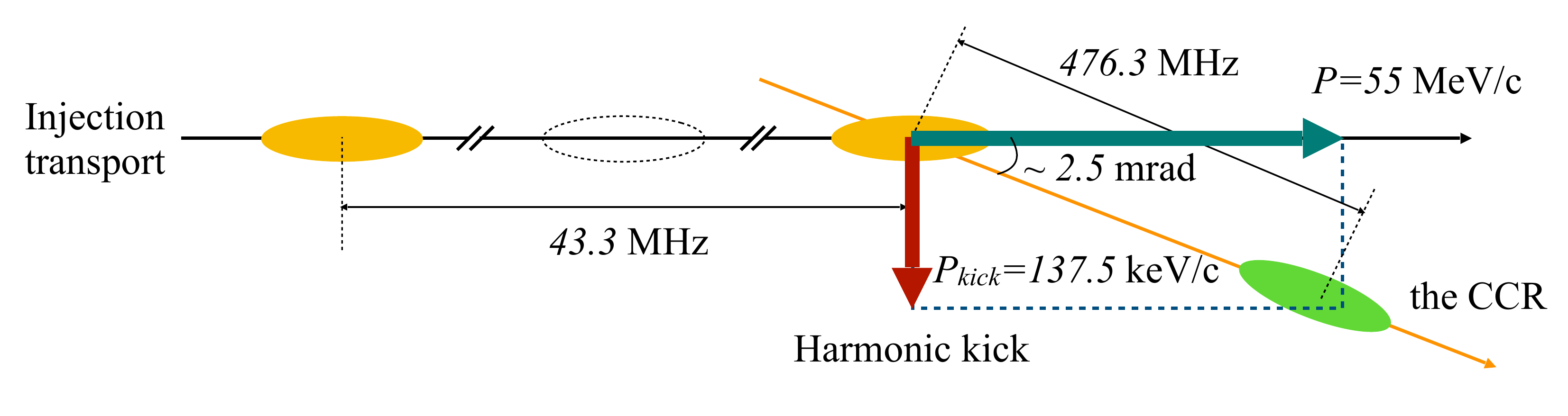}
   \caption{A schematic view of the kicker. The black arrowed line is injection transport, while the yellow line is the CCR. The kick is represented by a big red downward arrow. The yellow ellipses are injected bunches, while the green ellipse is a re-circulating bunch in the CCR. The dotted transparent ellipse is an empty bucket reserved for the upgrade.}
   \label{fig:schematic_kick}
\end{figure}

 The schematic view of the kicks on electron bunches in the injection scheme (the extraction scheme is analogous) is shown in Fig.\,\ref{fig:schematic_kick}. The injected bunch train at $f_k=43.3$\,MHz merges with the re-circulating bunches at $f_b=476.3$\,MHz via a vertically deflecting kick at the crosspoint. The $N_b=11$ bunches pass through the kicker in a single kick period $1/f_k$. Each bunch is indexed by $m=0,1,\cdots10$: $m=0$ is the injected bunch and the others ($m\neq0$) are re-circulating bunches. All the particles within the bunches are assumed to move at $\vec{v}=c\hat{z}$ ($c$ is speed of light). Given beam parameters from Table\,\ref{table:bp}, simple trigonometry in Fig.\,\ref{fig:schematic_kick} gives a good estimate for a kick voltage with $P_{kick}\approx \theta P\approx 137.5$\,keV/c for the given kick angle of $\theta\sim2.5$\,mrad. Furthermore, the kick must be applied on the electron bunches selectively, i.e., delivered on injected bunches only at kick frequency $f_k$ and not affect beam dynamics of the re-circulating bunches. This requires a temporal profile of a kick to be sharply peaked at kick frequency $f_k$ and drop to negligible value within $t_b=1/f_b=2.1$\,ns.

\par
\begin{figure}[hbt]
  \centering
    \subfigure[The temporal profile of a harmonic kick. The red dots correspond to electron bunches arrriving at the kicker every 2.1 ns. While the kick frequency is $86.6$\,MHz, the electron bunch in current scheme is injected into the CCR at a frequency of $43.3$\,MHz. We have introduced a DC kick to uniformly reduce the residual kicks to zero.]{\label{fig:hk-profile}\includegraphics[scale=0.42]{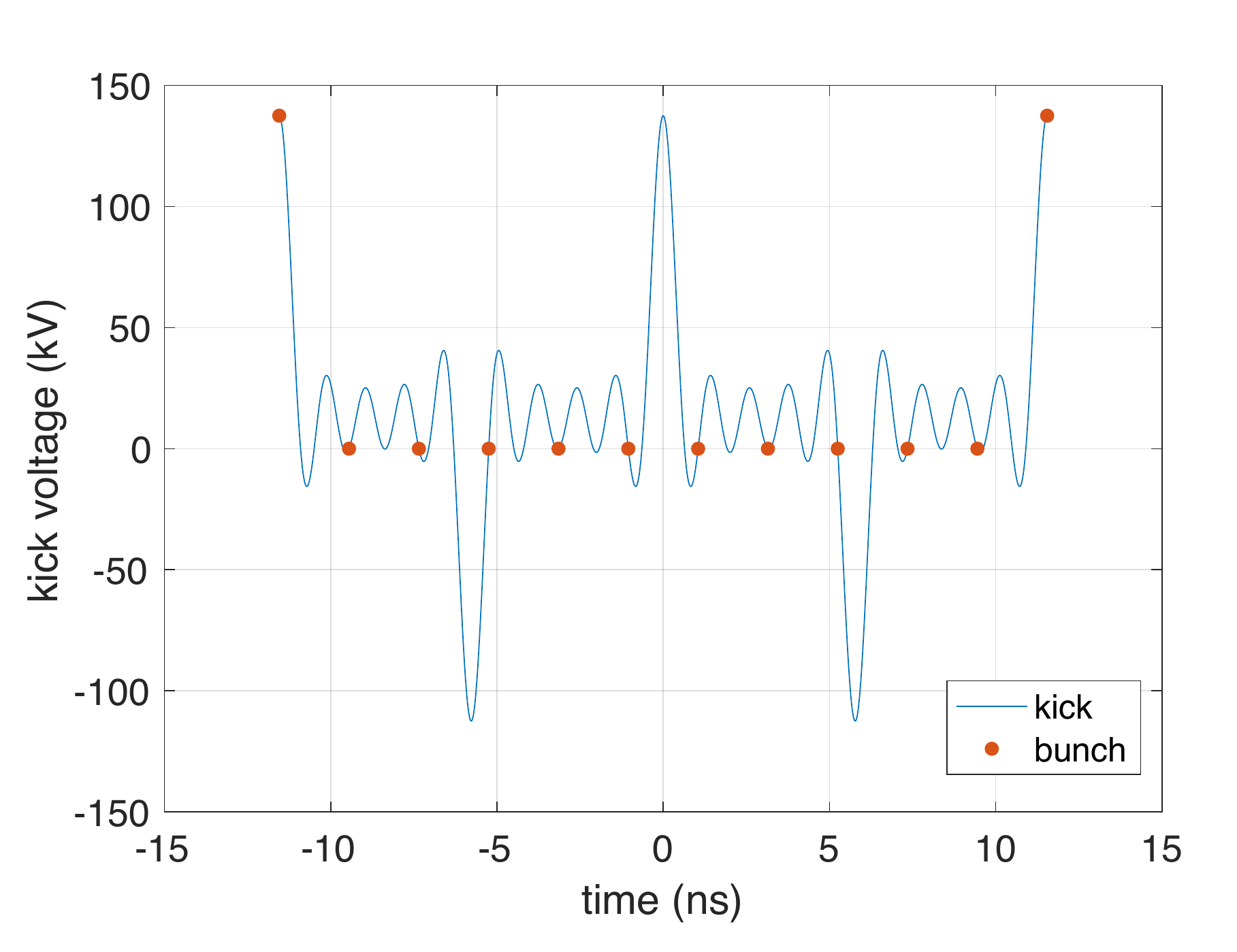}}\\
    \subfigure[The temporal profile of kick after addition of pre-/post-kick.]{\label{fig:PPK-profile}\includegraphics[scale=0.42]{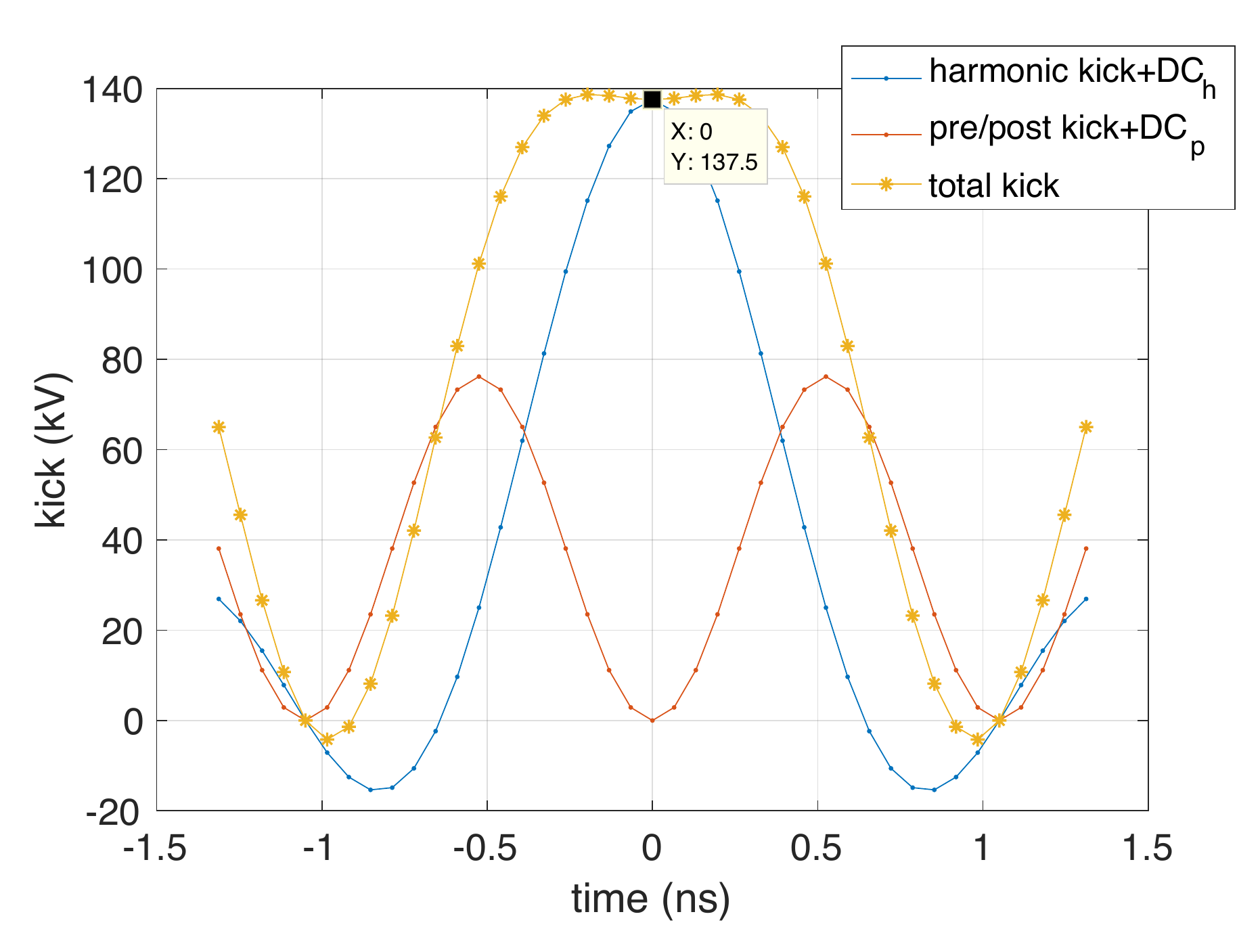}}
  \caption{The profiles of the designed kicks.}
  \label{fig:bd}
\end{figure}

 The required profile can be achieved using a harmonic kick, which is defined as a linear combination of harmonic modes with base frequency $f_k$. First, to define the relevant quantities more precisely, the coordinate system we will use is set up as follows. The longitudinal coordinate along the beam line is denoted by either $s$ or $z$ and the transverse coordinates by $x$ (vertical) and $y$ (horizontal), respectively. The origin $s=0$ is at the cavity center. The time the reference particle in the $m$th bunch arrives at the cavity center is set to be $t=m/f_b$. Consider a generic charged particle in the $m$th bunch that arrives at the cavity center at $t=m/f_b+\tau$. Then the trajectory is $z(t)=c(t-m/f_b-\tau)$. The fractional energy of the particle is $\delta=E/E_r-1$, with $E$ and $E_r$ being the energy of the particle and the reference particle, respectively. Next we assume the simple kick model used for the baseline design, i.e., $\vec{F}_L=eV_k\delta(z)\hat{x}$. Then the relevant equation of motion for the particle is given by
\begin{eqnarray}
\frac{dp_x}{dt}=F_x(z,t)=\sum_{n=0}^N\mathfrak{Re}\Bigg[eV_n\delta(z)e^{-i(\omega_n t+\phi_n)}\Bigg],\quad\quad
\label{eqn:eom0}
\end{eqnarray}
while dynamics in $y$ direction is trivial. In the second equality, the Lorentz force is given as a harmonic kick, i.e., a linear combination of $N$ harmonic modes (plus one DC mode) with each mode written as the product of the spatial (longitudinal) profile $V_n\delta(z)$ and the temporal (harmonic) profile. Also $\omega_n$ ($n>0$, with $\omega_0=0$) and $N$ are the angular frequency of the $n$th mode and the number of harmonic modes, respectively. The RF phase $\phi_n$ of the kicker is chosen so that the kick reaches the peak value at $t=0$ (on-crest phase), i.e., $\phi_n=0$ for all $n\geq 0$. Changing the independent variable from $t$ to $s$ and integrate the kick field over the cavity length, the kick voltage delivered by a harmonic kick is given as
 \begin{flalign}
  &V_x\left(\frac{m}{f_b}+\tau\right)=V_0\nonumber\\
  &+\sum_{n=1}^N V_n\cos{\Bigg\{2\pi (2n-1)f_k\left(\frac{m}{f_b}+\tau\right)\Bigg\}},\quad \quad
  \label{eqn:hk0}
 \end{flalign}
where $n=0$ refers to the static mode provided by the external DC magnet. Now let us apply the general formula (\ref{eqn:hk0}) to the requirements for the harmonic kickers. Whenever a reference particle in each exchanged bunch arrives at the kicker cavity, i.e., $\tau=0$ and $m=0$, the kick must deliver a kick voltage $V^{kick}$:
 \begin{eqnarray}
V_x\left(0\right)=V_0+\sum_{n=1}^N V_n=V^{kick}=137.5\,\hbox{kV}.\quad\quad
\label{eqn:voltage_sum0}
\end{eqnarray}
 In (\ref{eqn:hk0}), a set of constraints arise from the practical implementation of the kicker. In practice, a physical kicker cavity can accommodate only a few modes because of the kicker's power-to-deflecting voltage efficiency and RF wave controlling issues. Thus $N$ must be truncated at a number less than 10. In case a harmonic kicker is based on the quarter wave resonator (QWR), only odd harmonics appears in (\ref{eqn:hk0}) with $\omega_n=2\pi(2n-1)f_k$ because all the resonant frequencies supported by the QWR that has a coaxial structure with its closed end electrically shorted are odd harmonics. With the superposition of a few modes, the temporal profile of the kick would inevitably have ``ripples", i.e., non-zero residual kicks on the recirculating bunches between the kicks (For example, see Fig.\,\ref{fig:hk-profile}). These residual kicks could degrade the beam quality and should be minimized. More precisely, the kick should be close to zero with its slope also close to zero at temporal location of each recirculating bunch: from (\ref{eqn:hk0}), we must have for $m$th recirculating bunch ($m=1,2,\cdots, N_b-1$)
 \begin{eqnarray}
  V_x\left(\frac{m}{f_b}\right)=V_0+\sum_{n=1}^N V_n\cos{\Bigg\{2\pi (2n-1)f_km/f_b\Bigg\}}=0.\quad \quad
  \label{eqn:voltage_sum}
 \end{eqnarray}
 For stability of the kick with respect to arrival time (to the kicker) jitter, RF control errors, and the extent over the bunch length, the kick voltage slope must be close to zero as well
\begin{eqnarray}
  V'_x\left(\frac{m}{f_b}\right)=-\sum_{n=1}^N V_n 2\pi (2n-1) f_k\times\nonumber\\
  \sin{\Bigg\{2\pi (2n-1)f_km/f_b\Bigg\}}=0.
  \label{eqn:slope_sum}
\end{eqnarray}
 Note that the constraints defined in equations in (\ref{eqn:voltage_sum}), (\ref{eqn:slope_sum}), are not all independent due to the symmetry of trigonometric functions:
\begin{eqnarray}
\cos{\Bigg[2\pi (2n-1)m \frac{f_k}{f_b}\Bigg]}=\cos{\Bigg[2\pi (2n-1) (N_b-m)\frac{f_k}{f_b}\Bigg]},\quad \quad \\
\sin{\Bigg[2\pi (2n-1)m \frac{f_k}{f_b}\Bigg]}=-\sin{\Bigg[2\pi (2n-1) (N_b-m)\frac{f_k}{f_b}\Bigg]}.\quad \quad 
\end{eqnarray}
\par This reduces the number of the constraints in (\ref{eqn:voltage_sum}) and (\ref{eqn:slope_sum}) to $(N_b-1)/2$ $(N_b/2)$ for odd (even) $N_b$, i.e., the range of $m$ can reduce to $m=1,\cdots, 5$ for $N_b=11$. The total number of the constraints that includes the kick condition (\ref{eqn:voltage_sum0}) is then $11$. On the other hand, the effect of small residual kicks and their slopes on the re-circulating bunches can be minimized if a pair of the kickers is implemented into the CCR, i.e., one kicker (called injection kicker) for injection and the other kicker (called extraction kicker) for extraction as first introduced in \cite{huang1} (see Fig.\,\ref{fig:CCR}). In this 2-kicker system, the relative RF phase between the two kickers are set to zero 
so that the momentum changes due to the kickers are the same\textemdash with an impulsive kick model, offset changes are all zero. Moreover, the kickers are separated by a betatron phase advance of $\pi$, so the phase space variables transform as $x\rightarrow -x, x'\rightarrow -x'$. Then a straightforward computation of the overall momentum change shows that the residual extraction kick is cancelled by the corresponding injection kick and the beam dynamics requirements (\ref{eqn:slope_sum}) is satisfied via the two-kicker system without having to be imposed on the individual kicker. This leaves only the requirements (\ref{eqn:voltage_sum0}) and (\ref{eqn:voltage_sum}) relevant for profile construction with the total number of constraints reduced to 6. A system of requirements (\ref{eqn:voltage_sum0}) and (\ref{eqn:voltage_sum}) becomes critically determined with the non-zero DC mode and only 5 odd harmonic modes ($n=1,\cdots,5$), which can be easily accommodated in a single quarter wave resonator. The constraint (\ref{eqn:voltage_sum}) can be written as a $5\times 5$ matrix equation for the kicker voltages $V_n$ with $n=1,\cdots 5$
 \begin{eqnarray}
 \label{eqn:voltage_DC}
 V_0\textbf{1}+\textbf{M}\textbf{V}=\textbf{0},\qquad \qquad \\
\text{where } \textbf{1}=\left[\begin{array}{c}1\\\vdots\\1\end{array}\right], \quad \textbf{V}=\left[\begin{array}{c}V_1\\\vdots\\V_5\end{array}\right],\quad \textbf{0}=\left[\begin{array}{c}0\\\vdots\\0\end{array}\right],\quad \quad \nonumber\\
\quad \textbf{M}=\left[\begin{array}{ccc}\\& \mathcal{M}_{mn} & \\\\\end{array}\right],\;\mathcal{M}_{mn}=\cos{\Bigg\{2\pi nf_km/f_b+\phi_n\Bigg\}}.\quad\quad 
 \end{eqnarray}
 where the explicit computation of the rank of $\textbf{M}$ shows the matrix is non-singular. By solving equation (\ref{eqn:voltage_DC}) for $\textbf{V}$ with the inverse of $\textbf{M}$, we have $V_n=2V_0$ for $n=1,\cdots,5$, and combining with the constraint (\ref{eqn:voltage_sum0}), we obtain equal amplitude solutions, i.e., $V_n=25$\,kV and $V_0=12.5$\,kV (See Table\,\ref{table:fom}). The corresponding temporal profile is shown in Fig.\,\ref{fig:hk-profile}.

  \begin{table}[hbt]
   \centering
   \caption{Figures of merit for a harmonic kicker cavity. The middle box is for the harmonic kicker and the bottom box is for the pre-/post kicker. $n_h$ is harmonic number, $f$ is frequency, $V$ is kick voltage amplitude and $\phi$ is RF phase. $P_{wall}$ is wall loss, $Q_0$ is unloaded quality factor, $R_{sh, \perp}$ is the transverse shunt impedance of the harmonic kicker. DC$_h$ and DC$_p$ refer to DC magnet field associated with harmonic kicker and pre-/post-kicker and kick voltage translates to $0.1$ and $0.3$\,mT for a $0.4$\,m long magnet, respectively.}
   \vspace{5pt}
   \begin{tabular}{ccccccc}
       \toprule
              Modes & $f$& $V$& $\phi$ & $P_{wall}$ & $Q_0$  & $R_{sh, \perp}$ \\
       & MHz & kV & rad & kW & $-$  & M$\Omega$ \\
       \hline
       1 &$86.6$ & $25$& 0 & $0.43$ & $5785$ & $1.44$ \\
       2 & $259.8$ & $25$ & 0 & $0.80$ & $10026$ & $0.78$ \\
       3 &$433$ & $25$& 0 & $1.42$ & $13043$ & $0.44$ \\
       4 & $606.2$& $25$ &0 & $1.22$ & $15540$ & $0.51$ \\
       5 &$779.4$ & $25$ & 0 & $2.54$& $17452$ & $0.24$ \\
       0 & DC$_h$ &$12.5$ & - & - & - & - \\
       Total &  - &  $137.5$ & - &6.4 &  - & - \\
        \hline
        $0$  &  DC$_p$ & $38.1$  & - & - & - &- \\    		
        $11$ &  $952.6$ & $38.1$ & 0  \\
       \botrule
   \end{tabular}
   \label{table:fom}
\end{table}

\par Finally, we extend beam dynamics requirements on the kicker scheme\textemdash as determined by (\ref{eqn:voltage_sum0}),(\ref{eqn:voltage_sum}), and (\ref{eqn:slope_sum}) with respect to a reference particle\textemdash to the bunches. In particular with an exchanged bunch, the kick voltage in (\ref{eqn:hk0}) gained by a particle lagging the reference particle behind by $\Delta t=\tau$ would be less than $V_{kick}$ by a factor of $C_{RF}(\tau)=\sum_nV _n\cos{\omega_n\tau}/V^{kick}$, which is called the RF curvature term and shown in Fig.\,\ref{fig:PPK-profile}. Consequently, the temporal profile of harmonic kick around the reference particle will be ``imprinted" on the (vertical) angular distribution of the bunch over the bunch length, leading to a banana-shaped profile. This will result in a significant increase in the vertical normalized emittance and a reduction of the cooling rate. Moreover, the imprinted curvature at the first entry will persists through the cancellation scheme over the passes. To remove the curvature, a pre-kicker whose own RF curvature is designed to flatten the total curvature is introduced before the injection kicker. Similarly, in the case where the normalized emittance of the beam must be maintained small in the ERL, a post-kicker can be introduced in an extraction transport after the extraction kicker. A pre/post kicks are single frequency kickers designed so that the RF curvature effects of the injection/extraction kick are largely compensated as follows. Given that the betatron phase advance between the pre/post kicker and the harmonic kicker is set to be $\pi$, the total kick $V_{tot}$ as a combination of harmonic kick and pre-/post kick (which is identified as 6th mode) is written as
\begin{eqnarray}
V_{tot}=\sum_{n=0}^5V_n\cos{\omega_n \tau}-V_6\cos{\omega_6\tau}.
\label{eqn:Vtot}
\end{eqnarray}
From the parabola-shaped profile of the injection kick based on 5 harmonic modes in Fig.\,\ref{fig:PPK-profile}, we assume this total kick profile is also a parabola centered at the origin that can be described as $V_{tot}=\alpha \tau^2+V^{kick}$ for some constant $\alpha$. Then $\alpha$ is obtained by double differentiation with respect to time at origin:
\begin{eqnarray}
\alpha=\frac{1}{2}\frac{d^2}{d\tau^2}V_{tot}\Bigg|_{\tau=0}=-\frac{1}{2}\left(\sum_{n=1}^5V_n\omega_n^2-V_6\omega_6^2\right).
\end{eqnarray}
Now the smaller $\alpha$ is, the closer to a flat line the total kick profile becomes. For example $\alpha=0$ would be obtained by condition
\begin{eqnarray}
V_6=\frac{1}{\omega_6^2}\sum_{n=1}^5V_n\omega_n^2.
\end{eqnarray}
Although we could obtain a pre-kick amplitude for a more general harmonic kick by numerically adding up at this stage, we focus on the equal amplitude option, i.e., $V_n=V_h$ for $n=1,2,\cdots,5$:
\begin{eqnarray}
\frac{V_6}{V_h}=\left(\frac{f_k}{f_6}\right)^2\sum_{n=1}^5(2n-1)^2=165\left(\frac{f_k}{f_6}\right)^2.\quad \quad 
\label{eqn:pre_kick0}
\end{eqnarray}
Up to equation (\ref{eqn:pre_kick0}), we still have the freedom to choose $f_6$ and $V_6$, but a more practical choice would be a single-frequency $952.6$\,MHz QWR, which is the 6th odd harmonic of $f_k$.  Then $f_6=(2\times 6-1)f_k$ and $V_6/V_h=15/11$, which leads to $V_6=34.1$\,kV. Numerically adjusting for a flat field profile over the largest range possible shows that one can obtain a longer flat interval when a larger error is used. To obtain a flat interval of $\pm6l$ ($l$ is a bunch length in Table\,\ref{table:bp}), a slightly higher amplitude $V_6=38.1$\,kV will flatten the curve better\textemdash while adding the higher order terms in $V_{tot}=-\alpha t^2$. The specification of the pre/post kickers is listed in Table\,\ref{table:fom}. Because the pre/post kick is applied in the opposite direction to the harmonic kicks, its maximum amplitude $V_6$ must be compensated by an additional DC magnet of equal strength (see (\ref{eqn:Vtot}) equated to the expression below at $\tau=0$ in particular). The temporal profile of the RF fields with the pre-kicker introduced is shown to be flat in Fig.\,\ref{fig:PPK-profile}. 

\subsection{ Analytical description of the beam parameter change in the kicker: simple kick model \label{subsec:baseline_anal}} 
\par In this subsection, we focus on a specific kick model that will be used in the ELEGANT simulation and give an analytical description for some of the beam parameters. The longitudinal coordinate of an electron trailing the reference electron by time $\tau$ is $z(t)=c(t-\tau)$ and equation (\ref{eqn:eom0}) can be re-written as
\begin{eqnarray}
\Delta p_x=\sum_{n=1}^N \mathfrak{Re}\Bigg[\int_{-l/2c}^{l/2c}\!dt\,\mathcal{F}_n(c(t-\tau))e^{-i\omega_n t}\Bigg],
\label{eqn:delta_px}
\end{eqnarray}
where $l$ is an effective range of the kick fields. If the kick is given in an impulsive kick model, i.e., $\mathcal{F}_n(c(t-\tau))=eV_n\delta(c(t-\tau))$, where $V_n$ is a normalization constant that will be identified as the kick voltage of each mode, then (\ref{eqn:delta_px}) is straightforward to integrate:
\begin{eqnarray}
=\frac{e}{c}\sum_{n=1}^N\mathfrak{Re}\Bigg[V_ne^{i\omega_n  \tau}\Bigg].
\label{eqn:px_simple}
\end{eqnarray}
If $V_n=V_h$ for all $n$ in equal amplitude option, then (\ref{eqn:px_simple}) reduces to 
\begin{eqnarray}
=\frac{eV_h}{c}Re\Bigg[\frac{1-e^{2iN\omega_k\tau}}{1-e^{2i\omega_k\tau}}e^{i\omega_k\tau}\Bigg]\nonumber\\
=\frac{eV_h}{c}\frac{\sin{(\omega_k\tau)}\sin{(2N\omega_k\tau})}{1-\cos{(2\omega_k\tau})}=\frac{eV_h}{2c}\frac{\sin{(2N\omega_k\tau})}{\sin{(\omega_k\tau)}}.
\label{eqn:delta-kick}
\end{eqnarray}
(\ref{eqn:delta-kick}) is a relevant model for the ELEGANT simulation, where the implementation of the kick is limited to the impulsive kick model as its default option for a harmonic kicker.
\par Now we can compute the changes in some of the beam parameters. First, the motion of bunch is obtained by integrating over the bunch length weighted by longitudinal distribution function. With a top-hat distribution, the momentum change of the bunch whose reference electron arrives at $t=m/f_b$, $m=0,1,\cdots,11$ is given as
\begin{eqnarray}
\Delta \mathcal{P}_x\left(\frac{m}{f_b}\right)=\frac{eV_h}{c\tau_b}\sum_{n=1}^N\mathfrak{Re}\Bigg\{\int_{m/f_b-\tau_b/2}^{m/f_b+\tau_b/2}\!d\tau\,e^{i\omega_n  \tau}\Bigg\}\nonumber\\
=\frac{2eV_h}{c\tau_b}\sum_{n=1}^N\Bigg\{\frac{\cos{(\omega_nm/f_b)}}{\omega_n}\sin{\omega_n\tau_b/2}\Bigg\},
\end{eqnarray}
which reduces to $eV_h/c\sum_{n=1}^N\cos{(\omega_nm/f_b)}$ as $\tau_b \rightarrow 0$. In particular with $m=0$, $\Delta\mathcal{P}_x(0)\rightarrow NeV_h/c$, which is the design kick voltage. Secondly, 
\begin{eqnarray}
\label{eqn:p_sq}
\Delta \mathcal{P}_x^2\left(\frac{m}{f_b}\right)=\mathcal{N}\sum_{n,l=1}^N\int_{m/f_b-\tau_b/2}^{m/f_b+\tau_b/2}\!d\tau\,\cos{\omega_n  \tau}\cos{\omega_l \tau}\nonumber\\
=\mathcal{N}\sum_{n,l=1}^N\Bigg\{\frac{1}{\omega_-}\sin{\frac{\omega_-\tau_b}{2}}\cos{\frac{\omega_-m}{f_b}}+(\omega_-\rightarrow \omega_+)\Bigg\},\qquad\\
\text{where }\mathcal{N}=\frac{1}{\tau_b}\left(\frac{eV_h}{c}\right)^2, \quad \omega_{\pm}=\omega_n\pm\omega_l.\quad \quad 
\end{eqnarray}

Finally, the change in emittance can be computed perturbatively for a small momentum change as
\begin{eqnarray}
\delta \varepsilon=\frac{1}{2\varepsilon_0}\Big\{\langle x_0^2\rangle \langle (\delta x')^2 \rangle-2\langle x_0x_0'\rangle\langle x_0\delta x'\rangle\Big\}\nonumber\\
=\frac{1}{2\varepsilon_0}\langle x_0^2\rangle \langle (\delta x')^2\rangle=\frac{\langle x_0^2\rangle}{2\varepsilon_0m^2c^2\gamma^2}\Delta \mathcal{P}_x^2\left(\frac{m}{f_b}\right),
\end{eqnarray}
where $\alpha_0=\langle x_0x_0'\rangle=0$ at the kicker location has been used in the second equality. The last equality is obtained using $p_x=mc\gamma x'$ and (\ref{eqn:p_sq}). 

\subsection{A simulation study}
\par
\begin{figure}[hbt]
   \centering
   \includegraphics*[width=75mm]{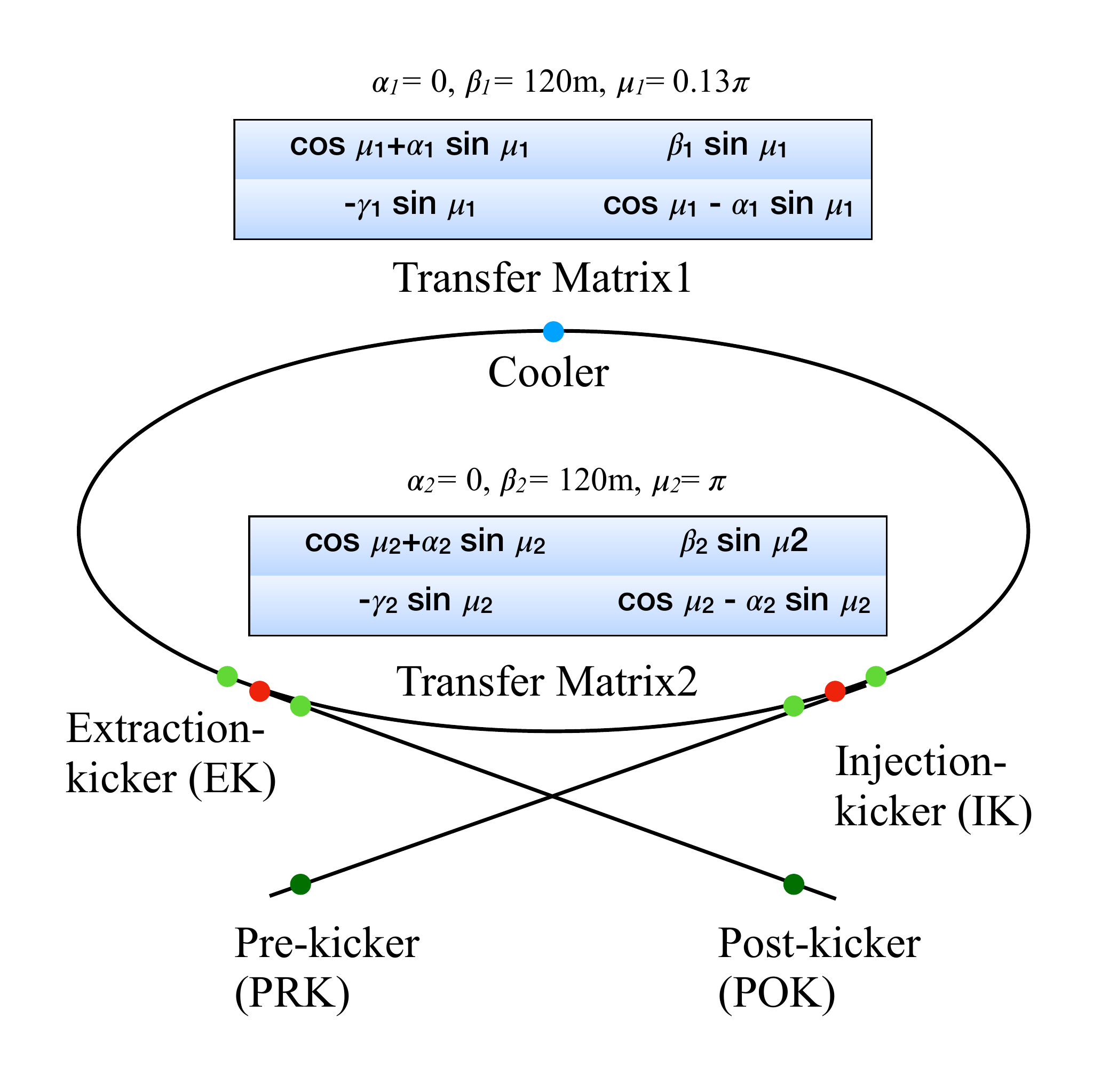}
   \caption{Schematic view of the CCR in the ELEGANT simulation. The red dots are the injection/extraction kickers, the dark green dots are pre/post kickers, the blue dot is the cooler, and the green dots are watch points. The beam line is represented by a pair of transfer matrices shown here.}
   \label{fig:el_sch}
\end{figure}

 The design of the kicker system in the CCR has been verified by simulations using  ELEGANT, and it was shown that the designed harmonic kickers can kick the bunches in/out of the CCR without degrading the beam quality for 11 turns. The simulation was done with the simplest setup, whose schematic is shown in Fig.\,\ref{fig:el_sch}. In Fig.\,\ref{fig:el_sch}, the circumference $l$ of the CCR was set to be $l=c/f_b=0.63$\,m for simplicity (with the real circumference being its multiple) so that a single bunch of electrons circulate each pass in $t=1/f_b=2.1$\,ns. Therefore, the bunch shots recorded in the monitor at the kickers can be viewed both as a bunch circulating 11\,turns of the CCR or equivalently 11 consecutive bunches passing the kickers. The beam line elements were represented by a pair of transfer matrices (see Fig.\,\ref{fig:el_sch}), with the matrix\,2 being $-1\times I$ ($I=$ identity matrix), corresponding to betatron phase advance of $\pi$. 
\par
\begin{figure*}[hbt]
  \centering
    \subfigure[\,Longitudinal profile of the angular divergence, $x_p$, for a circulating electron bunch at the cooler (without pre-kicker).]{\label{fig:C}\includegraphics[scale=0.22]{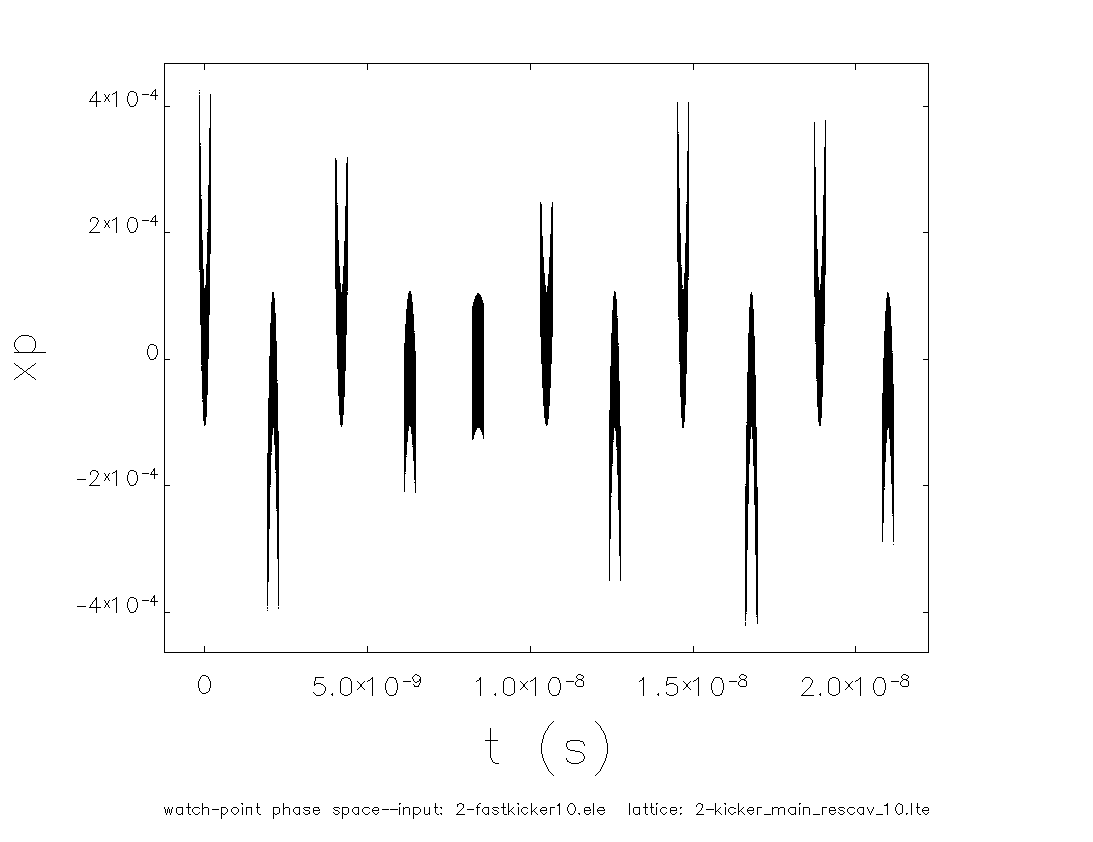}}%
    ~
    \subfigure[\,Same as in (a) but with pre-kicker.]{\label{fig:C-PK}\includegraphics[scale=0.22]{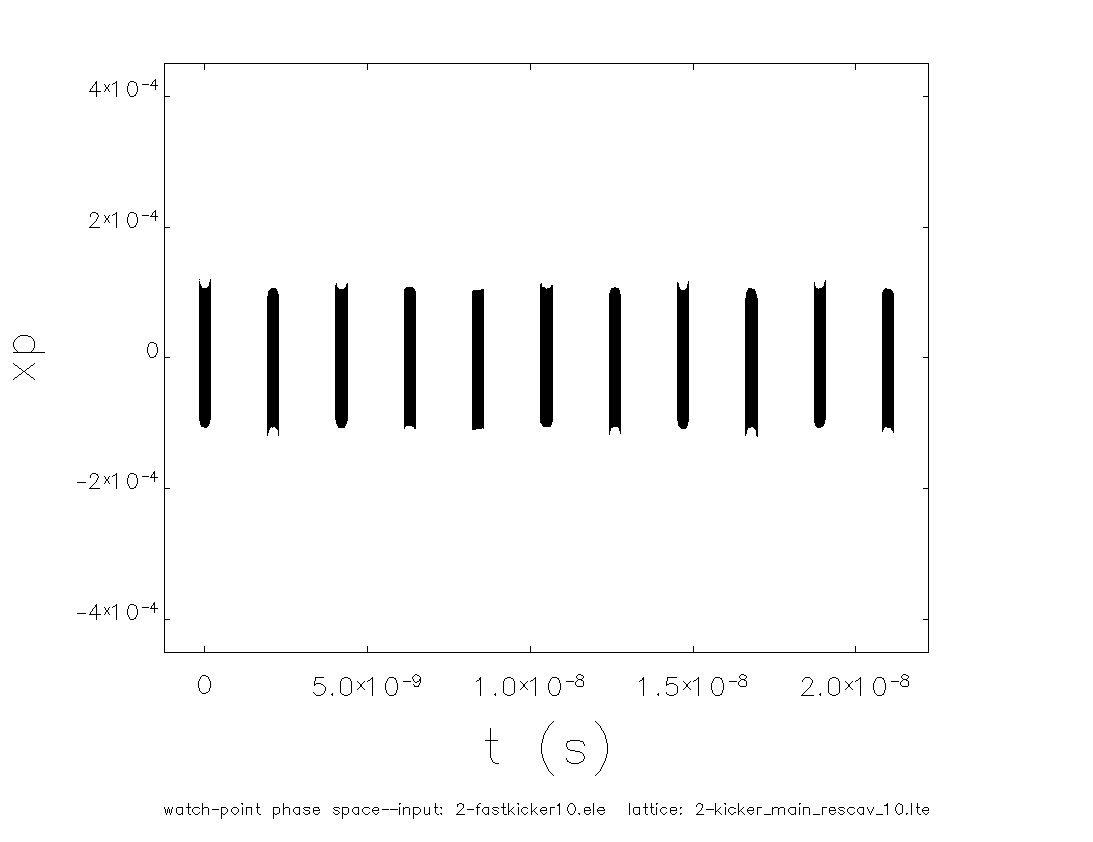}}\\
    \subfigure[\,Same as in (a) but just before the injection kicker.]{\label{fig:RCS}\includegraphics[scale=0.22]{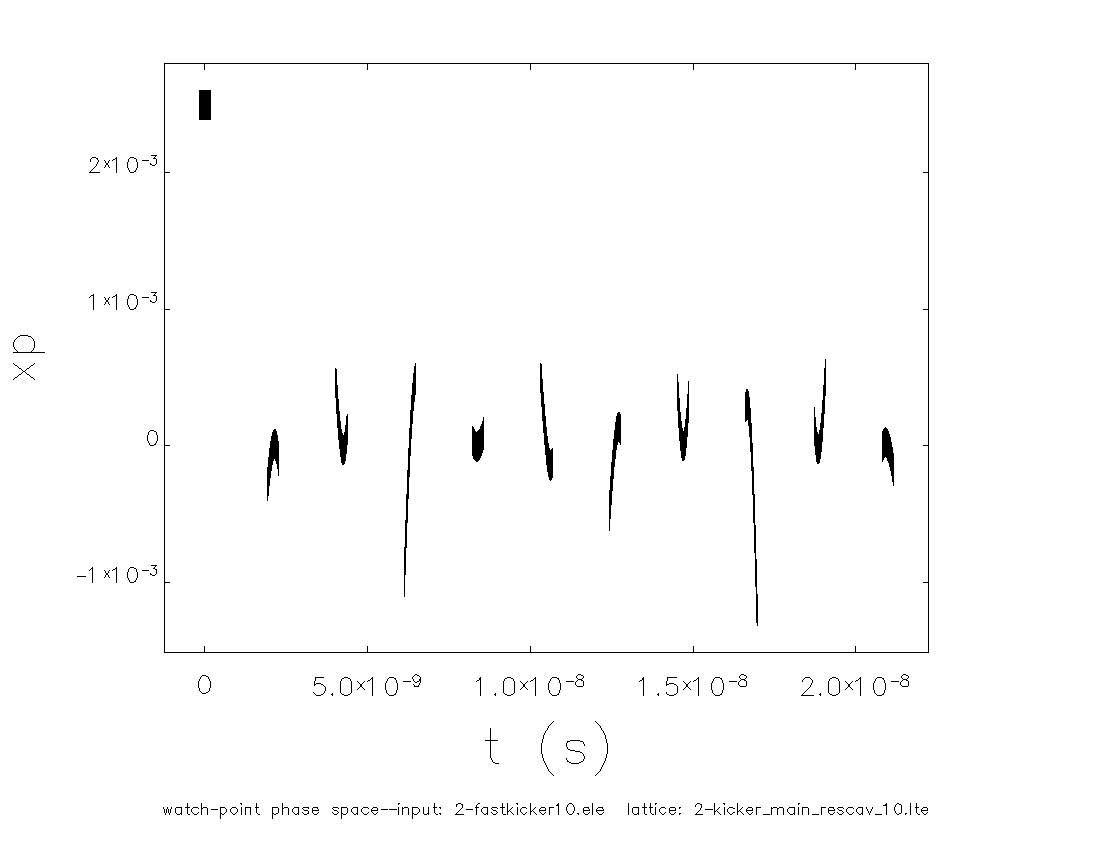}}%
    ~
    \subfigure[\,Same as in (c) but with-pre-kicker. The curvature of the first bunch is due to the prekick.]{\label{fig:RCS-PK}\includegraphics[scale=0.22]{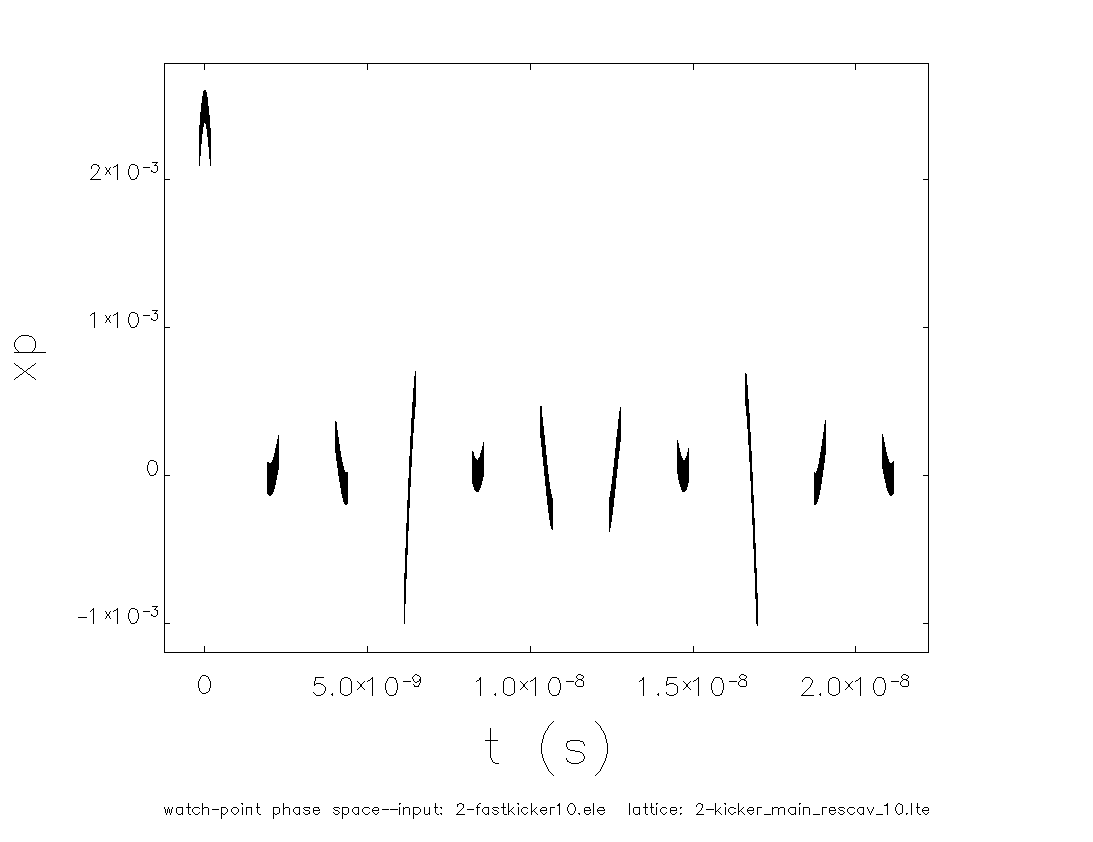}}
    \vspace{-10pt}
  \caption{Longitudinal profiles of an electron bunch circulating $11$ turns: at the CCR (top) and on the re-circulation transport before injection kick (bottom) with (left) and without (right) the use of a pre-kicker.}
  \label{fig:bd}
\end{figure*}

\begin{figure*}[hbt]
  \centering
    \subfigure[\,Emittance change without pre-/post-kickers. $x=1$ is the in-kicker (entrance) position, $x=8$ is the cooler position, and $x=11$ is the out-kicker (entrance) position.]{\label{fig:emittance_baseline_00}\includegraphics[scale=0.41]{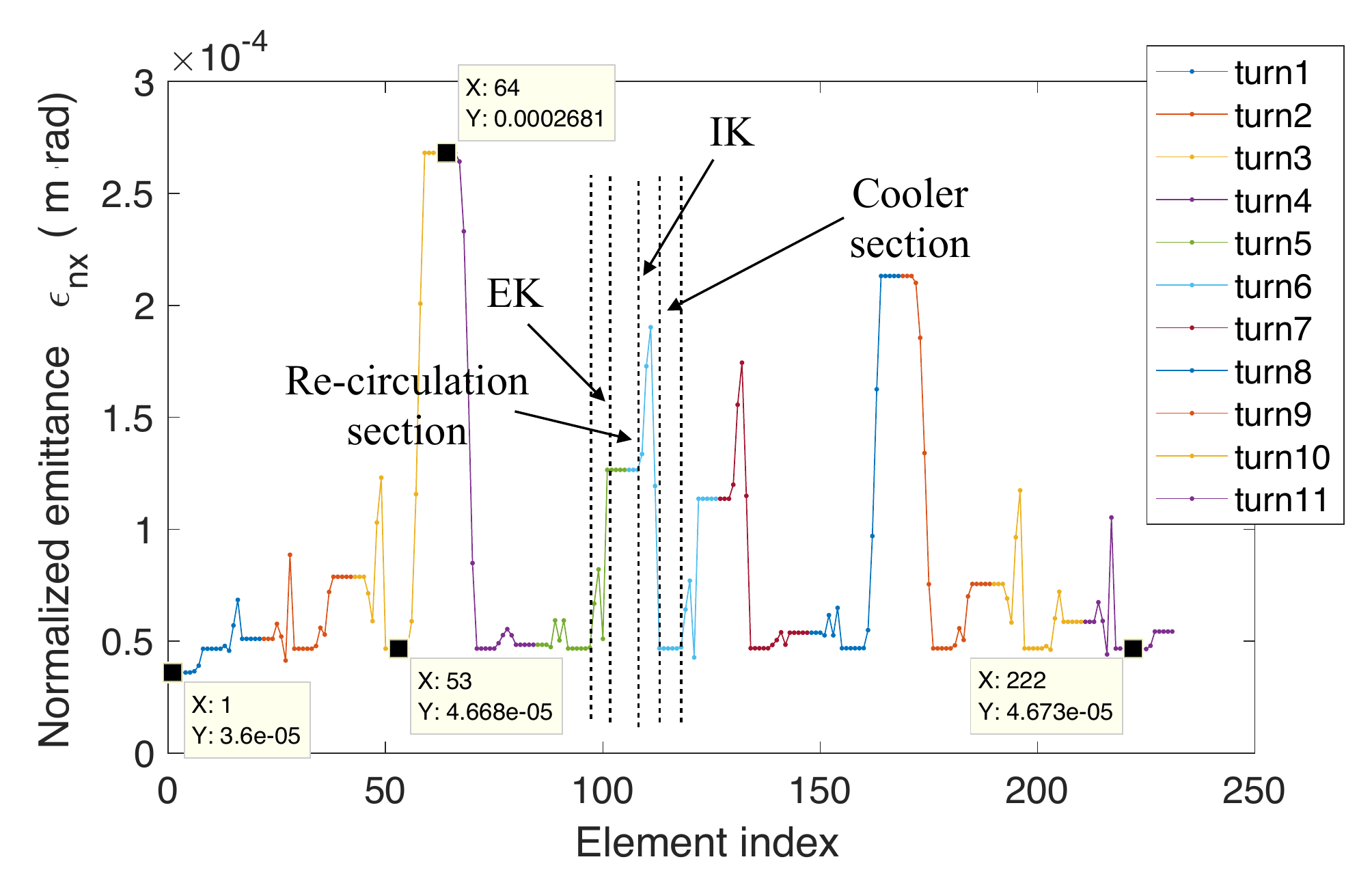}}%
~
    \subfigure[\,Emittance change with pre-/post-kickers. $x=1$ is in-kicker (entrance) position, $x=11$ is the cooler position, and $x=13$ is the out-kicker (entrance) position. The beamline in the simulation includes four extra elements per turn: pre-/post-kicker with the associated DC magnets, although they are turned off except for the first/last turn.]{\label{fig:emittance_baseline_PP}\includegraphics[scale=0.41]{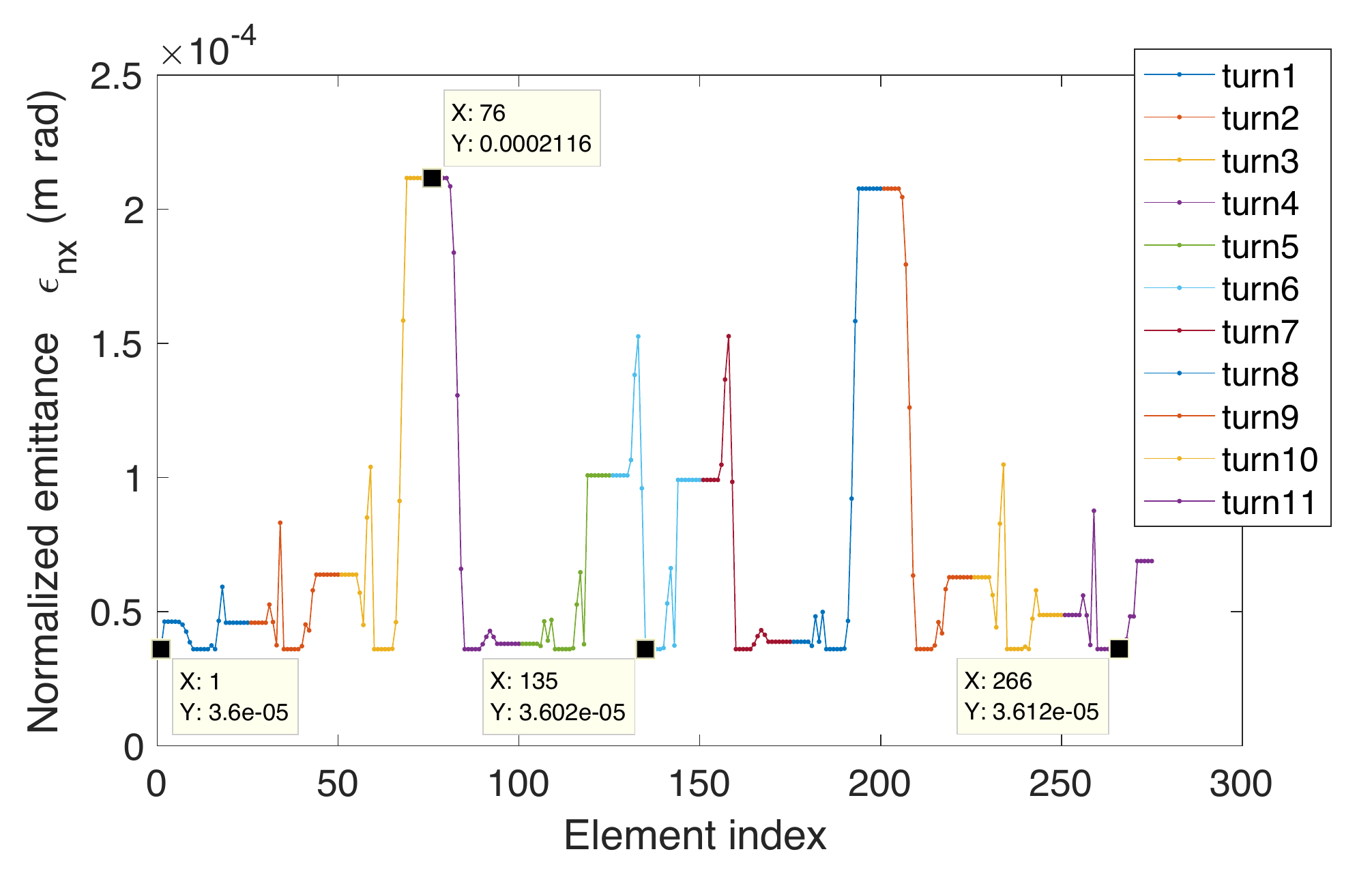}}\\
    \vspace{-10pt}
  \caption{The emittance growth through $11$\,turns in the CCR without (left) and with pre-/post-kickers (right).}
  \label{fig:emitt}
  \vspace{-13pt}
\end{figure*}

 There were some limitations implementing realistic beam dynamics into ELEGANT. Firstly the realistic 3D field map of the kickers could not be imported into ELEGANT. Instead, the kick was modeled as a transversely uniform and longitudinally impulsive kick. In the following sections and the appendix, a more realistic model for the kick will be implemented with multipole fields and a Gaussian longitudinal profile of the kick.  This will also be benchmarked against the 3D maps. Also the non-magnetized beam with the same beam parameter as magnetized beam but with minimal canonical angular momentum was propagated only to demonstrate the feasibility of the kicker system. The effect of the kick on the magnetized beam will be discussed in section~\ref{section:magnetization}. Finally, space-charge effects can not be implemented in ELEGANT. Therefore, only a small fraction of the bunch charge ($1.2$\,pC) was used to acheive a reasonably fast simulation. 
 First, a baseline simulation with a pair of the (injection/extraction) kickers only was done and the corresponding beam trajectory was examined in terms of the longitudinal (temporal) profiles of the angular distribution ($x'$) of the bunch. In Fig.\,\ref{fig:RCS}, the electron bunches at the entrance of the injection kicker are shown. The first bunch is an injected bunch kicked at kick angle of 2.5\,mrad. The subsequent re-circulating bunches are subject to the residual kicks of the extraction kicker (EK) upstream. For example, the $3$rd and $8$th bunch have large deformations with angular divergence up to $2.6\times10^{-4}$\,m-rad that are direct imprints from the steep RF slopes  of the kick profile on those bunches in Fig.\,\ref{fig:hk-profile}. These are largely eliminated by the injection kicker (IK) with the phase advance as shown in Fig.\,\ref{fig:C}, where the bunches at the exit of the injection kicker are shown. In Fig.\,\ref{fig:C}, the banana-shaped bunch profile with the side-wings to the edge reaching up to $4\times 10^{-4}$\,rad maintains its profile over the 11\,turns, which implies the effective cancellation by the IK for each turn. The banana shapes are imprinted from the RF curvature of the first injection kick and would decrease cooling efficiency significantly. The normalized emittance in a kick direction was tracked through 11\,turns and plotted in Fig.\,\ref{fig:emittance_baseline_00}. In Fig.\,\ref{fig:emittance_baseline_00}, the emittance is dramatically increased by the RF fields while the bunch is going through the EK and the phase advance, but decreases due to the aforementioned cancellation scheme by the IK. The emittance at the cooler is still larger than the injection value due to the RF curvature of the first IK. With the implementation of pre/post kicker (PRK/POK), the side-wings of the injected bunch is removed by a combination of the prekick and injection kick. Consequently, every bunch at the cooler has a flat angular distribution profile along the bunch length with the angular divergence reduced to $\pm0.2$\,mrad, as shown in Fig.~\ref{fig:C-PK}. The side-wings of the extracted bunch imprinted by the EK when the bunch goes back to the ERL after 11 turns is mostly eliminated with the use of the POK. The normalized emittance with PRK/POK at the cooler location in Fig.\,\ref{fig:emittance_baseline_PP} is much smaller than without PRK/POK (Fig.\,\ref{fig:emittance_baseline_00}) and is now almost the same as the initial emittance. The emittance growth between the EK and the IK, which are up to 2.4$\times 10^{-4}$\,m\,rad for the 3rd and the 8th turn, leads to a significant beam size increase, which was taken into account in the aperture design of the exchange region.  

\section{Beam propagation through more realistic kicks: an impulsive kick model with multipoles \label{section:realistic}}
\par In this section, the simple kick model used for the baseline design is generalized to a more realistic model of the actual field profiles within the QWR kicker cavity. We will compute the phase space transform through general field profiles for a beam whose transverse trajectory does not change over the effective field range. The transform can be expressed in terms of the multipole expansion of the fields via the Panofsky-Wenzel theorem. The multipole expansion coefficients for the QWR are obtained and fed into the ELEGANT simulations, where a modified cancellation scheme that includes the non-trivial multipole field contribution is demonstrated.

\subsection{Motion of electron bunch through a kick with general profiles}
\par Consider an arbitrary charged particle that passes through the kicker cavity whose RF fields are general. 
The relativistic equation of motion for the particle subject to the general Lorentz force is given as 
\begin{eqnarray}
\frac{d\vec{p}}{ds}=\frac{q}{c}(\vec{E}+\vec{v}\times\vec{B})\Bigg|_{t=s/c+\tau}.
\label{eqn:eom}
\end{eqnarray}
Here $\vec{v}$ is the velocity of the particle and $\vec{E},\vec{B}$ are the real electromagnetic fields. The coordinate system and the longitudinal initial conditions on the particle are described in \ref{subsec:baseline_anal}. For completeness, we add in transverse initial conditions. A trajectory of the particle is uniquely determined by a set of initial conditions at $s=-l/2$:
\begin{eqnarray}
\label{eqn:ic_r}
\vec{r}_\perp(s=-l/2)=\vec{r}_{\perp0},\\
\vec{p}_\perp(s=-l/2)=\vec{p}_{\perp0},\\
t(s=-l/2)=-\frac{l}{2c}+\tau,\\
\label{eqn:ic_E}
E(s=-l/2)=\mathcal{E}_0.
\end{eqnarray}
Here $\vec{r}_\perp$ is the transverse offset, $\vec{p}_\perp$ the transverse momentum, and $E$ the energy carried by a particle. In principle, the exact solutions to the non-linear equation (\ref{eqn:eom}) with the general initial condition (\ref{eqn:ic_r})-(\ref{eqn:ic_E}) could be given by a systematic iteration method based on the perturbative expansion of transverse phase space variables and fields. But with some physical constraints on the initial conditions and the fields, good approximate solutions are available. From physical consideration of the CCR beam dynamics and the geometry of the QWR, one can assume the motion is paraxial with $p_{x,y}\ll p_z\approx m_e \gamma c$ and the $\vec{E},\vec{B}$ are slowly varying fields. Consequently, the perturbation in the transverse trajectory of a particle within the kicker cavity is very small (order of submillimeter) due to fast longitudinal motion near $c$ and limited size of the kicker. Then the first order approximation to the solution is obtained by replacing $\vec{r}$ in $\vec{E}(\vec{r}\,), \vec{B}(\vec{r}\,)$ with $\vec{r}_0$ in (\ref{eqn:eom}). In components, (\ref{eqn:eom}) is written as
\begin{eqnarray}
\label{eqn:eom_c1}
\frac{d{p}_{x}}{ds}=\frac{e}{c}\Bigg\{E_x(\vec{r}_{\perp0},s)+\frac{B_z(\vec{r}_{\perp0},s)}{m_e\gamma}p_{y}-cB_y(\vec{r}_{\perp0},s)\Bigg\},\quad \quad \\
\label{eqn:eom_c2}
\frac{dp_{y}}{ds}=\frac{e}{c}\Bigg\{E_y(\vec{r}_{\perp0},s)-\frac{B_z(\vec{r}_{\perp0},s)}{m_e\gamma}p_{x}+cB_x(\vec{r}_{\perp0},s)\Bigg\},\quad \quad \\
\frac{dp_{z}}{ds}=\frac{e}{c}\Bigg\{E_z(\vec{r}_{\perp0},s)+\frac{B_y(\vec{r}_{\perp0},s)}{m_e\gamma}p_{x}-\frac{B_x(\vec{r}_{\perp0},s)}{m_e\gamma}p_{1,y}\Bigg\}.\quad \quad
\label{eqn:eom_c3}
\end{eqnarray}
Now we compute the phase space variable transform as solutions to (\ref{eqn:eom_c1})-(\ref{eqn:eom_c3}). First the transverse variables are computed. Introducing complex variables $\mathcal{P}_\pm=p_x\pm ip_y$, $\mathcal{E}_\pm=E_x\pm iE_y$, $\mathcal{B}_\pm=B_x\pm iB_y$, the equation of motion can be written as (by (\ref{eqn:eom_c1})+$i$(\ref{eqn:eom_c2}))
\begin{eqnarray}
\frac{d}{ds}\mathcal{P}_+=\frac{e}{c}\Bigg\{\mathcal{E}_+(\vec{r}_{\perp0},s)+ic\mathcal{B}_+(\vec{r}_{\perp0},s)-i\frac{B_z(\vec{r}_{\perp0},s)}{m_e\gamma}\mathcal{P}_+\Bigg\},\quad \quad
\end{eqnarray}
whose solution is given via 1D Green function $G(s,s')$ (for the operator $\frac{d}{ds}+ieB_z(s)/m_e\gamma c$) as
\begin{widetext}
\begin{eqnarray}
\mathcal{P}_+(s)=\mathcal{P}_0e^{-\frac{ie}{m_e\gamma c}\int_{-l/2}^s\!ds'\,B_z(\vec{r}_{\perp0},s)}+\frac{e}{c}\int_{-l/2}^s\!ds'\,G(s,s')\Bigg\{\mathcal{E}_+(\vec{r}_{\perp0},s)+ic\mathcal{B}_+(\vec{r}_{\perp0},s)\Bigg\},\\
\hbox{ where } G(s,s')=H(s-s')e^{-\frac{ie}{m_e\gamma c}\int_{s'}^s\!ds''\,B_z(\vec{r}_{\perp0},s'')}.
\end{eqnarray}
\end{widetext}
Here $H(s,s')$ is a Heaviside step function, which is 1 between $s$ and $-l/2$ and 0 otherwise. Therefore, the phase space variables at arbitrary $s$ are given as
 \begin{widetext}
\begin{eqnarray}
\label{eqn:px}
p_x(s)=p_{x0}\cos{\Theta(\vec{r}_{\perp0},s)}+p_{y0}\sin{\Theta(\vec{r}_{\perp0},s)}+\frac{e}{c}V_x(\vec{r}_{\perp0},s),\\
\label{eqn:py}
p_y(s)=p_{y0}\cos{\Theta(\vec{r}_{\perp0},s)}-p_{x0}\sin{\Theta(\vec{r}_{\perp0},s)}+\frac{e}{c}V_y(\vec{r}_{\perp0},s),\\
\label{eqn:x}
x(s)=x_0+p_{x0}\int_{-l/2}^s\!ds'\,\frac{\cos{\Theta(s')}}{m_ec\gamma}+p_{y0}\int_{-l/2}^s\!ds'\,\frac{\sin{\Theta(s')}}{m_ec\gamma}+\int_{-l/2}^s\!ds'\,\frac{eV_x(s')}{m_ec^2\gamma},\\
\label{eqn:y}
y(s)=y_0+p_{y0}\int_{-l/2}^s\!ds'\,\frac{\cos{\Theta(s')}}{m_ec\gamma}-p_{x0}\int_{-l/2}^s\!ds'\,\frac{\sin{\Theta(s')}}{m_ec\gamma}+\int_{-l/2}^s\!ds'\,\frac{eV_y(s')}{m_ec^2\gamma},\\
\label{eqn:Theta}
\hbox{where }\Theta(\vec{r}_{\perp0},s)=\frac{e}{m_ec\gamma}\int_{-l/2}^{s}\!ds'\,B_z(\vec{r}_{\perp0},s')\sin{\left\{\omega\left(\frac{s'}{c}+\tau\right)+\Phi\right\}},\\
\label{eqn:Vx}
V_x(\vec{r}_{\perp0},s)=\int_{-l/2}^{s}\!ds'\,\Bigg[E_x(\vec{r}_{\perp0},s')\cos{\left\{\omega\left(\frac{s'}{c}+\tau\right)\right\}}-cB_y(\vec{r}_{\perp0},s')\sin{\left\{\omega\left(\frac{s'}{c}+\tau\right)+\Phi\right\}}\Bigg]G(s,s'),\\
\label{eqn:Vy}
V_y(\vec{r}_{\perp0},s)=\int_{-l/2}^s\!ds'\,\Bigg[E_y(\vec{r}_{\perp0},s')\cos{\left\{\omega\left(\frac{s'}{c}+\tau\right)\right\}}+cB_x(\vec{r}_{\perp0},s')\sin{\left\{\omega\left(\frac{s'}{c}+\tau\right)+\Phi\right\}}\Bigg]G(s,s').
\end{eqnarray}
\end{widetext}
Here we assumed $\vec{E},\vec{B}$ are oscillating with harmonic frequency $\omega$ and phase $\Phi$. In a deflecting operation, the phase is set to be on-crest, i.e., $\Phi=0$.
\par
\begin{figure}[hbt]
   \centering
   \includegraphics*[width=75mm]{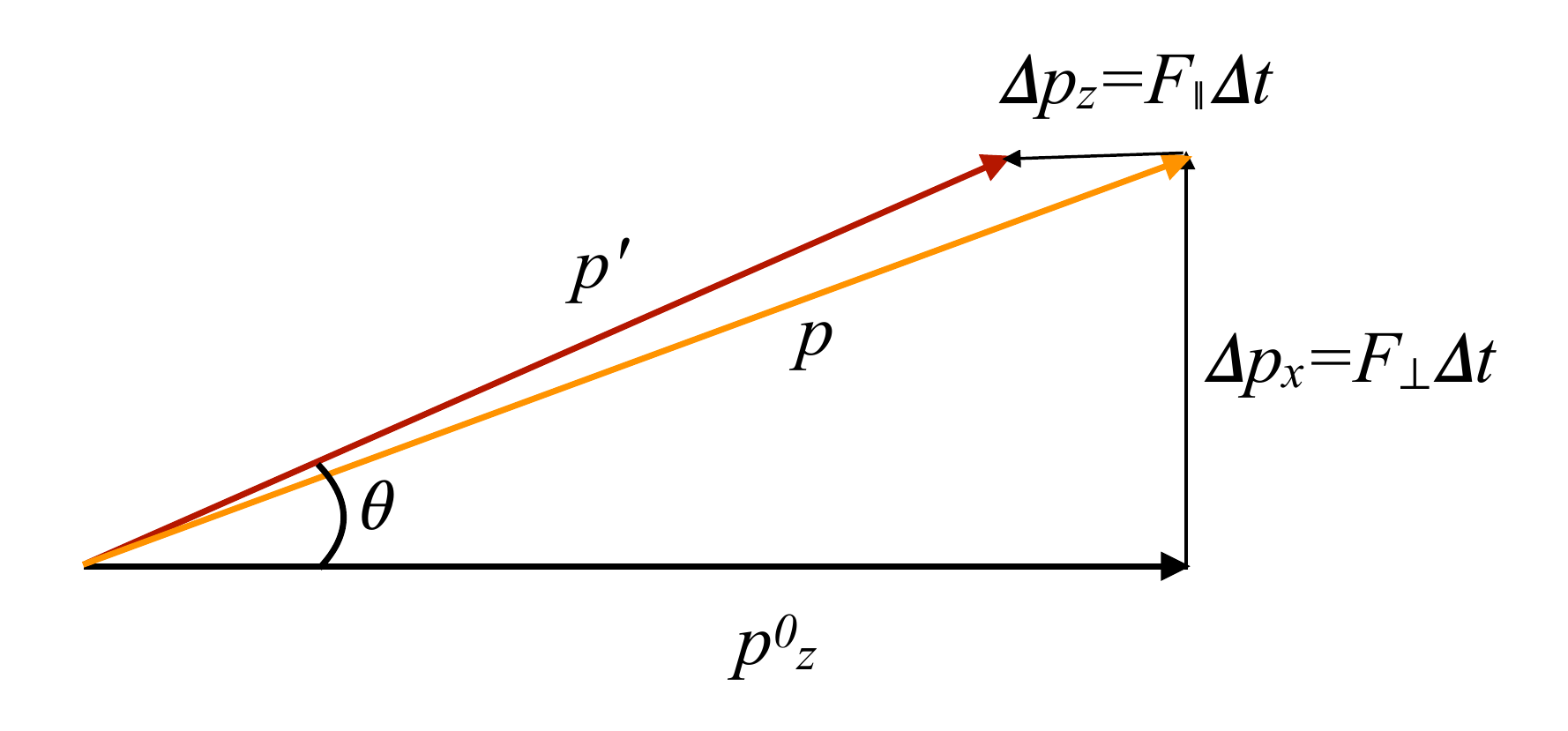}
   \caption{The diagram for momentum change through a harmonic kicker. The initial momentum is $p_z^0$ and the exit momentum is $p'$.}
   \label{fig:Delta_momentum}
\end{figure}

 To solve for the remaining (longitudinal) phase space variables, we first obtain a solution to (\ref{eqn:eom_c3}) as
\begin{widetext}
\begin{eqnarray}
\label{eqn:pz}
p_z(s)=p_{0z}+\Upsilon(\vec{r}_{\perp0},s),\\
\hbox{where }\,\Upsilon(\vec{r}_{\perp0},s)=\frac{e}{c}\int_{-l/2}^s\!ds'\,\Bigg[E_z(\vec{r}_{\perp0},s')\cos{\left\{\omega\left(\frac{s'}{c}+\tau\right)\right\}}\nonumber\\+\frac{p_x(s')}{m_e\gamma}B_y(\vec{r}_{\perp0},s')\sin{\left\{\omega\left(\frac{s'}{c}+\tau\right)\right\}}-\frac{p_{y}(s')}{m_e\gamma}B_x(\vec{r}_{\perp0},s')\sin{\left\{\omega\left(\frac{s'}{c}+\tau\right)\right\}}\Bigg].
\end{eqnarray}
\end{widetext}
With $s=l/2$, we have $p_z=p_{z0}+\Upsilon(\vec{r}_{\perp0})$ where $\Upsilon(\vec{r}_{\perp0})=\Upsilon(\vec{r}_{\perp0}, s=l/2)$. Notice that $\Upsilon(\vec{r}_{\perp0})$ becomes 0 with $\tau=0$ with the antisymmetric $E_z$ ($E_z(-s)=-E_z(s)$ within the kicker cavity, as suggested from the RF simulation of the kick fields), while $\Upsilon(\vec{r}_{\perp0})$ is non-zero with non-zero $\tau$. Then, the energy $\mathcal{E}$ at $s$ is obtained to the 2nd order perturbation using (\ref{eqn:px}), (\ref{eqn:py}), and (\ref{eqn:pz}) as
\begin{widetext}
\begin{eqnarray}
E(s)&=&\sqrt{p^2c^2+m_e^2c^4}\approx\sqrt{p_0^2c^2+m_e^2c^4}+\frac{c^2}{\mathcal{E}_0}\vec{p}_0\cdot\delta \vec{p}\,+\frac{c^2}{2\mathcal{E}_0}(\delta p)^2\nonumber\\
&\approx&\mathcal{E}_0+\frac{c^2}{\mathcal{E}_0}\Bigg\{-p_{x0}(1-\frac{eV_x}{c}\cos{\Theta}+\frac{eV_y}{c}\sin{\Theta})-p_{y0}(1-\frac{eV_x}{c}\sin{\Theta}-\frac{eV_y}{c}\cos{\Theta})\nonumber\\
&+&\frac{\mathcal{E}_0}{c}\Upsilon+p_{x0}^2+p_{y0}^2+\frac{1}{2}\left(\frac{eV_x}{c}\right)^2+\frac{1}{2}\left(\frac{eV_y}{c}\right)^2+\frac{\Upsilon^2}{2}\Bigg\}.
\label{eqn:energy_eq}
\end{eqnarray}
\end{widetext}
where the approximation in the first line was taken with paraxial momenta ($p_{x0},p_{y0}\ll p_{z0}$). For a rough estimation of the energy change, we compute the energy change through the kicker in case of vertical kick with zero initial slopes, i.e., $V_y=p_{x0}=p_{y0}=0$ and $s=l/2$, as illustrated in Fig.\,\ref{fig:Delta_momentum}. 
Then (\ref{eqn:energy_eq}) reduces in this limit to
\begin{eqnarray}
\Delta E=\sqrt{(p'c)^2+(m_ec^2)^2}-\sqrt{(p^0_zc)^2+(m_ec^2)^2}\nonumber\\
\approx c\left\{\Delta p_z+\frac{1}{2}\frac{\Delta p_x^2}{p^0_z}\right\}\approx c\Bigg\{\Upsilon+\frac{c}{2\mathcal{E}_0}\left(\frac{eV_x}{c}\right)^2\Bigg\}.
\label{eqn:E}
\end{eqnarray}
In the simple kick model, (\ref{eqn:E}) with $\Upsilon=0$ would reduce to
\begin{eqnarray}
\approx \frac{c^2}{2\mathcal{E}_0}\left(\frac{eV_h}{c}\right)^2\sum_{n,l=1}^N\cos{\omega_n\tau}\cos{\omega_l\tau}.
\end{eqnarray}
In particular with $\tau=0$, putting $p_z^0=55$\,MeV/c and $\Delta p_x=125$\,keV/c would lead to $\Delta E\approx 143$\,eV only, which is negligible. Finally, the relative time $t(s)$ (to a fiducial particle) it takes for an electron to arrive at $s$ along the beam line is obtained from inverting the relativistic definition of $p_z$:
\begin{eqnarray}
\frac{dt}{ds}=\frac{m_e\gamma}{p_z}-\frac{m_e\gamma}{p_z}\Bigg|_{\tau=0}.
\label{eqn:t_eqn}
\end{eqnarray}
With a narrow energy spread in the order of $10^{-4}$ and a small change in velocity over the effective field range, we have a Taylor-expansion as $\gamma\approx\gamma_0+\gamma_0^3\vec{\beta}\cdot \delta \vec{\beta}$ and $1/p_z\approx1/p_{z0}-\delta p_z/p_{z0}^2$ with $\delta$ denoting the derivative with respect to $\tau$ evaluated at $\tau=0$ and (\ref{eqn:t_eqn}) is approximated as
\begin{eqnarray}
\frac{dt}{ds}\approx-\frac{1}{c}\left[\frac{1}{p_{z0}}\delta p_z-\gamma_0^2\delta\beta_z\right]=-\frac{1}{c p_{z0}}\frac{\delta p_z}{1+\gamma_0^2}\nonumber\\
=-\frac{e\omega}{c^2 p_{z0}}\frac{1}{1+\gamma_0^2}\int_{-l/2}^{l/2}\!ds'\,E_z(\vec{r}_{\perp0},s')\sin{\frac{\omega s'}{c}}\ll 1,\quad 
\end{eqnarray}
where the second equality is obtained with $\delta p_z=m_ec\gamma_0^3\delta\beta_z+m_e\gamma_0\delta v_z=p_{z0}(\gamma_0^2+1)\delta \beta_z$. Therefore, the bunch length change is close to zero. 

\subsection{Multipole expansion  of the field \label{sec:multipoles}}
\par The phase space variable transforms through the cavity, obtained by setting $s=l/2$ in (\ref{eqn:px})-(\ref{eqn:Vy}), can be compactly re-written in terms of the multipole fields. In particular, for a beam in the extreme paraxial limit with $p_{x0}, p_{y0}\rightarrow 0$, $e.g.$, a non-magnetized beam that has nearly zero slopes at the kicker entrance, the Panofsky-Wenzel theorem~\cite{PW} holds accurately~\cite{browman} and is applicable to the $r.h.s$ of (\ref{eqn:Vx})\textemdash the $x,y$ motion of the beam is decoupled and we focus on the motion in the kick direction only. Consequently the transverse momentum change (\ref{eqn:px}) at the exit can be expressed in terms of the longitudinal field component $E_z$, making multipole evaluations much simpler compared to the full Lorentz force expansion:
\begin{eqnarray}
p_x=-\frac{e}{\omega}\int_{-l/2}^{l/2}\!dz\,\nabla_xE_z(\vec{r}_{\perp0},z)\sin{\Bigg\{\omega\left(\frac{z}{c}+\tau\right)\Bigg\}}.
\label{eqn:delta_p_E}
\end{eqnarray}
Hereafter, all the phase variables without arguments are understood to be evaluated at $s=l/2$. Now the integrand in (\ref{eqn:delta_p_E}) is expanded in terms of multipole fields of $E_z$ around the beam axis (analogous to static magnetic fields, see appendix of \cite{gpark3} for more details). First the complex version $\widetilde{E}_z$ of $E_z$ such that $E_z=\mathfrak{Re}[\widetilde{E}_z]$ is expanded over transverse plane in polar coordinates into
\begin{eqnarray}
\label{eqn:Ez_tilde}
\widetilde{E_z}(\vec{r}_{\perp0},z)=\sum_{n=0}^\infty\mathcal{C}_n(z)r^n e^{in\phi},\\
\hbox{where }\mathcal{C}_n(z)=\frac{1}{\pi r^n}\int_{0}^{2\pi}\!d\phi\,E_z(\vec{r}_{\perp0},z)e^{-in\phi},
\label{eqn:cnz}
\end{eqnarray}
By plugging the real part of (\ref{eqn:Ez_tilde}) into (\ref{eqn:delta_p_E}), we have
\begin{eqnarray}
p_x=-\frac{e}{\omega}\sum_{n=1}^\infty n\mathfrak{Re}\Bigg\{c_n\left(x_0+iy_0\right)^{n-1}\Bigg\},
\label{eqn:vx_mp}
\end{eqnarray}
where complex multipole expansion coefficients $c_n$'s are defined as
\begin{eqnarray}
c_n=\int_{-l/2}^{l/2}\!dz\,\frac{1}{\pi r^n}\Bigg[\int_0^{2\pi}\!E_z(\vec{r}\,)e^{-in\phi}d\phi\Bigg]\!\sin{\Bigg\{\omega\left(\frac{z}{c}+\tau\right)\Bigg\}}.\quad \quad
\label{eqn:mpc}
\end{eqnarray} 
The complex coefficients $c_n$'s are written as $c_n=b_n-ia_n$, where $b_n, a_n$'s are identified as normal and skew multipole coefficients, respectively. The coefficients in (\ref{eqn:mpc}) for the QWR are numerically evaluated by inserting the 3D field maps of the $E_z$, which is obtained from the RF field simulation by the CST-MWS (the details of the accurate evaluation of the 3D\,field maps in the QWR are found in the appendix of \cite{gpark3}). Finally, the phase space transforms (\ref{eqn:px})-(\ref{eqn:Vy}) after the kicker are simplified as 
\begin{widetext}
\begin{eqnarray}
\label{eqn:px_nm}
p_{x}=\frac{e}{c}V_x,\\
\label{eqn:py_nm}
p_{y}=\frac{e}{c}V_y,\\
\label{eqn:x_nm}
x=x_0+W_x,\\
\label{eqn:y_nm}
y=y_0+W_y,\\
\label{eqn:Vx_nm}
\text{where } V_x=\sum_{n,m=1}^5\frac{c}{\omega_m}\mathfrak{Re}\Bigg\{c_{nm}n(x_0+iy_0)^{n-1}\Bigg\},\\
\label{eqn:Vy_nm}
V_y=\sum_{n,m=1}^5\frac{c}{\omega_m}\mathfrak{Re}\Bigg\{ic_{nm}n(x_0+iy_0)^{n-1}\Bigg\},\\
\label{eqn:Wx_nm}
W_x=-\sum_{n,m=1}^5\frac{e}{m_e c\gamma \omega_m}\Bigg[w_x+\mathfrak{Re}\Big\{d_{nm}n(x_0+iy_0)^{n-1}\Big\}\Bigg],\\
\label{eqn:Wy_nm}
W_y=-\sum_{n,m=1}^5\frac{e}{m_e c\gamma \omega_m}\Bigg[w_y+\mathfrak{Re}\Big\{id_{nm}n(x_0+iy_0)^{n-1}\Big\}\Bigg],\\
\label{eqn:cnm}
\hbox{where } c_{nm}=\frac{\cos{\omega_m\tau}}{\pi r^n}\int_{-l/2}^{l/2}\,\Bigg[\int_0^{2\pi}\!E_z^{(m)}(\vec{r}_{\perp0},z\,)e^{-in\phi}d\phi\Bigg]\,\sin{\Bigg\{\frac{\omega_mz}{c}\Bigg\}}\,dz,\\
\label{eqn:wnm}
w_{x,y}=\sin{\omega\tau}\int_{-l/2}^{l/2}\!dz\,E_{x,y}^{(m)}(\vec{r}_{\perp0},z)\cos{\Bigg\{\frac{\omega_mz}{c}\Bigg\}},\\
\label{eqn:dnm}
d_{nm}=\frac{\cos{\omega_m\tau}}{\pi r^n}\int_{-l/2}^{l/2}\!dz\,\int_{-l/2}^{s}\!dz'\,\Bigg[\int_0^{2\pi}\!E_z^{(m)}(\vec{r}_{\perp0}, z'\,)e^{-in\phi}d\phi\Bigg]\sin{\Bigg\{\frac{\omega_mz'}{c}\Bigg\}}.
\end{eqnarray}
\end{widetext}
 Here we included all the harmonic modes (indexed with $m$) for completeness. With $W_x,W_y$ being very small, the phase space transform (\ref{eqn:px_nm})-(\ref{eqn:dnm}) can be effectively viewed as the impulsive kicks with multipole fields, which can be represented in ELEGANT as beamline elements. The resulting multipole coefficients up to decapole are listed in Table\,\ref{table:mpc}. The mode\,6 in the table refers to the multipole coefficients of the pre/post kickers. In the Table\,\ref{table:mpc}, skew multipoles are vanishingly small (compared to normal multipoles) because of horizontal symmetry (with respect to $xz$-plane) of the fields, while there is no apparent vanishing of even normal multipoles because of lack of vertical anti-symmetry (with respect to $yz$-plane) in the QWR structure. Also notice that the dipole coefficient for each mode agrees with the kick voltage on beam-axis (upon multiplying $c/\omega_m$'s according to the Panofsky-Wenzel theorem). 

\begin{table*}[hbt]
   \centering
   \caption{The multipole field coefficients of a harmonic kicker as evaluated based on the circle with 10\,mm radius in hexahedral meshing. The coefficients are evaluated based on kick voltage of $25$\,kV for each mode.}
   \vspace{5pt}
   \begin{tabular}{lcccccc}
       \toprule
    Multipoles       & Mode 1 & Mode 2 & Mode 3 &Mode 4 & Mode 5 & Mode 6\\ 
            $f$\,(MHz)  & 86.6 & 259.8 & 433 & 606.2 & 779.4 & 952.6\\
           \hline
           $b_1$ (V) &  $4.49\times10^{4}$ & $1.37\times10^5$ & $2.27\times10^5$ & $3.17\times10^5$ & $4.05\times10^5$ & $7.53\times10^5$\\
           $b_2$  (V/m) & $-4.16\times10^5$ & $-1.25\times^6$ & $-2.05\times10^6$ & $-2.79\times10^6$ & $-3.46\times10^6$ & $6.31\times10^6$\\
           $b_3$ (V/m$^2$) & $5.33\times10^6$ & $1.62\times10^7$ & $2.68\times10^7$ & $3.75\times10^7$ & $4.88\times10^7$ & $9.83\times10^7$\\
           $b_4$ (V/m$^3$) & $-3.66\times10^7$ & $-1.11\times10^8$ & $-1.82\times10^8$ & $-2.53\times10^8$ & $-3.28\times10^8$ & $6.68\times10^8$\\
           $b_5$ (V/m$^4$) & $2.08\times10^8$ & $6.31\times10^8$ & $1.05\times10^9$ & $1.48\times10^9$ & $2.03\times10^9$ & $4.85\times10^9$\\
           $a_1$ (V) & $-2.19$ & $3.26$ & $6.76\times10$ & $4.07\times10^2$ & $2.57\times10^3$ & $3.09\times10^3$\\
           $a_2$ (V/m) & $8.09\times10$ & $4.45\times10$ & $-1.18\times10^3$ & $-7.86\times10^3$&$-4.99\times10^4$ &$5.8004\times10^4$ \\
           $a_3$ (V/m$^2$) & $-4.48\times10^3$ & $-1.10\times10^4$ & $-2.71\times10^3$& $6.86\times10^4$ & $5.18\times10^5$ & $5.68\times10^5$\\
           $a_4$ (V/m$^3$) & $-3.85\times10^4$ & $7.80\times10^4$ & $-1.03\times10^5$ & $-1.28\times10^6$ & $-8.80\times10^6$ & $9.72\times10^6$\\
           $a_5$ (V/m$^4$) & $6.51\times10^6$ & $-2.01\times10^7$ & $3.54\times10^7$& $6.06\times10^7$ & $1.53\times10^8$ & $2.11\times10^8$\\
        \botrule
   \end{tabular}
   \label{table:mpc}
\end{table*}

\subsection{Cancellation scheme for multipole effects}
\par
\begin{figure}[hbt]
  \centering
    \subfigure[ \, Cancellation schematic. The red arrow is the design kick, i.e., dipole fields. The blue arrow refers to the velocity of an electron along its trajectory.]{\label{fig:cs}\includegraphics[scale=0.25]{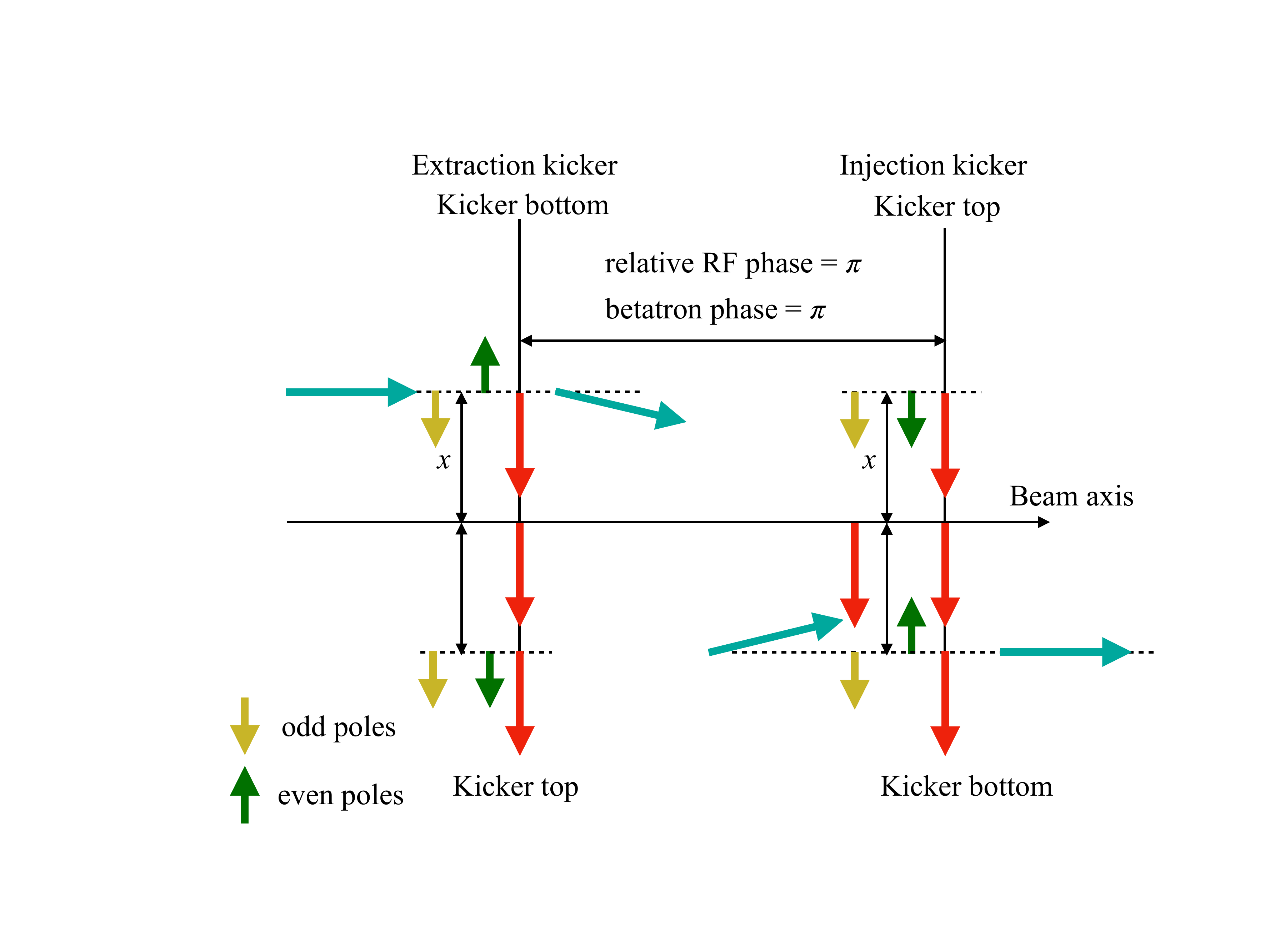}}\\
    \subfigure[ \,Beamline schematic with the magnets in injection transport. The green dots are monitors.]{\label{fig:csc}\includegraphics[scale=0.35]{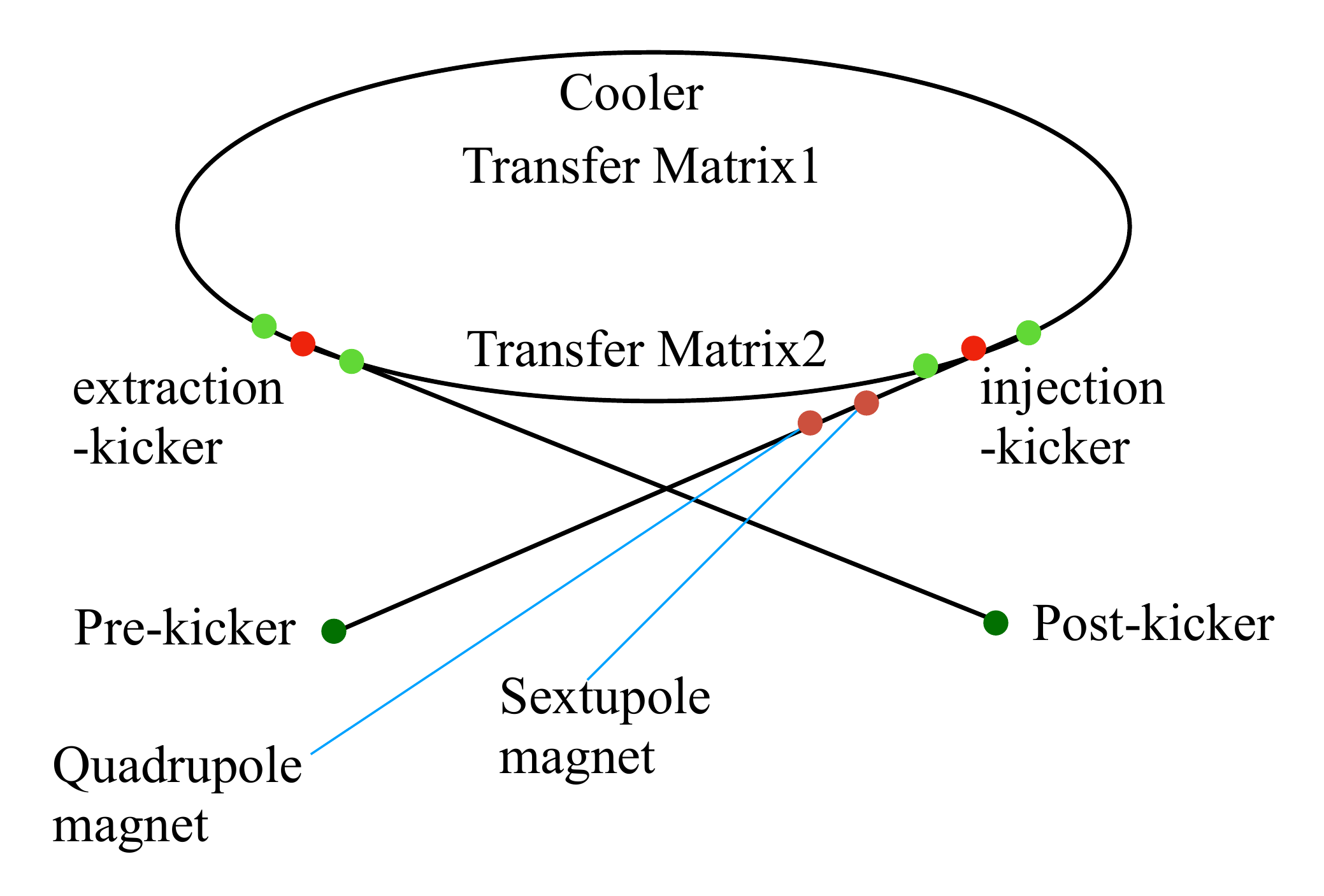}}\\
    \vspace{-10pt}
  \caption{Multipole cancellation scheme at $E_e=55$\,MeV.}
  \label{fig:emitt00}
\end{figure}

 The betatron phase advance cancellation scheme implemented in section\,\ref{section:baseline}, based on a uniform transverse profile, does not cancel  kicks with non-trivial multipole fields: at each turn, the effects of the even order multipoles in (\ref{eqn:mpc}) are not cancelled between the injection and extraction kicks but are doubled. 
To achieve multipole cancellation, the kickers within the kicker system were re-arranged as illustrated in Fig.\,\ref{fig:cs}. In Fig.\,\ref{fig:cs}, the EK is displaced from the IK by betatron phase advance of $\pi$ as in baseline, but the kicker is now flipped upside down with its RF phase set to be $\pi$ relative  to the IK. With respect to this re-arrangement, the odd multipoles (dipole, sextupole, $\cdots$) are invariant, while the even multipoles (quadrupoles, octopoles,$\cdots$) flip the signs. Then, the extraction kick as a vector sum of all the relevant multipoles at arbitrary $x>0$ is identical with the injection kick at $-x$. This configuration leads to the desired cancellation with a betatron phase advance of $\pi$: An electron entering the EK at offset of $x>0$ and slope $x'$ will be subject to a certain kick (a sum of all the multipoles) from the EK ending up with $x'+\Delta x'$, and then move down to $-x$ with its slope flipped upside down, i.e., $-(x'+\Delta x')$. Then at $-x$ in the IK, multipoles are exactly the same as those in EK at $x$ and an electron will pickup $\Delta x'$ from the IK, leading to $-x'$, which preserves the beam matrix with symmetric beam distribution in $x$ direction.
\par This can be stated more compactly as follows. If we label an electron before the extraction kicker by $0$, after the extraction kicker by $1$, after a betatron phase advance of $\pi$ by $2$, and after the injection kicker by $3$, we have
\begin{eqnarray}
\label{eqn:px_cancel}
p_{x3}=-p_{x0}-\frac{e}{c}V_{Ex}(x_0)+\frac{e}{c}V_{Ix}(-x_0),\\
p_{y3}=-p_{y0}-\frac{e}{c}V_{Ey}(x_0)+\frac{e}{c}V_{Iy}(-x_0),
\label{eqn:py_cancel}
\end{eqnarray}
where the suffices $E,I$ refer to extraction and injection kicker, respectively. On the other hand, for arbitrary $x,y$, the extraction kick voltage $V_E$ is related to injection voltage $V_I$ using (\ref{eqn:mpc}) as follows:
\begin{eqnarray}
V_E(x)=-\frac{e}{\omega}\Bigg\{\sum_{n=odd}n\mathfrak{Re}\Big[c_n(x+iy)^{n-1}\Big]\nonumber\\
+\sum_{n=even}n\mathfrak{Re}\Big[-c_n(x+iy)^{n-1}\Big]\Bigg\}\nonumber\\
=-\frac{e}{\omega}\Bigg\{\sum_{n=odd, r=0}^{r=n-1}n\mathfrak{Re}\Big[c_nC_{n-1,r}x^r(iy)^{n-1-r}\Big]\nonumber\\
+\sum_{n=even,r=0}^{r=n-1}n\mathfrak{Re}\Big[-c_nC_{n-1,r}x^r(iy)^{n-1-r}\Big]\Bigg\}\nonumber\\
=-\frac{e}{\omega}\Bigg\{\sum_{n=odd, r=even}^{r=n-1}n\Big[c_nC_{n-1,r}x^ry^{n-1-r}\Big]\nonumber\\
-\sum_{n=even,r=odd}^{r=n-1}n\Big[c_nC_{n-1,r}x^ry^{n-1-r}\Big]\Bigg\}\nonumber\\
=-\frac{e}{\omega}\Bigg\{\sum_{n=odd, r=even}^{r=n-1}n\Big[c_nC_{n-1,r}(-x)^ry^{n-1-r}\Big]\nonumber\\
+\sum_{n=even,r=odd}^{r=n-1}n\Big[c_nC_{n-1,r}(-x)^ry^{n-1-r}\Big]\Bigg\}\nonumber\\
=-\frac{e}{\omega}\Bigg\{\sum_{n=odd}n\mathfrak{Re}\Big[c_n(-x+iy)^{n-1}\Big]\nonumber\\
+\sum_{n=even}n\mathfrak{Re}\Big[c_n(-x+iy)^{n-1}\Big]\Bigg\}=V_I(-x).
\label{eqn:m_cancel}
\end{eqnarray}
Here $c_n$'s are the multipole expansion coefficients of the IK and $C_{n-1,r}=(n-1)!/((n-1-r)!r!)$ are binomial expansion coefficients. Then making use of (\ref{eqn:m_cancel}) leads (\ref{eqn:px_cancel}) and (\ref{eqn:py_cancel}) to complete cancellation. 
\par Although this configuration cancels the multipole effects of the EK by the IK, the multipole effects of the IK on the electron entering the CCR in the first pass, i.e., the multipoles of the injection deflecting kick are not cancelled and survive through all 11 turns. To eliminate these effects, we introduce DC magnets whose field strength is adjusted against $n$th multipole of the kicker according to the formulae
\begin{eqnarray}
\sum_m\frac{n}{\omega_m}b^m_n=-\frac{B\rho L}{n-1}k_{n-1},
\end{eqnarray}
 where $(B\rho)$ is particle rigidity, i.e., $p/q$ with $p,q$ being total momentum and charge of an electron, respectively and $L$ is the length of the magnet.  In practice, $n$ is limited to $n=2,3$ (quadrupole and sextupole). This adjustment cancels the multipoles of the PRK as well. The DC magnets are installed in the injection transport line (see Fig.\,\ref{fig:csc}) without having to modify the CCR lattice, which would involve non-trivial tune adjustments. 
 
\subsection{ELEGANT simulation results with multipole fields}
 \par
\begin{figure}[hbt]
  \centering
       \subfigure[ \,Longitudinal profiles of electron bunches with multipoles (without any cancellation scheme), showing a blow-up in angular divergence.]{\label{fig:MP_xp_no_cancel}\includegraphics[scale=0.25]{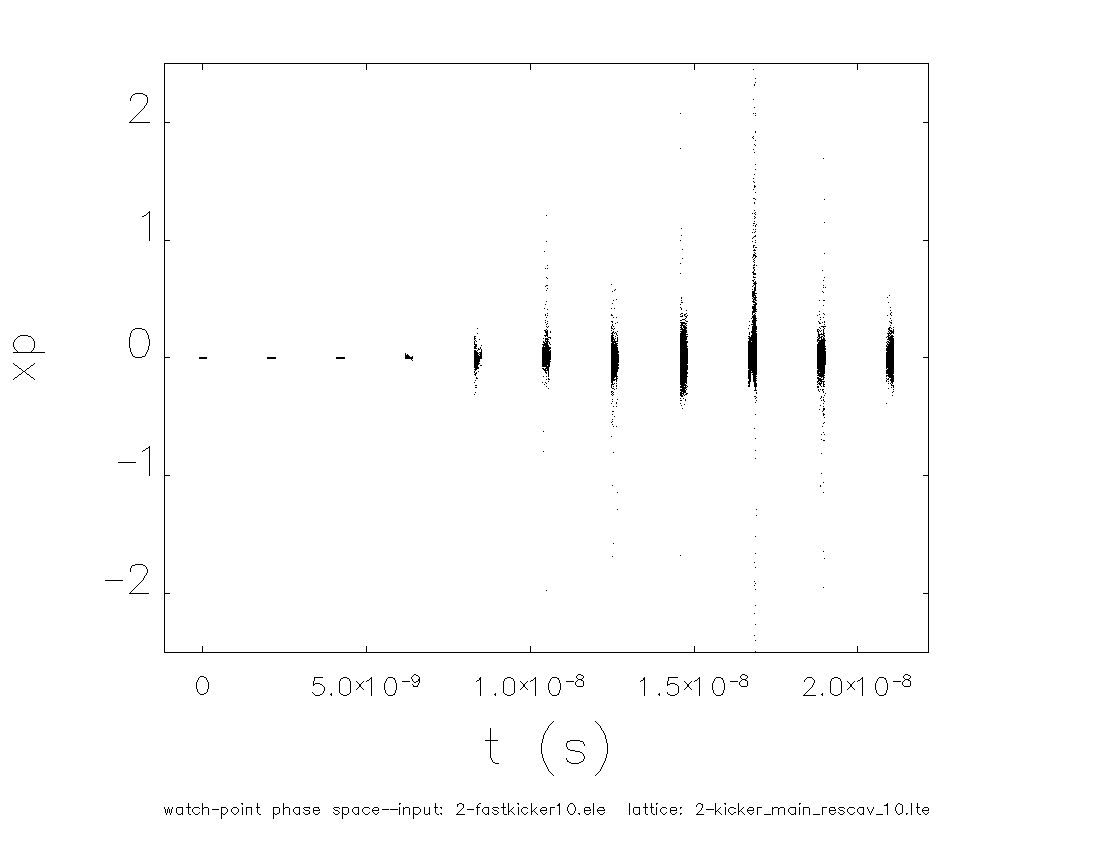}}\\
    \subfigure[ \,Integrated Lorentz force at various offsets. The red (blue) curve is along the vertical (horizontal) axis.]{\label{fig:Kick_vol_int}\includegraphics[scale=0.41]{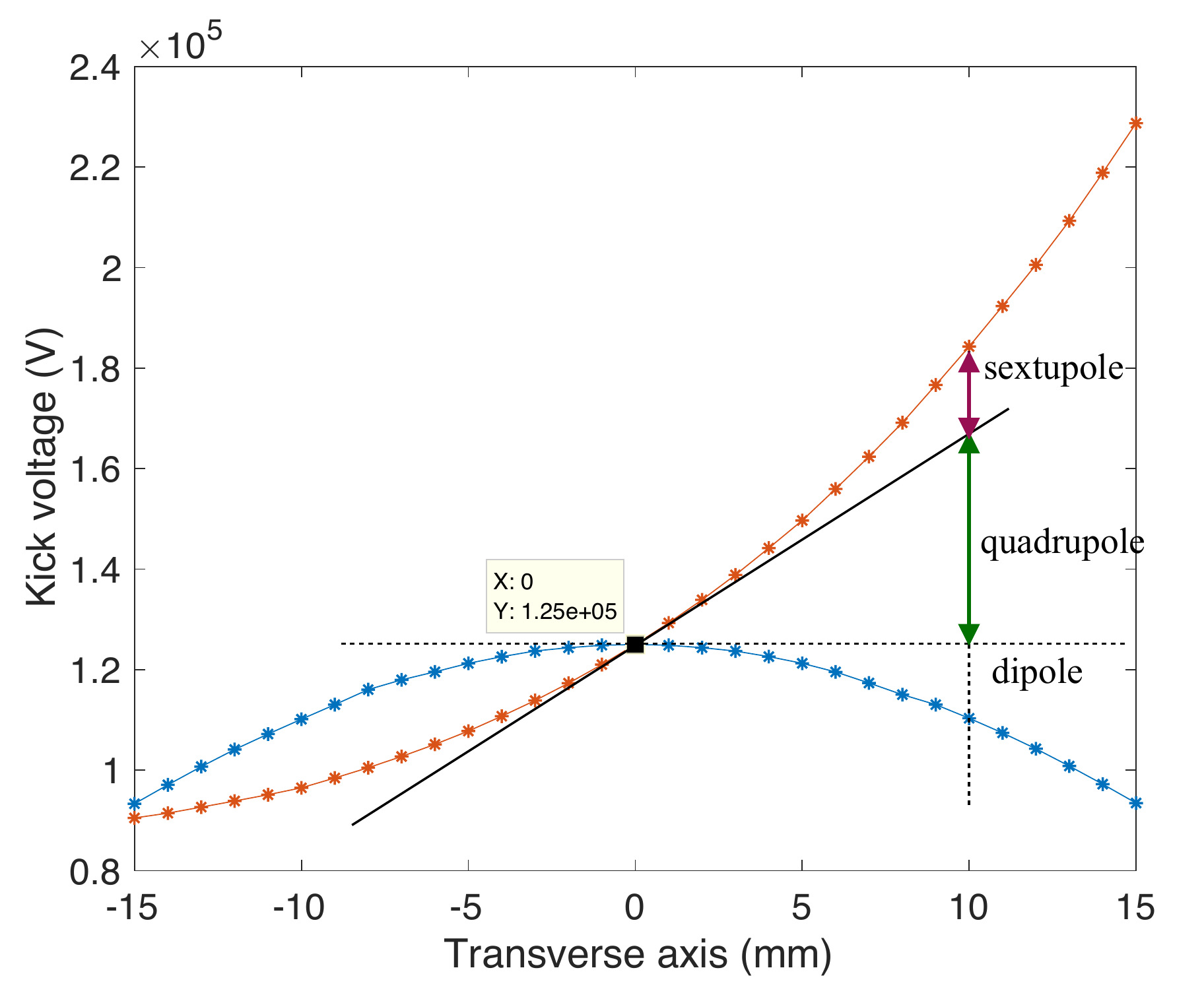}}
    \vspace{-10pt}
  \caption{Multipole effects of the kicker without a cancellation scheme.}
  \label{fig:multipole}
\end{figure}

 An implementation of multipoles of the kickers\textemdash including pre/post kickers\textemdash 
in the ELEGANT simulation without modifying a baseline cancellation configuration leads to beam blow-up before completing 11\,turns, as illustrated in Fig.\,\ref{fig:MP_xp_no_cancel}. The direct effects of multipoles fields on the beam distribution would be an increase in angular divergence, which introduces a mismatch to the beam line lattice leading to an accumulating increase of beam size over the turns. The consecutive excitation of each multipole in the simulation suggests the quadrupole fields in all the modes are main contributions to the beam degradation, with a smaller contribution from the sextupoles. This is consistent with the transverse profile (in the kick direction) of the kick voltage in Fig.\,\ref{fig:Kick_vol_int}, where the slope at the origin corresponds to the quadrupole and the deviation of the voltage curves from linear extension of the slope are mostly accounted for by the sextupole term. 

\par
\begin{figure}[hbt]
  \centering
    \subfigure[The angular distribution over 11\,turns with cancellation scheme implemented at $E_e=55$\,MeV.]{\label{fig:Ca}\includegraphics[scale=0.41]{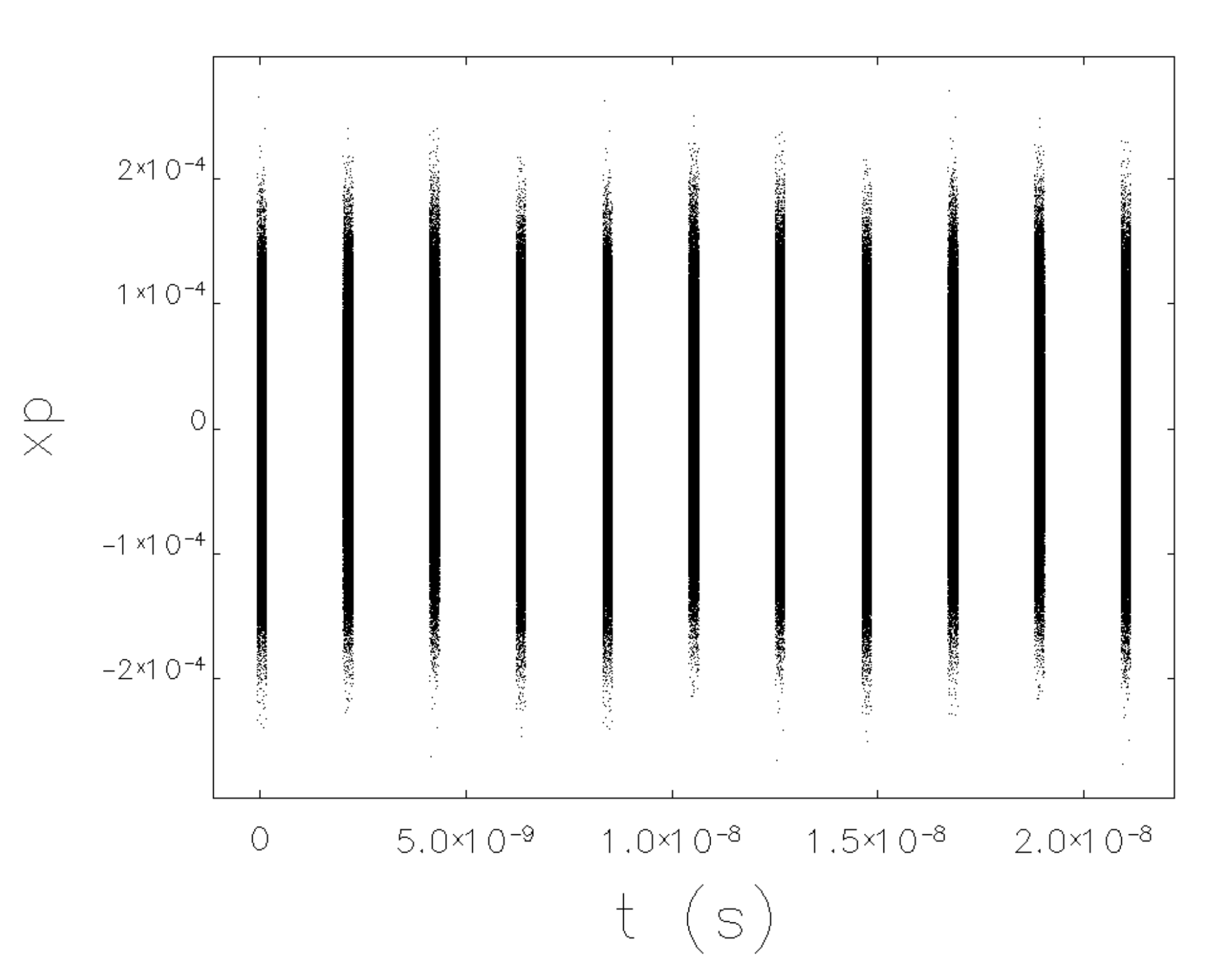}}\\  
 \subfigure[The emittance growth over 11\,turns with cancellation scheme implemented at $E_e=55$\,MeV.]{\label{fig:Cb}\includegraphics[scale=0.41]{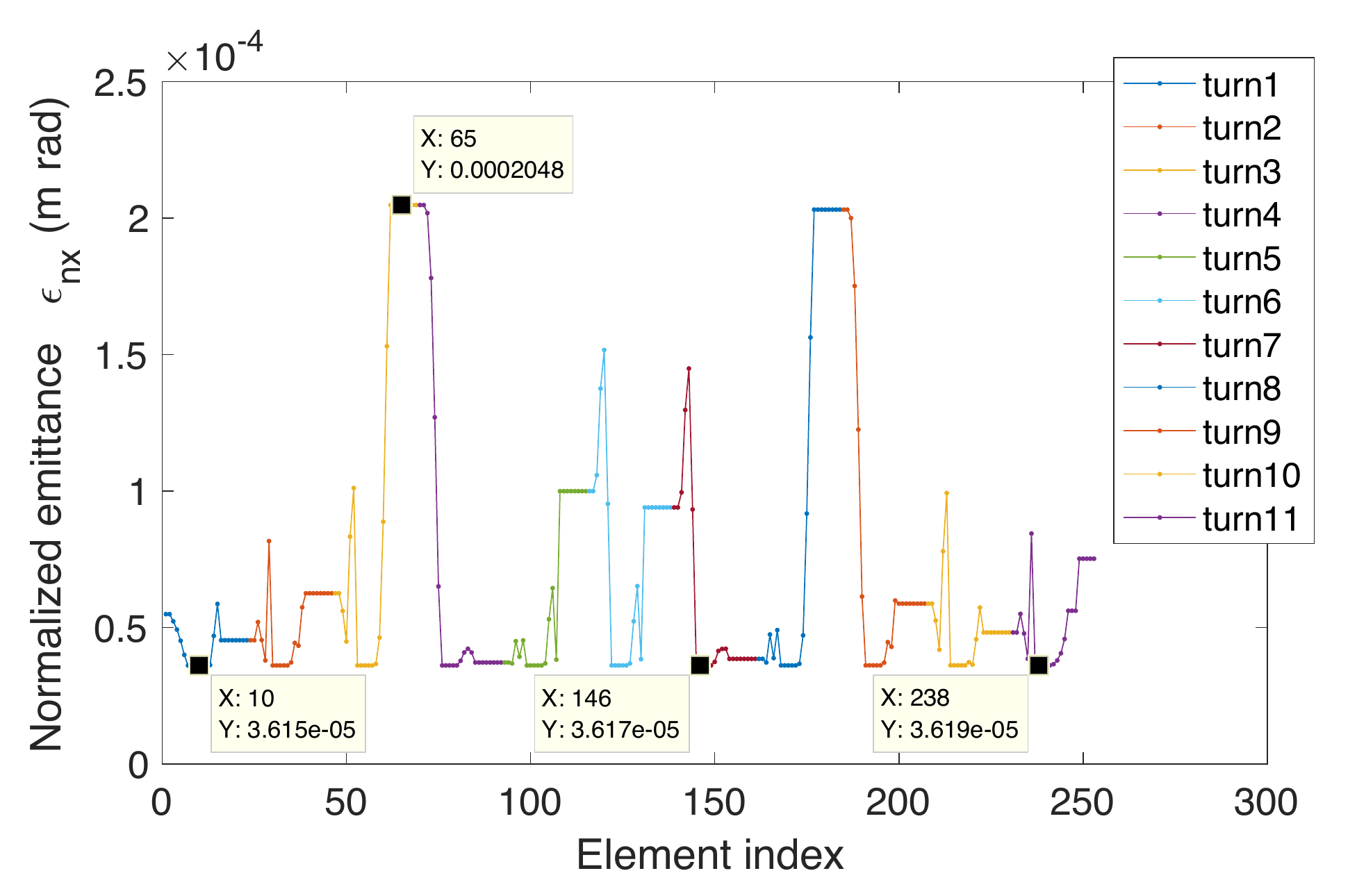}}
  \caption{The angular distribution evolution over the turns with multipoles in cancellation scheme.}
  \label{fig:ad}
\end{figure}

 With the modified cancellation scheme implemented, the simulation results show that most of the multipole effects are compensated between the EK and the IK. In Fig.\,\ref{fig:Ca}, the bunches at the exit of the IK are shown to be well-aligned along the beam axis with the minimum centroid fluctuations and without significant increase in angular divergence. The remaining small fluctuations in centroid and angular distribution come from the uncancelled higher order (octopoles and decapoles) multipoles in the first IK and the prekick. The emittance tracking in Fig.\,\ref{fig:Cb} also shows no significant increase over the turns, indicating the effectiveness of the modified scheme. The growth in the emittance, (36.15\,mm\,mrad) slightly larger than without multipole case (36.02\,mm\,mrad), is due to the first kick and prekick. 

\section{Propagation of a magnetized beam \label{section:magnetization}}
\par In this section, the propagation of a magnetized beam through the CCR is studied. In subsection\,\ref{subsec:wo_kicker}, we will describe the design principle of the CCR without a harmonic kicker system that achieves the optimal cooling with a magnetized beam. In subsection\,\ref{subsec:w_kicker}, we add the kicker system to the CCR and describe the interaction of the kick with a magnetized beam, re-examining the cancellation scheme. In subsection\,\ref{subsec:mb_simulation}, a simulation study with a magnetized beam and a realistic kick model is presented. The cooling characteristics, including the Larmor emittance, are tracked to verify that their values stay within allowed limits. 

\subsection{Magnetized beam in a CCR without kickers \label{subsec:wo_kicker}}
\par The beamline design of the CCR appearing in \cite{pCDR} was optimized for high cooling efficiency without a kicker system. The design is based on the principle of using magnetized beam and a properly matched beamline, which was first proposed in \cite{burov1} (see also \cite{burov2} and \cite{kim}, which we will summarize in this section). \par
\begin{figure}[htb]
   \centering
   \includegraphics*[width=90mm]{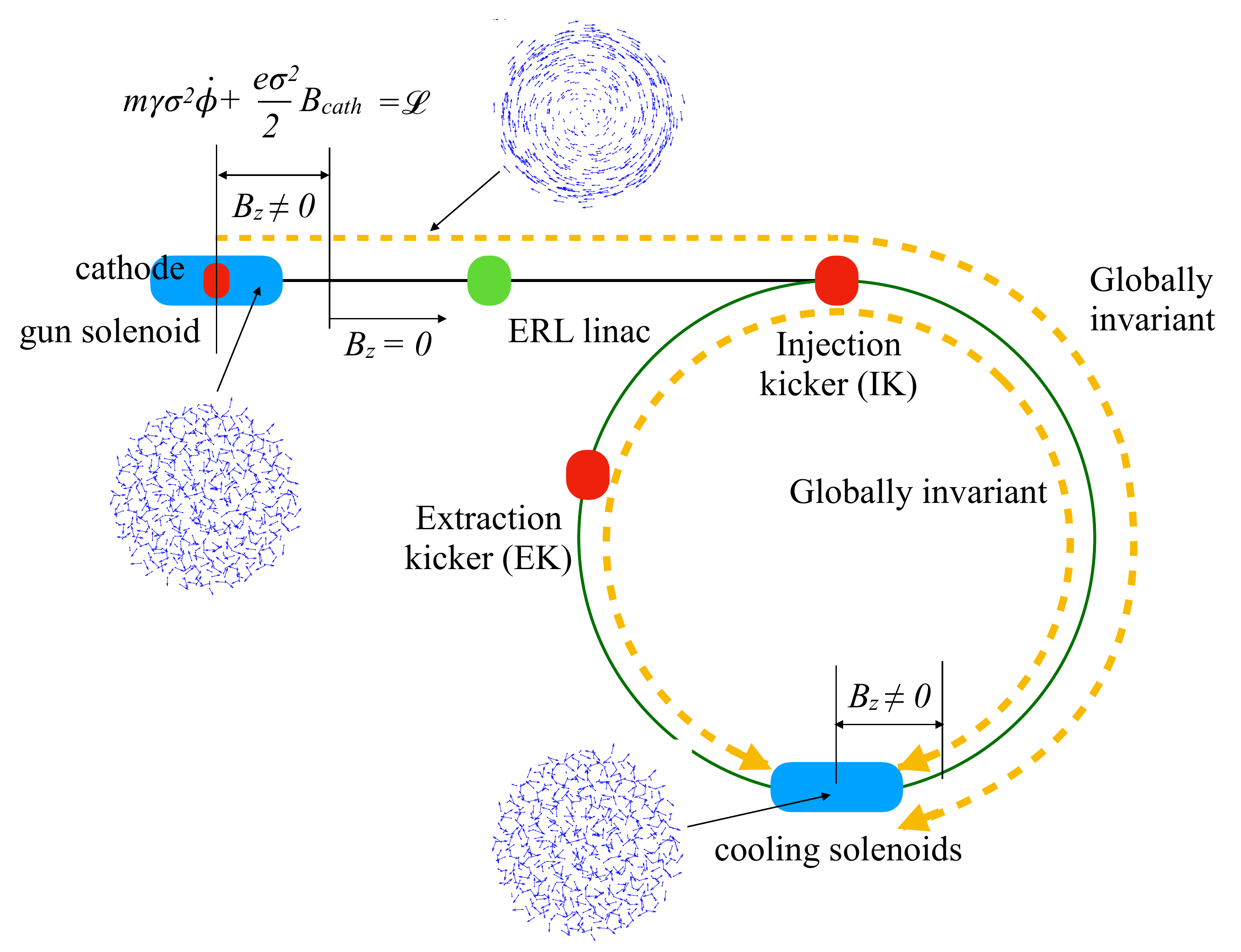}
   \caption{Transport of the magnetized beam to the cooler. The ERL linac mostly consists of an axisymmetric accelerating field and is a rotationally invariant, decoupled beamline.}
   \label{fig:in_scheme}
 \end{figure}

 The cooling efficiency within the cooling solenoids is inversely proportional to the cube of relative velocity and proportional to overlap of an electron and ion beam. Assuming the ion beam is on the beam axis,  the relative velocity increases rapidly as the electron transverse velocity characterized by cyclotron motion increases, while the overlap depends on various factors such as the transverse aspect ratio (i.e., roundness of the cross section), the offset of the centroid, angular divergence, and arrival time jitter. In nominal operation of the CCR without a centroid offset, angular divergence, and arrival time jitter, the optimal cooling efficiency would be achieved with a round magnetized (i.e., calm without cyclotron motion) beam in cooling solenoids. Such a beam can be obtained straightforwardly if the beam is generated at the cathode as a round magnetized beam and transported properly, i.e., through a globally invariant decoupled beamline~\cite{burov1}, \cite{burov2} (See Fig.\,\ref{fig:in_scheme})\textemdash We have found that the beamline does not have to be locally rotationally symmetric but only has to be globally symmetric. The globally invariant beamline conserves both the canonical angular momentum (CAM) and the decoupling of cyclotron motion so that the magnetized beam for optimal cooling rate can be recovered at the cooler. In Fig.\,\ref{fig:in_scheme}, the laser beam with the round transverse profile is applied to the cathode in a photocathode gun embedded in a Helmholtz coil to generate a round magnetized beam. Due to the uniform longitudinal magnetic fields provided by the Helmholtz coils, the motion of the beam remains largely longitudinal. 
The $4\times 4$ beam matrix of a round beam generated at the cathode, with a rotationally invariant distribution, is given in a Cartesian lab frame as
\begin{eqnarray}
\label{eqn:Sigma0}
\Sigma_{cath}=\left[\begin{array}{cc}
\mathcal{A} & 0\\
0 & \mathcal{A}
\end{array}
\right],\,\hbox{where }\mathcal{A}=\left[\begin{array}{cc}
\sigma_L^2+\beta\epsilon_{th} & 0\\
0 & \epsilon_{th}/\beta
\end{array}
\right],
\end{eqnarray}
where $\sigma_L$ is $\textit{rms}$ size of the laser spot, $\beta$ is the betatron function at the cathode, and $\epsilon_{th}$ is the thermal emittance of the gun, which in this case is negligibly small. Within the solenoid, the convenient canonical variables capable of describing the small transverse motion of the electrons are the displacement $\vec{d}$ of the Larmor center and the cyclotron motion $\vec{k}_\perp$ around the center, which are given as
\begin{eqnarray}
\label{eqn:k}
\vec{k}_\perp=\vec{p}_\perp+\frac{e}{2c}\left(\vec{B}\times\vec{\rho}\right),\\
\vec{d}=\vec{\rho}-\vec{\rho}_L=\frac{1}{2}\vec{\rho}-\frac{c}{eB^2}\left(\vec{p}\times\vec{B}\right),
\label{eqn:d}
\end{eqnarray}
where $\vec{B}=B\hat{z}$ and $\vec{\rho}_L=c(\vec{k}\times\vec{B})/eB^2$ is the relative displacement with respect to Larmor center. The circular basis $\hat{r}$ is defined based on the coordinates (\ref{eqn:k}) and (\ref{eqn:d}) and related to Cartesian basis $r$ as
\begin{eqnarray}
\hat{r}=\left[\begin{array}{c}
\kappa=\sqrt{\frac{c}{eB}}\left[\begin{array}{c}
k_y\\k_x
\end{array}
\right]
\\
\\
\xi=\sqrt{\frac{eB}{c}}\left[\begin{array}{c}
d_x\\d_y
\end{array}
\right]
\end{array}
\right], \quad \hat{r}=\mathcal{K}r,
\end{eqnarray}
where $2\times1$ vectors $\kappa, \xi$ are called cyclotron and drift degree of freedom, respectively. The explicit expression for $4\times4$ matrix $\mathcal{K}$, if needed, can be obtained from (\ref{eqn:k}) and (\ref{eqn:d}). 
The electrons subsequently evolve in the magnetic field of the solenoid (including the fringe field) to a round beam with rotation (i.e., non-trivial physical angular momentum) as described in Cartesian basis as
\begin{eqnarray}
\label{eqn:Sigma_m}
\Sigma_0&=&\left[\begin{array}{cc}
\mathcal{B} & L\mathcal{J}\\
-L\mathcal{J} & \mathcal{B}
\end{array}
\right], \quad  \hbox{where }  L =\frac{B_s}{2B\rho}(\sigma_L^2+\beta\epsilon_{th}),\quad\quad \\
\mathcal{B} &=& \varepsilon_{eff}\left[\begin{array}{cc}
\beta & 0\\
0 &1/\beta
\end{array}
\right],\quad 
\mathcal{J}=\left[\begin{array}{cc}
0 & 1\\
-1 & 0
\end{array}
\right],
\end{eqnarray} 
where $B_s$ and $B\rho$ are magnetic field of the solenoid and beam rigidity, respectively, and $L$ defines the magnetization of the beam $M=\gamma L$ and is identified as half of (physical) angular momentum. The two (degenerate) eigenvalues of $\Sigma_0$ have been computed in (\ref{eqn:Sigma_m}) and can be represented by the eigenemittances
\begin{eqnarray}
\varepsilon_\pm=\frac{\varepsilon_{eff}}{2}(\beta+\frac{1}{\beta})\Bigg\{1\pm\sqrt{1-\frac{4(\varepsilon_{eff}^2-L^2)}{\varepsilon_{eff}^2(\beta+1/\beta)^2}}\Bigg\},\quad 
\label{eqn:eigen_emit}
\end{eqnarray}
 These eigenemittances are often called the Larmor and drift emittances, respectively. Notice that within a solenoid, a purely longitudinal motion of homogeneous beam implies $\beta\approx1$ and (\ref{eqn:eigen_emit}) reduces to
\begin{eqnarray}
\varepsilon_\pm=\varepsilon_{eff}\pm L.
\label{eqn:eigenemittance}
\end{eqnarray}

\par The dynamics of a round magnetized beam along the beamline is most conveniently described in terms of beam matrix $\mathcal{S}$ and transfer matrices $\mathcal{T}$ of the beamline in the circular basis, which is obtained from those ($\Sigma, T$) in Cartesian basis via similarity transform, i.e., $\mathcal{K}(s_0)\Sigma \mathcal{K}^{-1}(s_0)$ and $\mathcal{K}(s)T(s, s_0)\mathcal{K}^{-1}(s_0)$ for an arbitrary initial coordinate $s_0$ and final coordinate $s$. The beam matrix $\Sigma_{0}$ (\ref{eqn:Sigma0}) at the beamline entrance is diagonalized by a set of symplectic circular bases, i.e., 
\begin{eqnarray}
\mathcal{S}_0=\mathcal{K}\Sigma_{0}\mathcal{K}^{-1}=\left[\begin{array}{cc}
\varepsilon_+\textbf{1} &0\\
0 &\varepsilon_-\textbf{1}
\end{array}
\right],
\label{eqn:S_0}
\end{eqnarray}
where $\varepsilon_\pm$ corresponds to the emittance associated with cyclotron and drift motion, respectively. Associated to the beam matrix, there exist two invariants under symplectic transform first found in \cite{kim}: If effective emittance $\varepsilon_{eff}$ is defined as square root of the determinant of $\mathcal{B}$ and ``4D emittance" $\varepsilon_{4D}$ as square root of the determinant of $\Sigma_0$, then they are related via (\ref{eqn:Sigma_m}) as
\begin{eqnarray}
\varepsilon_{4D}^2=\hbox{det}(\Sigma_0)=(\varepsilon_{eff}^2-L^2)^2.
\label{eqn:det}
\end{eqnarray}
Although $L$ and $\varepsilon_{eff}$ may not be invariant under an arbitrary symplectic transform, $\varepsilon_{4D}$ is always an invariant. In addition, there exists a trace invariant $I_2$ defined as
\begin{eqnarray}
I_2(\Sigma_0)=-\frac{1}{2}Tr(J_4\Sigma_0J_4\Sigma_0)=2(\varepsilon_{eff}^2+L^2).
\label{eqn:trace}
\end{eqnarray}
 Because the basis conversion matrix $\mathcal{K}$ to circular basis is symplectic, the 4D\,emittance and the trace invariant are conserved. Then using (\ref{eqn:S_0}), (\ref{eqn:det}), (\ref{eqn:trace}) and calculating det$(\mathcal{S}_0)$, $I_2(\mathcal{S}_0)$, we have the relation in equation (\ref{eqn:eigenemittance}). The beam transport between the two (cathode and cooler) solenoids is designed so that the corresponding transfer matrix $\mathcal{T}$ in a circular basis is globally rotation-invariant and decoupled, i.e., block-diagonal:
\begin{eqnarray}
\mathcal{T}=\left[\begin{array}{cc}
\mathcal{C}_+ &0\\
0 & \mathcal{C}_-
\end{array}
\right], \quad \mathcal{C}_\pm=\left[\begin{array}{cc}
\cos{\psi_\pm} &\sin{\psi_\pm}\\
-\sin{\psi_\pm} & \cos{\psi_\pm}
\end{array}
\right].
\label{eqn:beamline_gi}
\end{eqnarray}
Here $\mathcal{C}_\pm$ are $2\times2$ rotationally invariant sub-matrices parametrized by $\psi_\pm\in [0,2\pi]$. Notice that $det(\mathcal{C}_\pm)=1$ and $\mathcal{C}_\pm\mathcal{C}_\pm^T=1$. Then the transported beam matrix $\Sigma_c$ in the cooler is given as
\begin{eqnarray}
\mathcal{S}_{cool}=\mathcal{T}\mathcal{S}_{0}\mathcal{T}^T=\left[\begin{array}{cc}
\varepsilon_+\mathcal{C}_+\mathcal{C}_+^T & 0\\
0 & \varepsilon_-\mathcal{C}_-\mathcal{C}_-^T
\end{array}
\right]\nonumber\\
=\left[\begin{array}{cc}
\varepsilon_+\textbf{1} & 0\\
0 & \varepsilon_-\textbf{1}
\end{array}
\right],
\end{eqnarray}
which implies the conservation of the canonical angular momentum (CAM) and the eigenemittances (Larmor and drift) of the beam. Upon matching Twiss parameters at the cooler entrance, this would physically correspond to restoration of the round beam and negligibly small cyclotron motion in the cooler, if the beam starts with small cyclotron motion in the cathode.  
\par While the Busch's theorem holds throughout a globally rotation-invariant beamline, the theorem simplifies if the beamline is matched so that the cyclotron motion in the cooler is zero: According to the theorem, the canonical angular momentum (CAM) $\mathcal{L}$ for an electron within the (gun and cooler) solenoids is a constant of motion given as
\begin{eqnarray}
\mathcal{L}=m_e\gamma\rho^2\dot{\phi}+\frac{1}{2}e\rho^2B,\quad \hbox{in [SI]}
\end{eqnarray}
where $m_e$ is the mass of electron, $\gamma$ is the electron energy at the gun, $\rho$ is radial offset from the axis, and $B$ is a uniform magnetic field. With negligible transverse motions, i.e., $\dot{\phi}=0$, in both solenoids, the CAM for the beam can be computed as
\begin{eqnarray}
\mathcal{L}=\frac{1}{2}eB_{cath}\sigma_{cath}^2=\frac{1}{2}eB_{cool}\sigma_{cool}^2,
\label{eqn:matching1}
\end{eqnarray}
where $\sigma_{cath}, \sigma_{cool}$ are the $rms$ beam size of the electron and $B_{cath}, B_{cool}$ are the magnetic field at the cathode and the cooler, respectively. From (\ref{eqn:matching1}), the solenoid fields at the cooler can be adjusted to match the electron beam size to the ion beam size. According to Table\,\ref{table:bp}, the beam radius of electron bunch at the cooling channel and the cathode are $r_{cooler}=0.35$\,mm and $r_{cath}=1.1$\,mm ($\sigma_L=2r_{cath}$), respectively, leading to $B_{cool}=1$\,T for the given $B_{cath}=0.1$\,T. 
\par
\begin{figure*}[htb]
 \centering
 \includegraphics*[width=180mm]{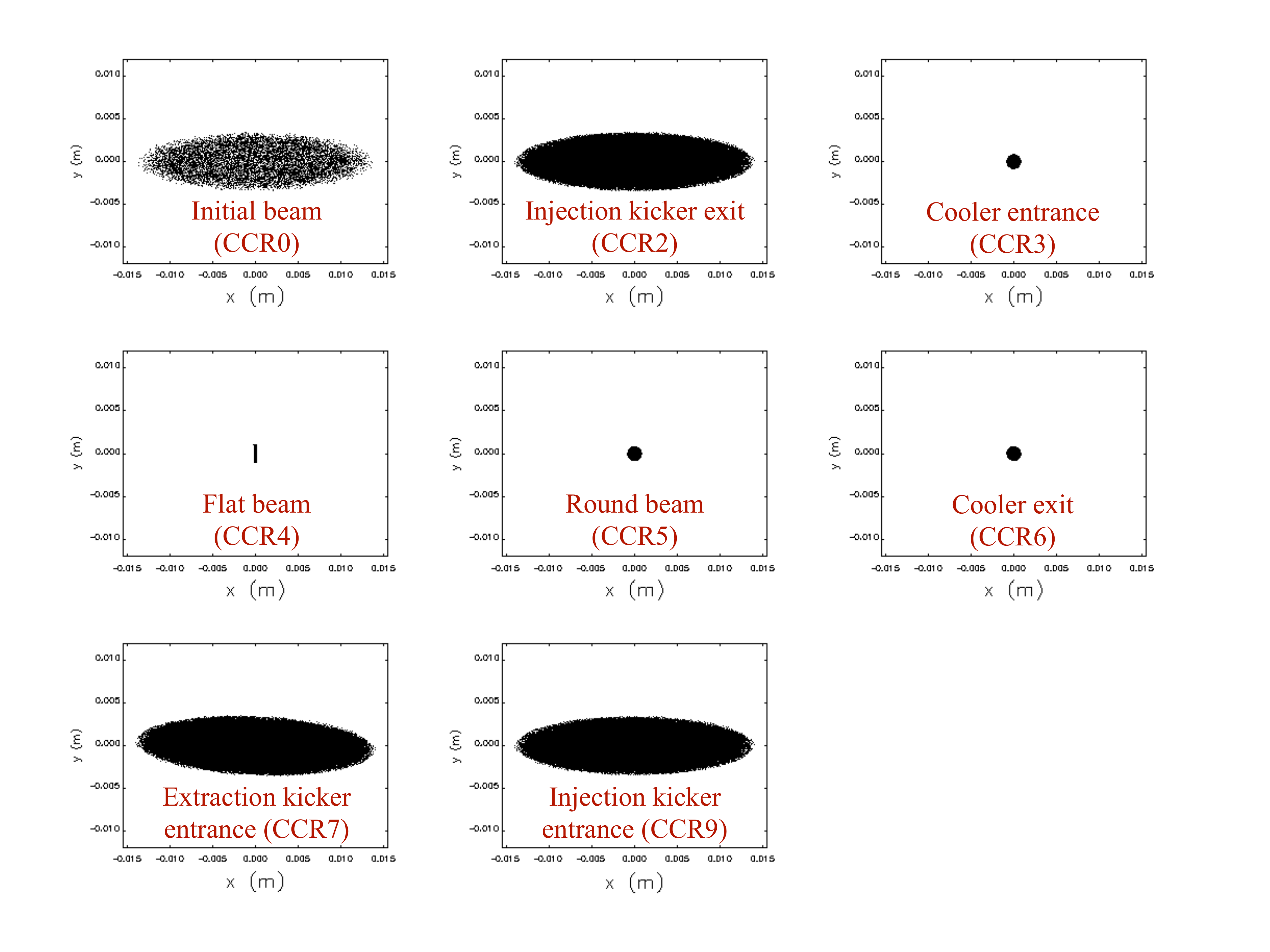}
 \caption{The transverse ($x-y$) profiles of the magnetized beam around the CCR without kickers. The profiles were overlapped over 11\,passes and they are close to single pass profiles due to stable beam dynamics. }
 \label{fig:x-y-no_kicker}
\end{figure*}

 As a benchmark to the CCR design in~\cite{pCDR}, the propagation of a round magnetized beam through the CCR, optimized without kickers, was simulated in ELEGANT, tracking the characteristics of cooling efficiency to define a baseline for the modification that includes a kicker system. In the beamline setup of the simulation, a round-to-flat beam transformation (RTFB)~\cite{douglas1} was inserted at the cooler entrance (at ``flat beam" (CCR4) in Table\,\ref{table:Sigma-no_kicker}) for beam diagnostic purpose: half the vertical (horizontal) effective emittance $\varepsilon_x$ ($\varepsilon_y$) of the artificially created flat beam corresponds to the Larmor (drift) emittance of the round beam, respectively. The flat beam was transformed back to the round beam by a flat-to-round beam transformer (FTRB)~\cite{douglas1}, before propagating into the cooling solenoids. The beam parameters at important watch points around the CCR are listed in Table\,\ref{table:Sigma-no_kicker}. In Fig.\,\ref{fig:x-y-no_kicker}, the transverse ($x$-$y$) beam profiles (overlapped over 11 passes) around the CCR are shown. With the beamline designed to be only globally invariant, the transverse profiles at the kickers are not round but elliptical. The round magnetized beam profile is recovered at the cooler entrance in every pass with aspect ratios very close to 1, implying that the CAM is conserved and beamline is indeed globally invariant. There is no significant degradation in the Larmor emittance over 11\,passes, indicating the beamline between the two solenoids is properly decoupled. 

\begin{table*}[hbt]
   \centering
   \caption{The averaged beam parameters of the magnetized beam along the CCR over 11\,passes. WIK is injection kicker entrance, WIC the cooler entrance, WCE cooler exit, WEK extraction kicker exit. There is a mirror symmetry in the beam parameters with respect to the cooler.}
   \vspace{5pt}
   \begin{tabular}{llcccc}
       \toprule
           Beam parameters& unit &WIK & WIC & WCE & WEK \\ 
           \hline
           Twiss parameter $\alpha_x$ & - & 0 & 0  & 0 & 0\\
           Twiss parameter $\beta_x$ & m & 110 & 0.37 & 0.37 & 109\\
           Twiss parameter $\alpha_y$ & - & 0 & 0  & 0 & 0\\
           Twiss parameter $\beta_y$ & m & 6.7 & 0.37 & 0.36 & 6.8\\
           $rms$ size $\sigma_x$ & mm & 6.13 & 0.36 & 0.36 & 6.14\\
           $rms$ size $\sigma_y$ & mm & 1.51 & 0.36 & 0.36 & 1.51\\
           Aspect Ratio $\eta$ & - & 4.06 & $\sim$1 & $\sim$1 & 4.06\\
           $rms$ angular divergence $\sigma_{x'}$ & mrad & 0.056 & 0.97 & 0.97 & 0.056\\
           $rms$ angular divergence $\sigma_{y'}$ & mrad & 0.23 & 0.97 & 0.97 & 0.23\\
           Drift normalized emittance $\varepsilon_{n+}/2$ & mm\,mrad & - & 36 & 36 & -\\
           Larmor normalized emittance $\varepsilon_{n-}/2$ & mm\,mrad & - & 1 & 1 & -\\
           Normalized CAM $\mathcal{L}/m_e\gamma c$ & mm\,mrad & -1.4 & -0.65 & -0.65 & -1.4\\
        \botrule
   \end{tabular}
   \label{table:Sigma-no_kicker}
\end{table*}

\subsection{Interaction of harmonic kicks with a magnetized beam\label{subsec:w_kicker}}
\par When a deflecting kick in a specific direction is inserted into the beamline of the CCR, the beamline is neither globally invariant nor decoupled anymore, leading to the possibility of beam quality degradation. The analysis of the effects of deflecting RF kick on the magnetized beam was done in \cite{douglas} in a simpler context of TM type deflecting cavity\textemdash the deflection by the TM mode is purely rotational by the magnetic fields and leads to a momentum change dependent on the factor $\cos{\omega\tau}$. In \cite{douglas}, the transformation of the phase space variables was explicitly computed in linear optics with the kick modeled as an impulsive kick and its effects linearized. In addition, the cancellation scheme was shown to be effective, which would imply that the cooling efficiency does not degrade except for a small degradation from the injection kick on the first pass. However, it was subsequently demonstrated through numerical simulations that the cancellation scheme fails when an extended kick model is used. Here we take a more general and realistic approach, making use of a general phase space transform (\ref{eqn:px})-(\ref{eqn:y}), where (1) the deflecting kicks are based on TEM modes with the momentum change depending on the factor $\sin{\omega\tau}$, (2) the kicks include non-trivial multipole fields so that the computed transforms include a non-linear contribution, and (3) the longitudinal profile is extended over the effective field range. 
\par Unlike the case of a non-magnetized beam, the initial transverse momenta $p_{x0}, p_{y0}$ of a magnetized beam are small but significant as determined by the angular momentum. Subsequently, this leads to a transverse deviation of the beam from the zero-slope trajectory for a small perturbation. In view of (\ref{eqn:px})-(\ref{eqn:y}), the non-zero transverse momenta are now coupled with other components of the fields, for example $B_z$, that would change the angular momentum. Also the transverse coordinates of the fields within the integrals are not constant $r_{\perp0}$ anymore. This is inconsistent with the impulsive kick model based on the zero-slope trajectory as defined at the kicker entrance. To account for the trajectory deviation, an extended kick model must be introduced based on a realistic 3D\,field profiles as obtained from the CST simulations. Assuming that the spatial profile $F_{nm}$ of a generic $n$th-multipole field in the $m$th mode is separable as $F_{nm}(\vec{r}_\perp, s)=F_{\perp,nm}(\vec{r}_\perp)F_{\parallel,nm}(s)$ with temporal profile $e^{-i\omega_m (t+\tau)}$ oscillating at the frequency $\omega_m$, the longitudinal profile $F_{\parallel,nm}$ is fitted to the CST-generated kick profile, which is a combination of Gaussian profile and its derivative:
\begin{eqnarray}
F_{\parallel,nm}(s)=\mathcal{N}_{nm}e^{-s^2/2\sigma_{nm}^2}(1+|s|B_{nm}), \\
\text{where } \mathcal{N}_{nm}=\frac{1}{\sqrt{2\pi\sigma_{nm}^2}+2\sigma_{nm}^2B_{nm}}.
\label{eqn:Gaussian}
\end{eqnarray}
Here $\sigma_{nm}$ is the standard deviation of the Gaussian profile, $B_{nm}$ is a fitting constant to the CST-generated profiles, and $\mathcal{N}_{nm}$ is a normalization constant so that the integral of $F_{\parallel,m}$ is 1. With the transverse profile $F_{\perp,nm}$ identified with the multipole expansion at the fixed transverse coordinate $r_{\perp}$, this profile produces the nominal kick voltage on beam axis. Collecting all the multipoles and modes, the total field profile $F$ can be approximated as a series of impulsive kicks separated by drift spaces:
\begin{widetext}
\begin{eqnarray}
\label{eqn:profile_long}
F_{e}^{X,Y}(r_\perp, s)=\sum_{k=-N}^N\sum_{n,m=1}^5\Bigg\{1+B_{nm}^{X,Y}|k|d\Bigg\}A_{nm}^{X,Y}(r_{\perp k}, k)\delta(s-kd),\\
\label{eqn:profile_long1}
\text{where } A_{nm}^{X,Y}(r_{\perp k}, k)=\frac{1}{Z_{nm}^{X,Y}}e^{-k^2d^2/2\sigma_{nm}^2}\mathfrak{Re}\Bigg[c_{nm}^{X,Y}n(x_k+iy_k)^{n-1}\Bigg],\\
\label{eqn:profile_long2}
Z_{nm}^{X,Y}=\sum_{k=-N}^N(1+B_{nm}^{X,Y}|k|d)e^{-k^2d^2/2\sigma_{nm}^2},\quad c_{nm}^X=c_{nm},\quad c_{nm}^Y=ic_{nm}.
\end{eqnarray}
\end{widetext}
where $F_e$ is a Lorentz force in transverse direction per charge, $d$ vacuum (drift) spacing between $\delta$ functions, $N$ roughly half the number of $\delta$ functions in the field profile range ($(2N+1)d=l$), $B_{nm}^{X,Y}$ a fitting constant to the CST-generated profiles, and $A(r_{\perp,k},k)$ the field amplitude of the $\delta$ function at the $k$th displacement. In this model, we used $r_{\perp}(s)=r_{\perp k}\delta (s-kd)$ with $r_{\perp k}$ recursively determined from initial value $r_{\perp0}$ via $r_{\perp k}=r_{\perp k-1}+p_{\perp k-1}d/(m_e\gamma c)=r_{\perp k-1}+d/(m_e\gamma c)\sum_{n,m=1}^5nc_{nm}(r_{\perp k-1})\,^{n-1}$. Then making use of (\ref{eqn:profile_long})-(\ref{eqn:profile_long2}) as the transverse component of the Lorentz force, the $V_{x,y},W_{x,y}$ appearing in (\ref{eqn:px})-(\ref{eqn:y}) are modified as follows. First, the coordinates $r_{0\perp}$ in the arguments of the fields in (\ref{eqn:px})-(\ref{eqn:y}) are replaced by $r_\perp$, and the integration is broken up into a series of segments containing $\delta$-functions. Then each integration over the $k$th segment $(k-1/2)d\leq s <(k+1/2)d$ is done in the similar way as in an impulsive kick model with a different $r_{\perp k}$ at each $s=kd$. Consequently, we have for a kick direction
\begin{widetext}
\begin{eqnarray}
\label{eqn:px_ext}
p_{x}&=&p_{x0}\cos{\Theta}+p_{y0}\sin{\Theta}+\frac{e}{c}V_{x},\\
\label{eqn:Vx_ext}
\text{where }V_x&=&\int_{-l/2}^{l/2}\!ds\,\sum_{k=-N}^N\Big\{1+B_{nm}^X|k|d\Big\}A_{nm}^X(k)\delta(s-kd)\cos{(\omega_n\tau)}\nonumber\\
&=&\sum_{k=-N}^N\sum_{n,m=1}^5\Big\{1+B_{nm}^X|k|d\Big\}A_{nm}^X(k)\cos{(\omega_n\tau)},\\
\label{eqn:x_ext}
x&=&x_{0}+Cp_{x0}+Sp_{y0}+W_{x},\\
\label{eqn:Wx_ext}
\text{where }W_x&=&\int_{-l/2}^{l/2}\!ds\,\frac{e}{m_e c^2\gamma}\int_{-l/2}^{s}\!ds'\,\sum_{k=-N}^N\Big\{1+B^X_{nm}|k|d\Big\}A_{nm}(k)\delta(s'-kd)\nonumber\\
&=&\frac{e}{m_e c^2\gamma}\Bigg[\int_{-l/2}^{-l/2+d}\!ds+\cdots+\int_{-l/2+k'd}^{-l/2+(k'+1)d}\!ds+\cdots+\int_{l/2-d}^{l/2}\!ds\Bigg]\sum_{k=-N}^{N(s)}\Big\{1+B^X_{nm}|k|d\Big\}A_{nm}(k)\nonumber\\
&=&\frac{e}{m_e c^2\gamma}\sum_{k'=0}^{2N}\sum_{k=-N}^{-(N-k')}\,\Big\{1+B^X_{nm}|k|d\Big\}A_{nm}(k),
\end{eqnarray}
\end{widetext}
where the $B_z$ field in $\Theta$ has a similar expansion as (\ref{eqn:profile_long}). The phase space transform in $y$ components is similarly obtained with the replacement of $A_{nm}^X, B_{nm}^X, Z_m^X, c_{nm}^X$ with $A_{nm}^Y, B_{nm}^Y, Z_m^Y, c_{nm}^Y$, respectively. For the beam dynamics simulation with magnetized beam in the next subsection, this extended kick model will be implemented into the ELEGANT and $x,p_x$ in (\ref{eqn:px_ext}) and (\ref{eqn:x_ext}) will be used to numerically evaluate the phase space transform for comparison. In (\ref{eqn:px_ext}), (\ref{eqn:x_ext}), the extended kick model has three contributions to the changes in phase space variables. One comes directly from the initial momentum $p_{0x,0y}l/(m_e\gamma c)$ as determined by the CAM of the magnetized beam. This contribution is accounted by the first term in (\ref{eqn:x}) and (\ref{eqn:y}), which would vanish in an impulsive kick model. The second is from the multipole fields at non-zero displacement as in impulsive kick model, leading to the changes in $V_{x,y}, W_{x,y}$: the electrons at different initial offset will get different kicks. The third is from combination of momentum change at each impulsive kick, its corresponding displacement over each drift space, and a different kick (due to multipoles) at the next impulsive kick. Consequently, both momentum and the displacement change, but this change is relatively small due to compact length of the kicker profile. With the second contribution almost cancelled in the cancellation scheme, the first contribution becomes a leading order contribution to the change, i.e., no significant momentum change, but the more subtle parameters such as the Larmor emittance are affected via the change of transverse beam size.  
\par Subsequently, the cancellation scheme with an extended kick model is examined based on phase space variable transform (\ref{eqn:px_ext}), (\ref{eqn:x_ext}) with (\ref{eqn:Vx_ext}), (\ref{eqn:Wx_ext}). With a magnetized beam whose $4$D phase space variables are denoted as $V_k=(x_k, p_{xk}, y_k,p_{yk})^T$\textemdash the subscripts in $V_k$'s, i.e., $k=0,a,b,c$ denote the phase space variables before the EK, after the EK, before the IK, and after the IK, respectively\textemdash the betatron phase advance is given by a generalized (block-diagonal) transform matrix, i.e., 
\begin{eqnarray}
\left[\begin{array}{c}x_b \\ p_{xb}\\y_b\\ p_{yb}\end{array}\right]=\left[\begin{array}{cccc}P_x&  Q_x& 0 & 0 \\R_x & S_x &0& 0\\0 & 0 & P_y & Q_y \\0 & 0 & R_y &S_y\end{array}\right]\left[\begin{array}{c}x_a \\ p_{xa}\\y_a \\ p_{ya}\end{array}\right].
\label{eqn:betatron_ext}
\end{eqnarray}
Here we assumed the longitudinal phase space transform of the betatron phase advance is identity, i.e., the beamline is isochronous and monoenergetic~\cite{douglas}. Then through a pair of the kickers with betatron phase advance (\ref{eqn:betatron_ext}), using the general formula (\ref{eqn:px_ext})-(\ref{eqn:Wx_ext}), phase space variables transform from $V_0$ to $V_c$ as
\begin{widetext}
\begin{eqnarray}
\label{eqn:px_gen}
p_{xc}=R_x\cos{\widetilde{\Theta}}x_0+R_y\sin{\widetilde{\Theta}}y_0+\Bigg[\cos{\widetilde{\Theta}}(S_x\cos{\Theta}+CR_x)-\sin{\widetilde{\Theta}}(S_y\sin{\Theta}+SR_y)\Bigg]p_{x0}\nonumber\\
+\Bigg[\cos{\widetilde{\Theta}}(S_x\sin{\Theta}+SR_x)+\sin{
\widetilde{\Theta}}(S_y\cos{\Theta}+CR_y)\Bigg]p_{y0}+R_x\cos{\widetilde{\Theta}}W_x+R_y\sin{\widetilde{\Theta}}W_y\nonumber\\
+\Bigg[\frac{e\widetilde{V}_x}{c}+\frac{eV_x}{c}S_x\cos{\widetilde{\Theta}}\Bigg]+S_y\sin{\widetilde{\Theta}}\frac{eV_y}{c},\\
\label{eqn:py_gen}
p_{yc}=R_y\cos{\widetilde{\Theta}}y_0-R_x\sin{\widetilde{\Theta}}x_0+\Bigg[\cos{\widetilde{\Theta}}(S_y\cos{\Theta}+CR_y)-\sin{\widetilde{\Theta}}(S_x\sin{\Theta}+SR_x)\Bigg]p_{y0}\nonumber\\
-\Bigg[\cos{\widetilde{\Theta}}(S_y\sin{\Theta}+SR_y)+\sin{\widetilde{\Theta}}(S_x\cos{\Theta}+CR_x)\Bigg]p_{x0}+R_y\cos{\widetilde{\Theta}}W_y-R_x\sin{\widetilde{\Theta}}W_x\nonumber\\
+\Bigg[\frac{e\widetilde{V}_y}{c}+\frac{eV_y}{c}S_y\cos{\widetilde{\Theta}}\Bigg]-S_x\sin{\widetilde{\Theta}}\frac{eV_x}{c},\\
\label{eqn:x_gen}
x_c=(P_x+\widetilde{C}R_x)x_0+\Bigg[C(P_x+\widetilde{C}R_x)-S\widetilde{S}R_y+\cos{\Theta}(Q_x+\widetilde{C}S_x)-\widetilde{S}S_y\sin{\Theta}\Bigg]p_{x0}\nonumber\\
+\widetilde{S}R_yy_0+\Bigg[S(P_x+\widetilde{C}R_x)+C\widetilde{S}R_y+\sin{\Theta}(Q_x+\widetilde{C}S_x)+\widetilde{S}S_y\cos{\Theta}\Bigg]p_{y0}\nonumber\\
+(P_x+\widetilde{C}R_x)W_x+\widetilde{W}_x+(Q_x+\widetilde{C}S_x)\frac{eV_x}{c}+\widetilde{S}R_yW_y+\widetilde{S}S_y\frac{eV_y}{c},\\
\label{eqn:y_gen}
y_c=(P_y+\widetilde{C}R_y)y_0+\Bigg[C(P_y+\widetilde{C}R_y)-S\tilde{S}R_x+\cos{\Theta}(Q_y+\widetilde{C}S_y)-\widetilde{S}S_x\sin{\Theta}\Bigg]p_{y0}\nonumber\\
-\widetilde{S}R_xx_0-\Bigg[S(P_y+\widetilde{C}R_y)+C\widetilde{S}R_x+\sin{\Theta}(Q_y+\widetilde{C}S_y)+\widetilde{S}S_x\cos{\Theta}\Bigg]p_{x0}\nonumber\\
+(P_y+\widetilde{C}R_y)W_y+\widetilde{W}_y+(Q_y+\widetilde{C}S_y)\frac{eV_y}{c}-\widetilde{S}R_xW_x-\widetilde{S}S_x\frac{eV_x}{c},
\end{eqnarray}
\end{widetext}
where $\Theta, C,S, V_{x,y}, W_{x,y}$ are evaluated at extraction kicker, while $\widetilde{\Theta}, \widetilde{C}, \widetilde{S}, \widetilde{V}_{x,y}, \widetilde{W}_{x,y}$ are at injection kicker. In this general phase space transform (\ref{eqn:px_gen})-(\ref{eqn:y_gen}), the cancellation can not be made perfect for any choice of transform matrix in (\ref{eqn:betatron_ext}). Notice that with an impulsive kick model with $B_z=0$, the kicks are completely cancelled through the multipole cancellation scheme. Firstly $l\rightarrow 0$ and $\widetilde{\Theta}, \Theta\rightarrow 0$ implies $C,\widetilde{C}\rightarrow l/m_e c\gamma_0\rightarrow 0, S,\widetilde{S}\rightarrow 0$, $W_x, \widetilde{W}_x, W_y, \widetilde{W}_y\rightarrow 0$ and with the choice of $P_x=S_x=P_y=S_y=-1$ and $Q_x=R_x=Q_y=R_y=0$ in (\ref{eqn:betatron_ext}), the transforms (\ref{eqn:px_ext})-(\ref{eqn:x_ext}) reduce to
\begin{eqnarray}
\label{eqn:px_red}
p_{xc}=-p_{x0}+\Bigg\{\frac{e\widetilde{V}_x}{c}-\frac{eV_x}{c}\Bigg\},\\
\label{eqn:py_red}
p_{yc}=-p_{y0}+\Bigg\{\frac{e\widetilde{V}_y}{c}-\frac{eV_y}{c}\Bigg\},\\
\label{eqn:x_red}
x_c=-x_0,\quad \\
\label{eqn:y_red}
y_c=-y_0.\quad 
\end{eqnarray}
Now the kicks are completely cancelled with $V_x=\widetilde{V}_x, V_y=\widetilde{V}_y$. Then the only remaining effect is by the injection kick at the first pass, which is relatively small.
\par The realistic 3D\,field profiles of the kick within the kicker cavity, as suggested from the CST simulations, have negligibly small $B_z, E_y, B_x$ fields in the vicinity of the beam axis. This implies $V_y,W_y\rightarrow 0$, $\Theta\rightarrow 0$, $C\rightarrow l/m_ec\gamma, S\rightarrow 0$. 
Now with the choice of $P_x=-1, R_x=0, Q_x=l/m_ec\gamma_0, S_x=-1$ in (\ref{eqn:betatron_ext}), we can eliminate many contributing terms in general phase space transforms (\ref{eqn:px_gen})-(\ref{eqn:y_gen}) leading to 
\begin{eqnarray}
\label{eqn:xp_simple}
p_{xc}=-p_{x0}+\frac{e\widetilde{V}_x}{c}-\frac{eV_x}{c},\\
\label{eqn:py_simple}
p_{yc}=-p_{y0}+\frac{e\widetilde{V}_y}{c}-\frac{eV_y}{c},\\
\label{eqn:x_simple}
x_c=-x_0-\frac{l}{m_ec\gamma_0}p_{x0}-W_x+\widetilde{W}_x,\\
\label{eqn:y_simple}
y_c=-y_0-\frac{l}{m_ec\gamma_0}p_{y0}-W_y+\widetilde{W}_y.
\end{eqnarray}
In (\ref{eqn:xp_simple})-(\ref{eqn:y_simple}), the momentum cancellation can not be exact because the multipole effects in $V_x(x_0,y_0)$ and $\widetilde{V}_x(x_b,y_b)$ are not the same. However, these effects are not significant to angular distribution of the beam, as shown in Fig.\,\ref{fig:MB-angular_dist}. On the other hand, with negligibly small $W_x, \widetilde{W}_x, W_y, \widetilde{W}_y$ over the short field range, the offset evolutions approximate to $x_c\approx-x_0-lp_{x0}/m_e\gamma$, $y_c\approx-y_0-lp_{y0}/m_e\gamma$, increasing the Larmor emittance. 

\subsection{Beam dynamics simulations for the CCR with kickers \label{subsec:mb_simulation}}
\par
\begin{figure}[htb]
   \centering
   \includegraphics*[width=85mm]{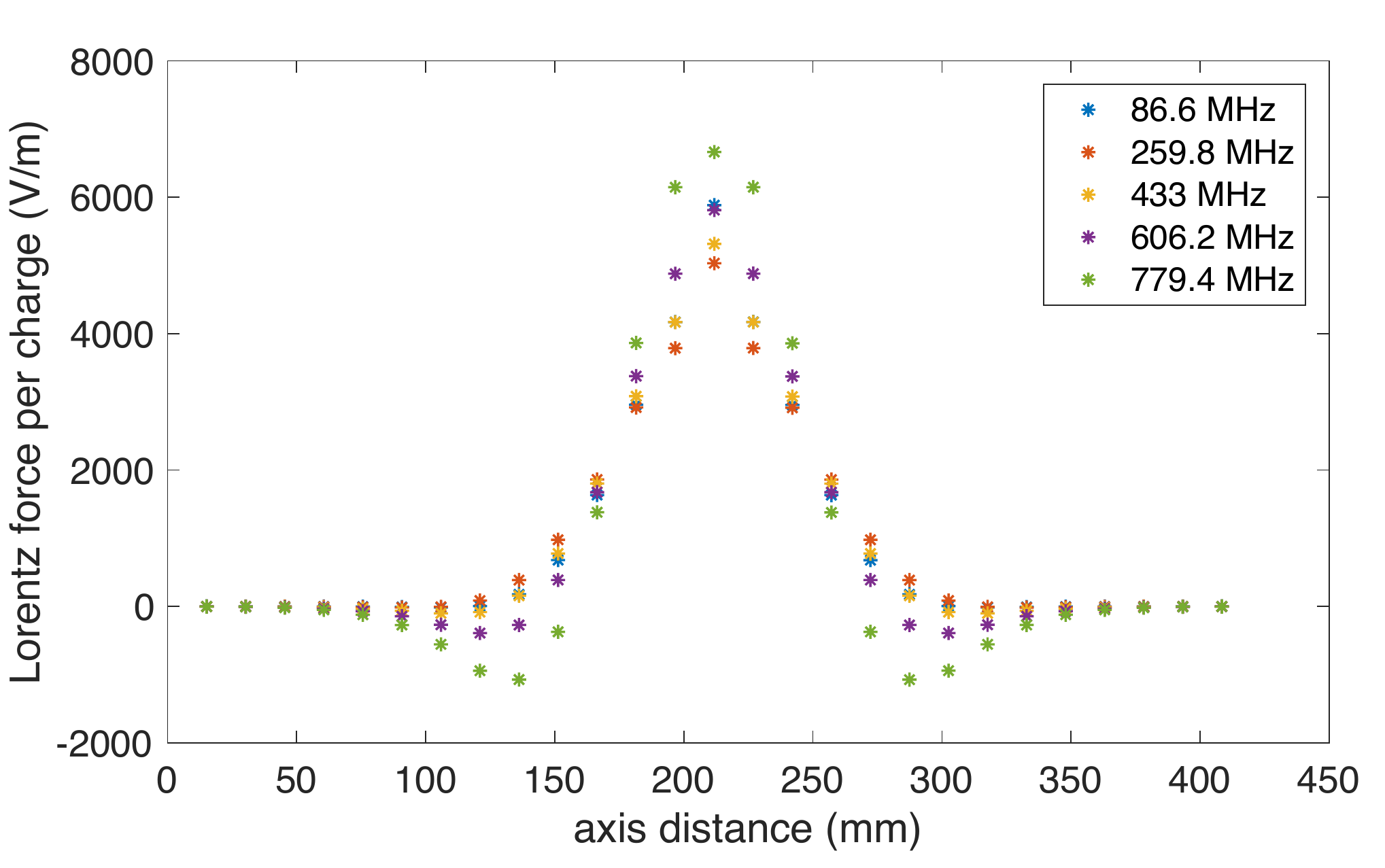}
   \caption{The longitudinal field profile in a long kick model for the Lorentz force. The model consists of a series of 27 $\delta$-function kicks separated by drift spaces. The envelope of the amplitudes is Gaussian.}
   \label{fig:long_kick_profile}
 \end{figure}

 In this subsection, we present simulation results for the CCR beam dynamics with harmonic kicks implemented, tracking a few critical parameters for cooling efficiency. The propagation of the magnetized beam would be most accurately simulated with the 3D field maps from the CST-MWS\textemdash whose direct import into the beam dynamics simulation is available in the tracking code GPT~\cite{gpt}. However, GPT, designed as a tracking code for linear accelerators, does not support multipass-tracking or transfer matrix implementation, which makes it inadequate for full-fledged CCR beam dynamics simulations. In this paper, we use GPT only to benchmark the ELEGANT simulation and analytical computations, by simulating a single pass through the cavity. The benchmarking is described in Appendix \ref{apx:gpt}. The results from the ELEGANT, GPT, and analytical computation show that the beam parameters, beam matrix elements, and centroid trajectories agree with one another to within 5\,$\%$. Therefore, we used ELEGANT to carry out the full simulations with an appropriate kick model for the 3D field maps: a harmonic kicker was realized in an extended kick model, i.e., a train of $\delta$ functions spaced with drift space as shown in Fig.\,\ref{fig:long_kick_profile}.  
 The amplitude of each $\delta$ function is scaled according to the pseudo-Gaussian (sum of Gaussian and its derivative)  longitudinal kick profile as obtained from the CST-MWS simulation. 
\par
\begin{figure*}[hbt]
  \centering
    \subfigure[Eigenemittance evolution.]{\label{fig:MB-emittances}\includegraphics[scale=0.37]{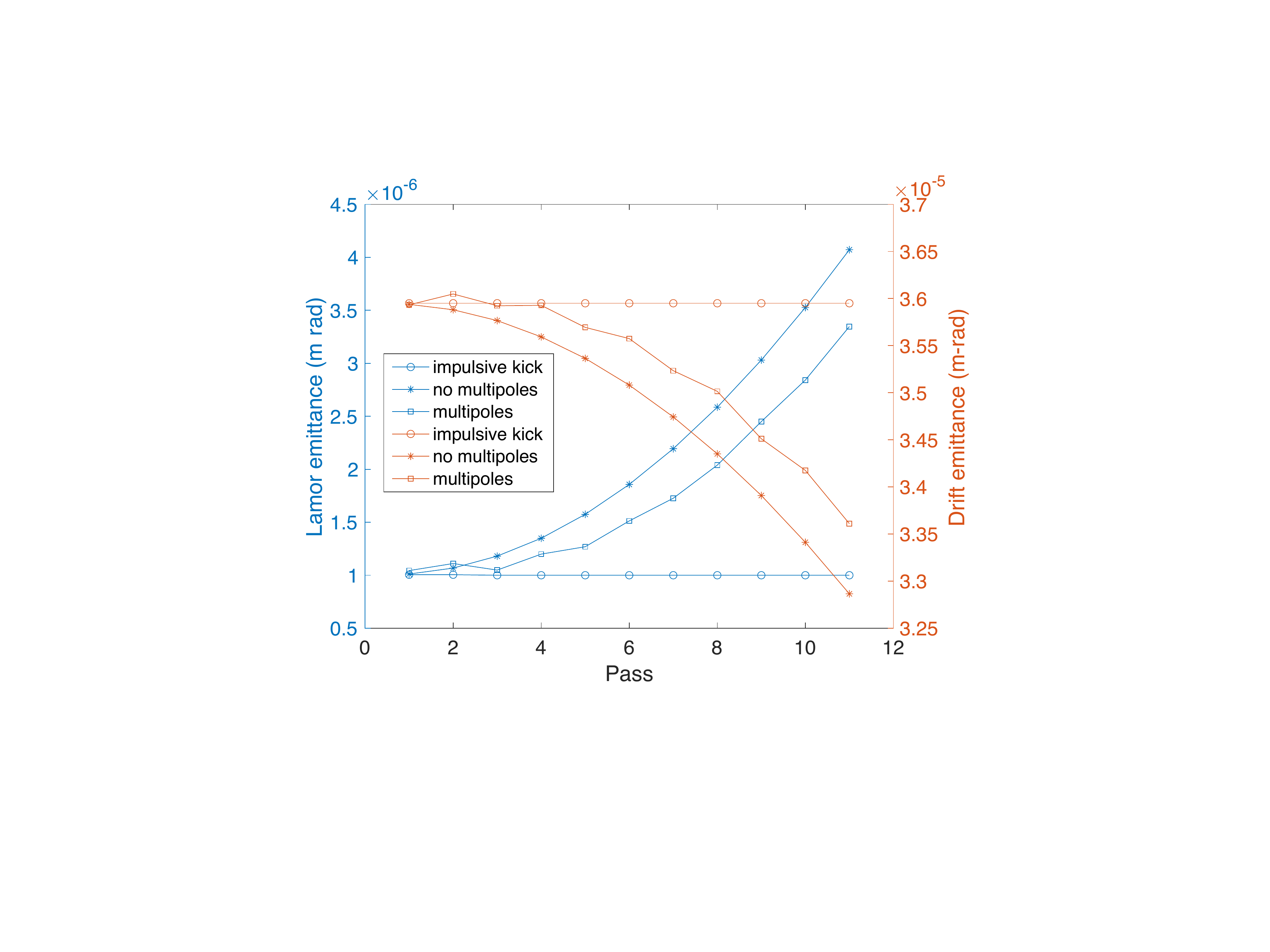}}%
    ~
    \subfigure[Evolution of CAM and aspect ratio.]{\label{fig:MB-CAM-AR}\includegraphics[scale=0.395]{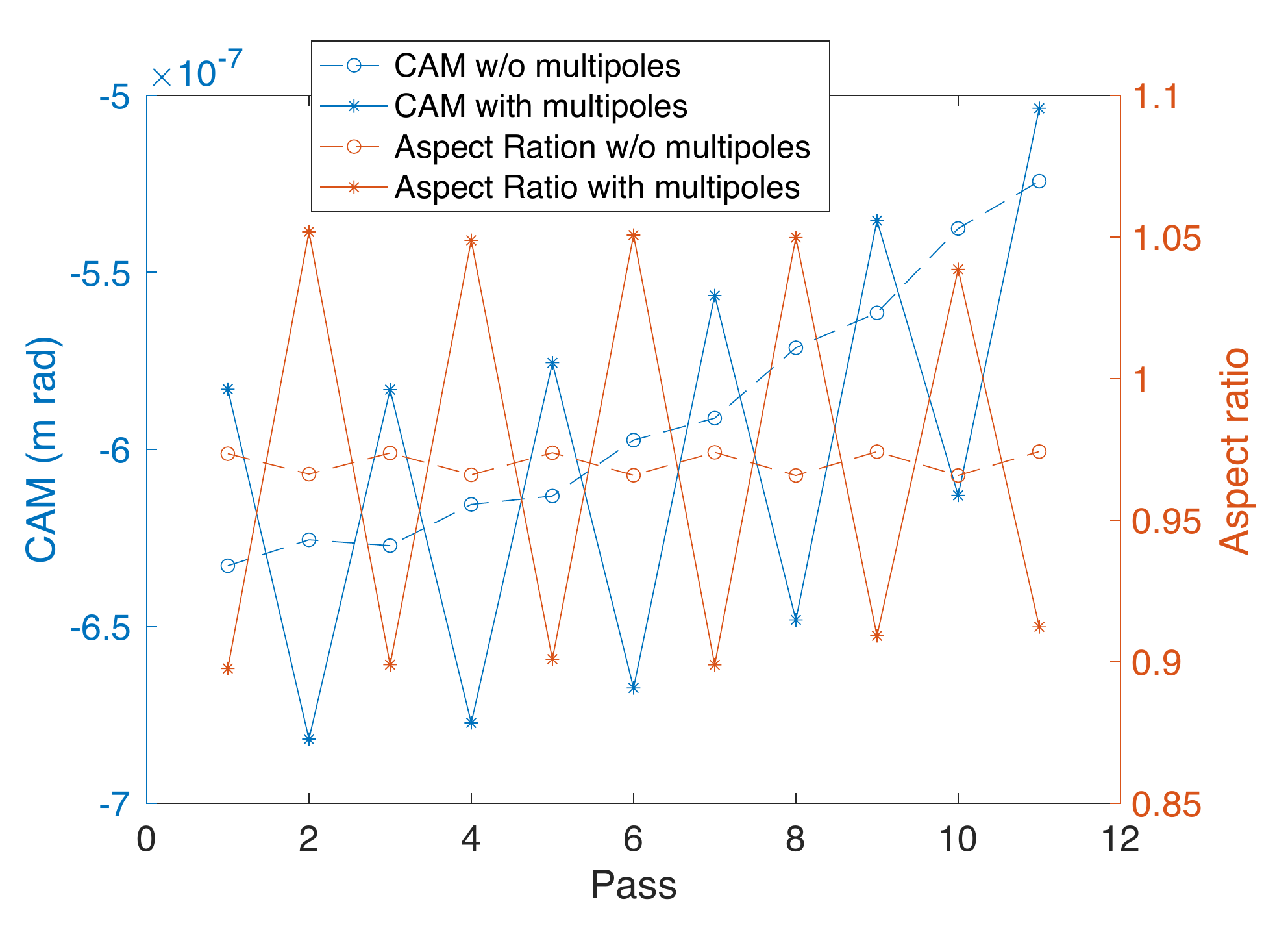}}\\
    \subfigure[Centroid evolution.]{\label{fig:MB-centroid}\includegraphics[scale=0.375]{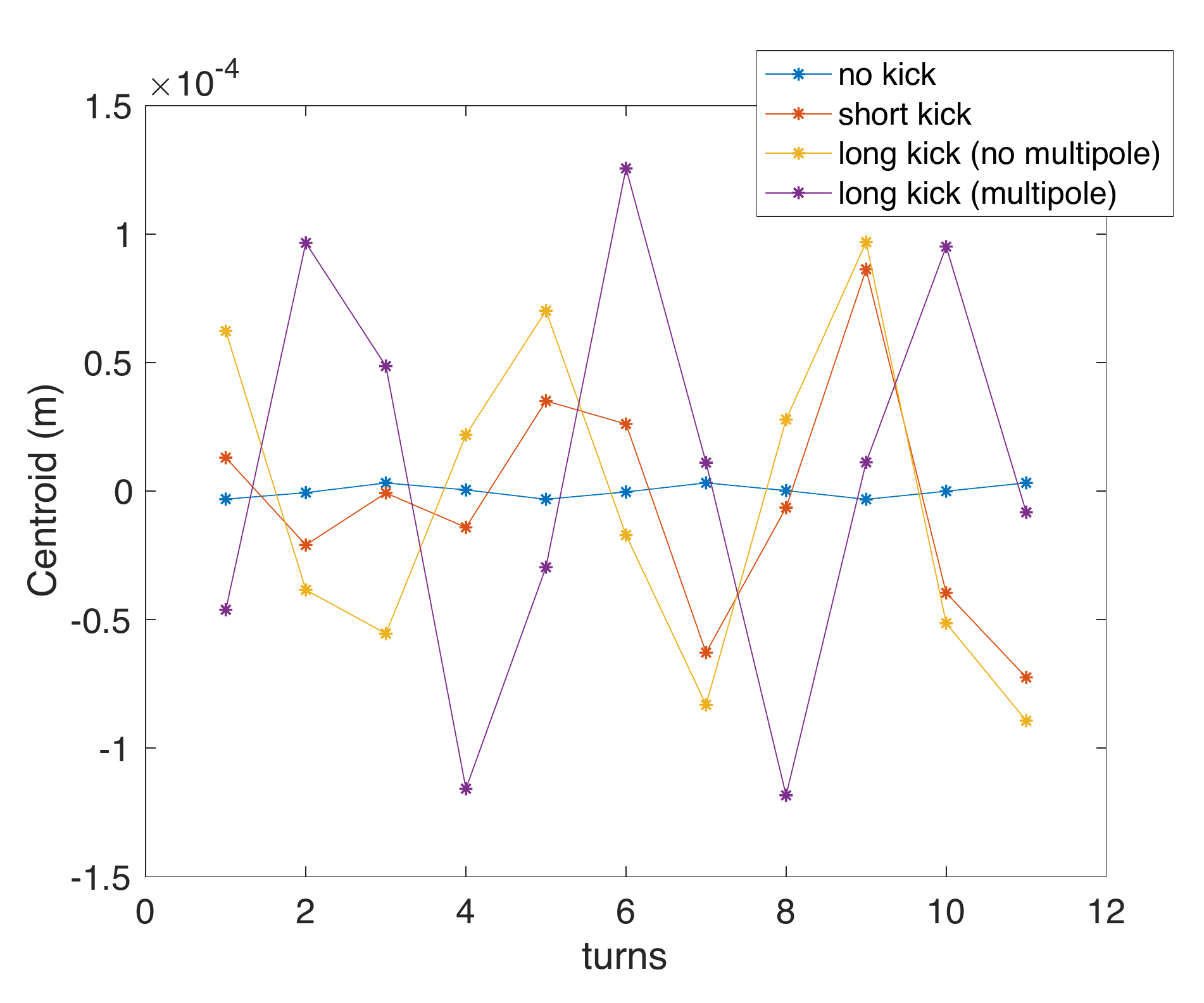}}%
    ~
    \subfigure[Slope of centroid evolution.]{\label{fig:MB-centroid_slope}\includegraphics[scale=0.38]{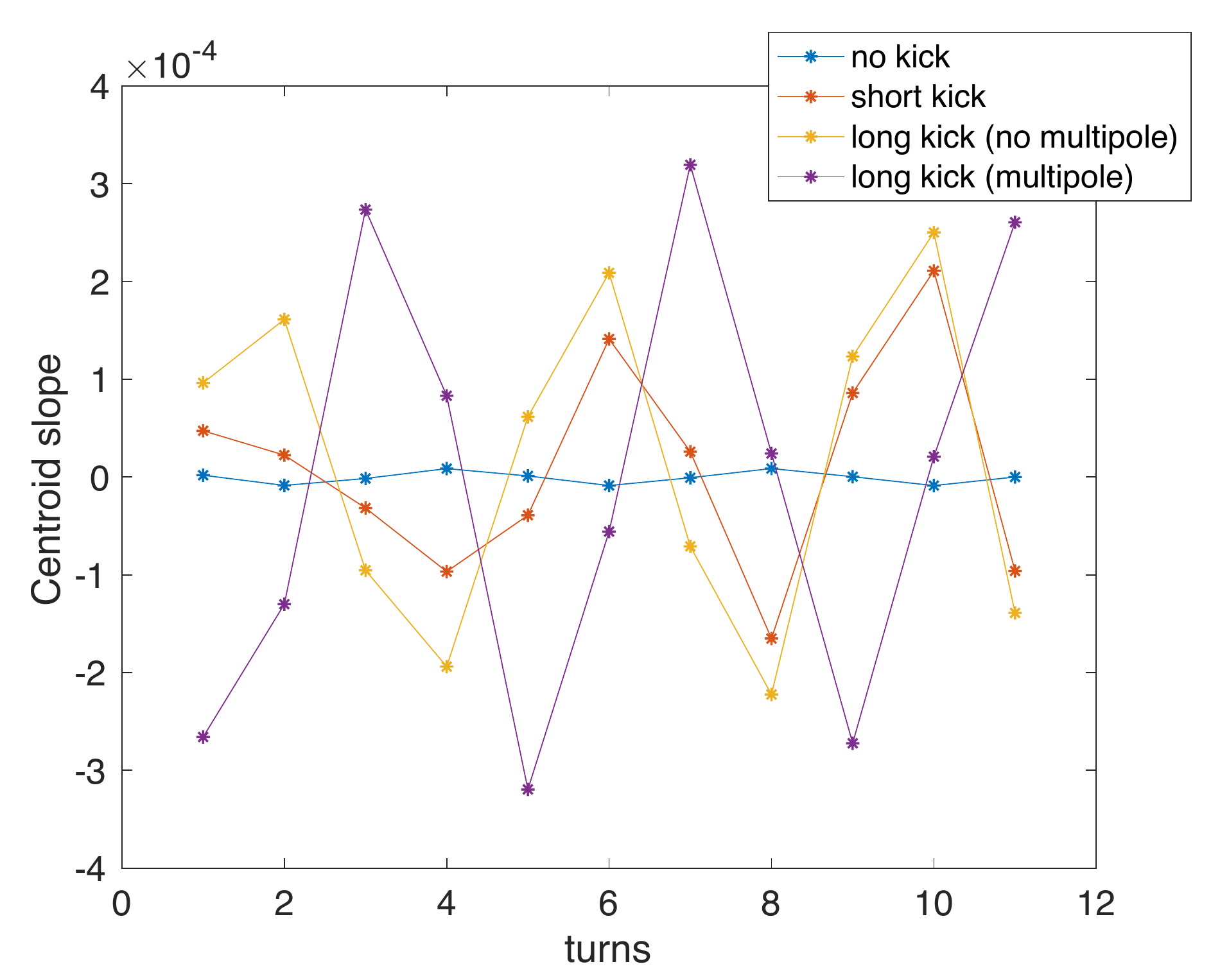}}\\
    \subfigure[The evolution of the angular distribution with an extended kick and multipoles.]{\label{fig:MB-angular_dist}\includegraphics[scale=0.5]{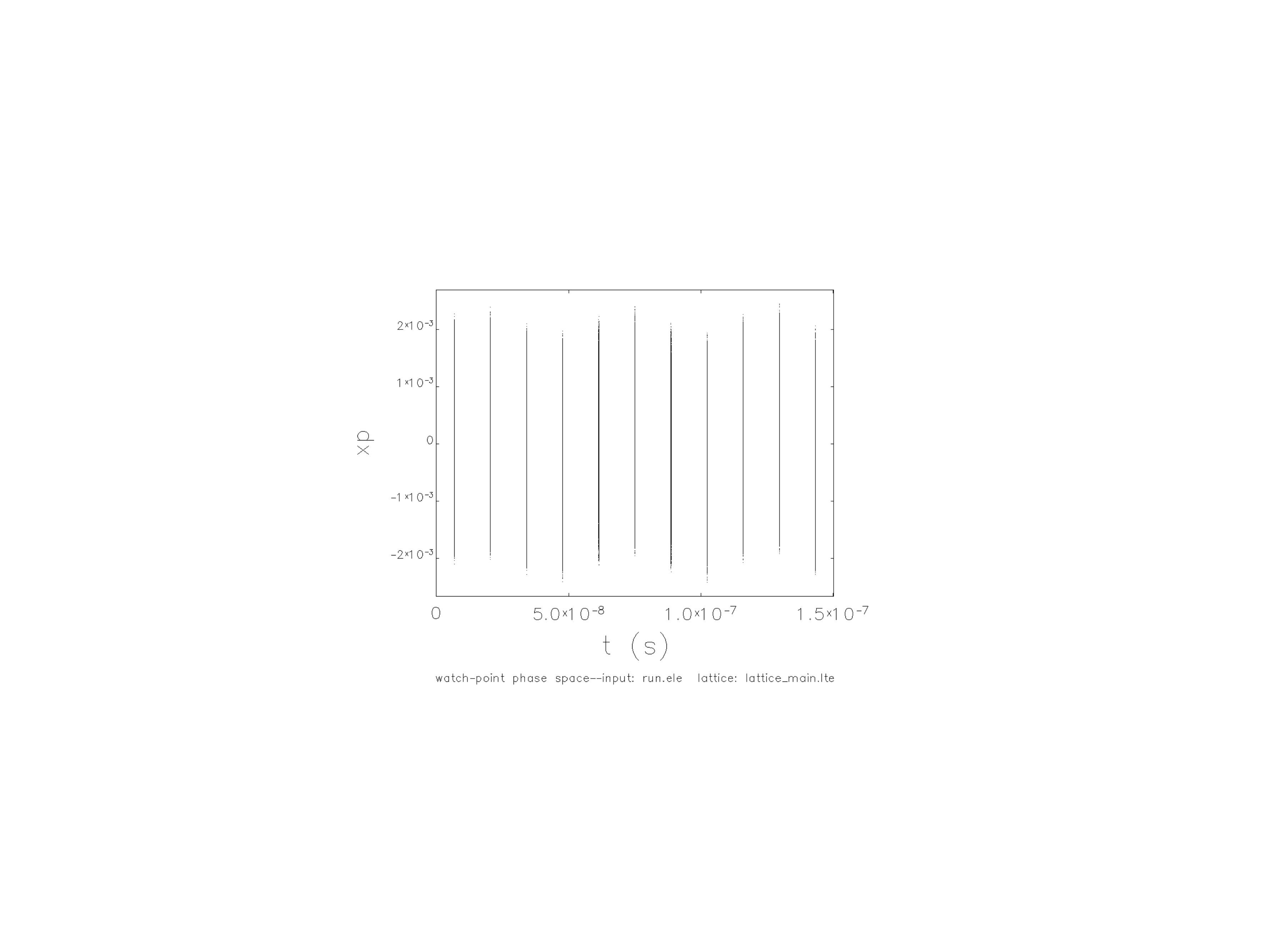}}
 \caption{Evolution of the beam parameters over 11\,turns with extended kicks.}
 \label{fig:MB-bp}
\end{figure*}

 The betatron phase advance (equal to $\pi$ ) between the EK and the IK is adjusted to a non-diagonal matrix (according to the parameter choices above (\ref{eqn:xp_simple})-(\ref{eqn:y_simple})) to account for magnetized beam propagation. To compare the effects of the extended kick, the simulations were run with both the impulsive and extended kick models. With an impulsive kick model, as theoretically predicted, there is little degradation in all the beam parameters except for a very small degradation from the first injection kick. All the subsequent kicks are completely cancelled. The resulting beam parameter evolutions at the cooler entrance are shown in Fig.\,\ref{fig:MB-bp}. As the results of the perfect cancellation with the impulsive kick, there are very little change in eigenemittances (Fig.\,\ref{fig:MB-emittances}) and  only small fluctuations in the aspect ratio and the CAM (Fig.\,\ref{fig:MB-CAM-AR}), implying that the beam remains round over 11\,turns. In addition, the evolution of the centroid and its slope is shown in Fig.\,\ref{fig:MB-centroid} Fig.\,\ref{fig:MB-centroid_slope} with little contribution from the kicker. 
\par
\begin{figure*}[htb]
   \centering
   \includegraphics*[width=175mm]{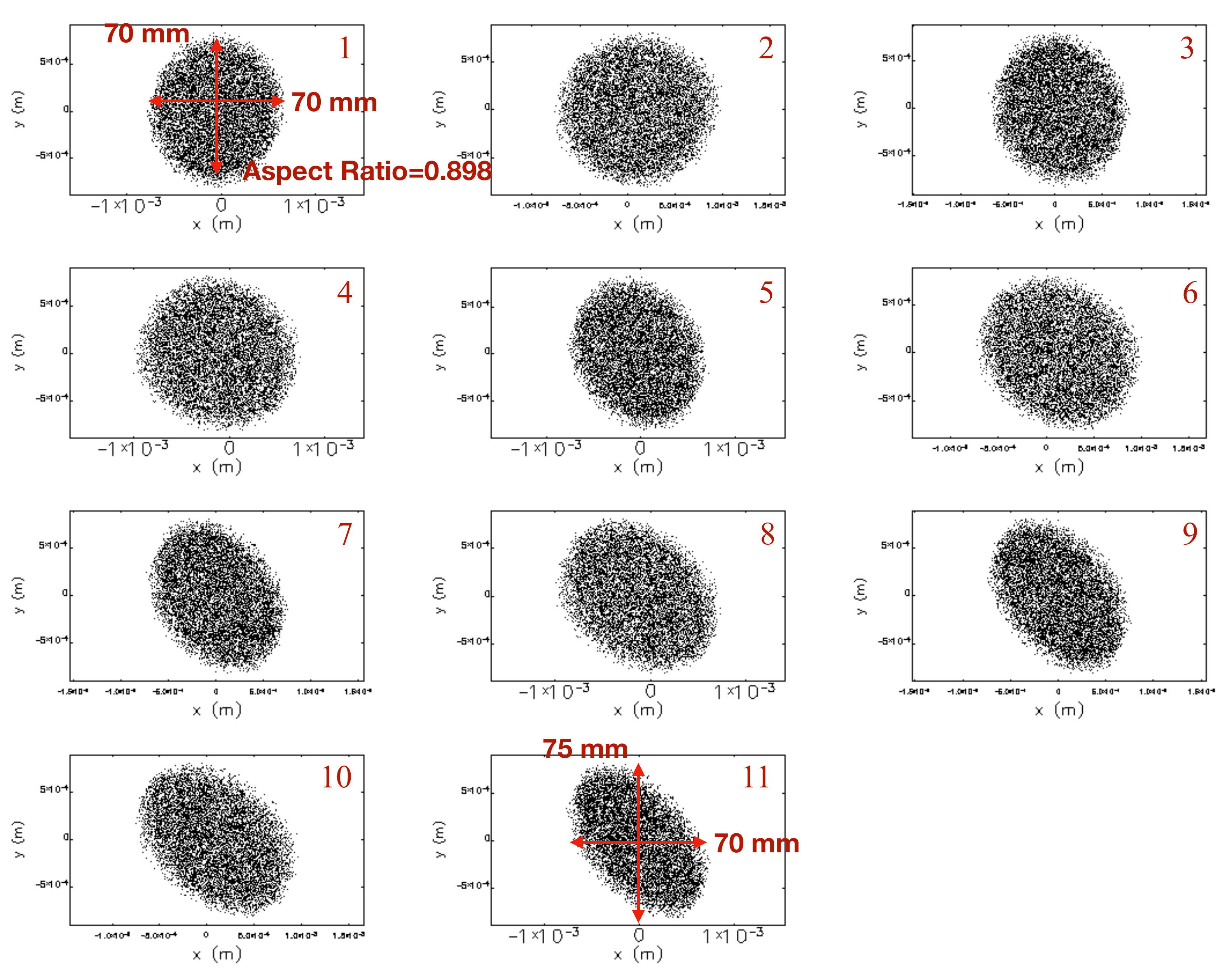}
   \caption{The evolution of transverse profiles at the cooler over 11\,turns.}
   \label{fig:xy-long_kick-cooler}
\end{figure*}

\begin{figure*}[htb]
   \centering
   \includegraphics*[width=175mm]{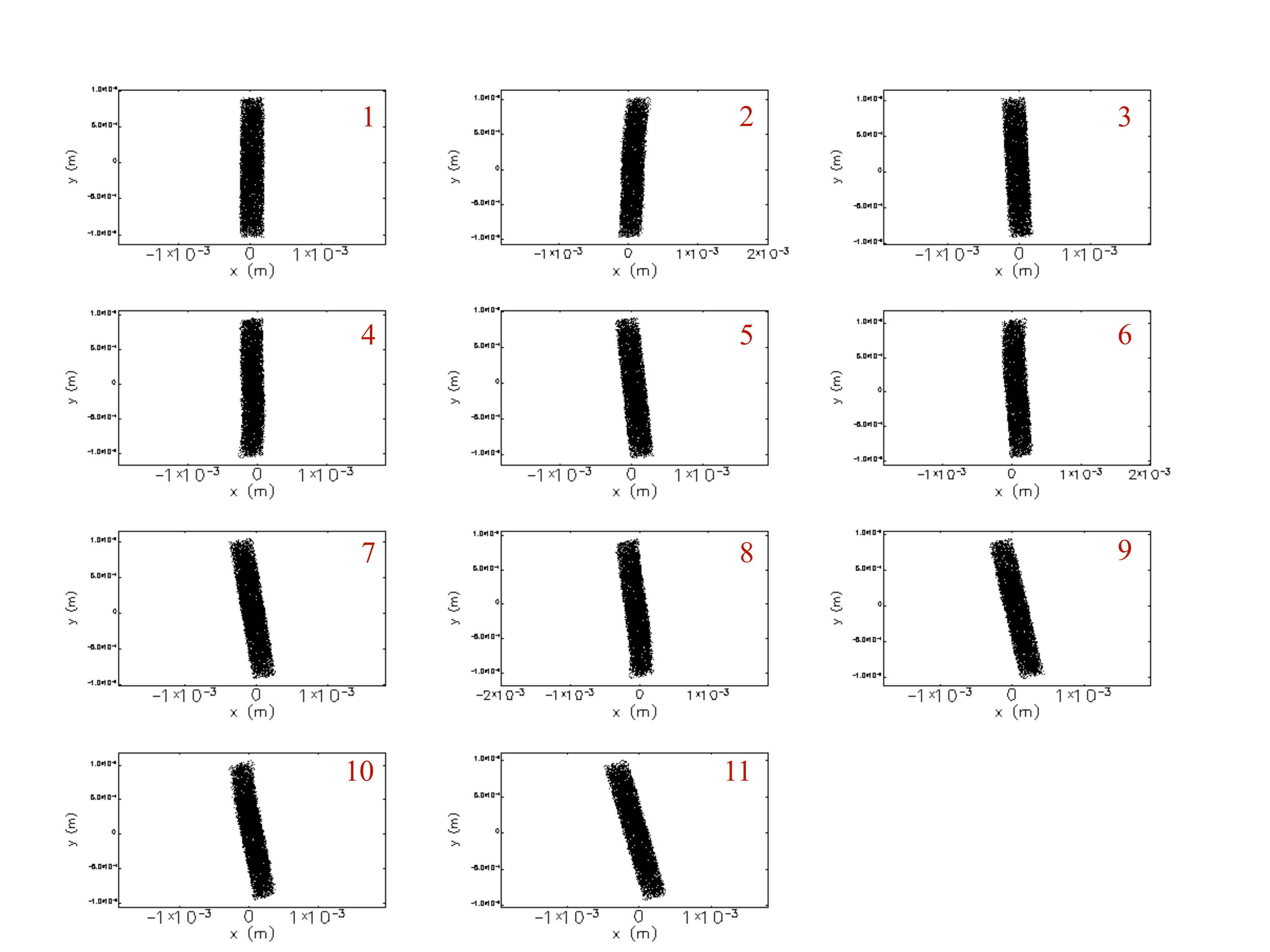}
   \caption{The evolution of flat beam profiles at the cooler over 11\,turns.}
   \label{fig:flat-long_kick-cooler}
\end{figure*}

 With the extended kicks however, tracking of the beam parameters (Fig.\,\ref{fig:MB-bp}) shows significant degradations. The transverse profiles ($x$-$y$ plot) at the cooler entrance shown in Fig.\,\ref{fig:xy-long_kick-cooler} show the deformations gradually evolving over 11\,passes, implying imperfect cancellation between the kicks. The solid round circle gradually turns into a tilted ellipse with the aspect ratio degrading significantly as the passes accumulate. 
More specifically, the deformation is a combined effect of non-conservation of the CAM (Fig.\,\ref{fig:MB-CAM-AR}) and centroid deviations (Fig.\,\ref{fig:MB-centroid})\textemdash caused by the harmonic kick. As expected from the analytical expression, the Larmor emittance of the beam at the cooler increases significantly over passes, while the drift emittance drops (Fig.\,\ref{fig:MB-emittances}). This can also be seen from Fig.\,\ref{fig:flat-long_kick-cooler}, where the flat beam distributions at the cooler are shown to deform\textemdash tilted with an increased horizontal emittance\textemdash as the passes accumulate. Further analysis based on (\ref{eqn:eigenemittance}) and parameter inspection shows that the increase in the Larmor emittance over the passes comes mostly from the increased CAM, while a decrease in the Larmor emittance with multipoles on is caused by a quadrupole focusing contribution that reduces output beam size $\langle x_0^2 \rangle$ in $\varepsilon_{eff}$ downstream at the cooler.  
Overall, after 11\,passes, the Larmor emittance is still much smaller than 19\,mm\,mrad, tolerance for the cooling efficiency. The centroid trajectories and slopes were tracked over the passes in Fig.\,\ref{fig:MB-centroid} and Fig.\,\ref{fig:MB-centroid_slope}, respectively. The fluctuating deviations from zero line are a little larger than with the impulsive kick, i.e., up to $\pm2.5$\,mrad and $\pm0.13$\,mm, respectively. The resulting angular distribution of the bunch is shown in Fig.\,\ref{fig:MB-angular_dist} with small fluctuations in centroid slopes while its angular divergence (along the bunch length) remains roughly the same. The slope fluctuation indicates the incomplete cancellation of the slopes due to the offset evolution combined with the multipole fields\textemdash one can see in Fig.\,\ref{fig:MB-centroid_slope} that the extended kick without multipoles is comparable to the impulsive kick (with the multipoles, but relatively strong cancellation) in terms of the slope fluctuations. 

\section{Conclusion}
\par A harmonic kicker system for the CCR of the JLEIC is designed satisfying all the requirements of beam dynamics of the CCR. The kick is constructed out of the 5 odd harmonic modes of kick frequency $f_k=86.6$\,MHz and can be accommodated in a single QWR cavity. Similar to \cite{huang1}, the residual kicks on the passing bunches are cancelled based on the betatron phase advance, while the kicks on the exchanged bunches were made flat with pre/post kickers. The baseline design was confirmed with a more realistic kick model based on the the RF simulation of the QWR cavity, which includes non-uniform transverse profile and longitudinally extended profile. Accordingly, the phase advance cancellation scheme was modified to cancel the effects of the multipole fields as well. The effects of a harmonic kicker on the magnetized beam was studied with a longitudinally extended kick model and we have demonstrated that the degradation in terms of the Larmor emittance is within our accepted tolerance. The results were also benchmarked against the GPT simulations that uses 3D field maps directly. 
\appendix
\begin{appendices}
\section{Benchmarking with GPT using 3D field maps}
 \label{apx:gpt}
\par The simulation by ELEGANT and analytic computations were benchmarked against GPT simulations. Due to the inability of GPT to loop over the multiple passes and include a transfer matrix as a beamline element, the comparison was made only for a single pass through the kicker cavity. For a fair comparison, the extended kick model with multipoles was used in ELEGANT and analytic computations and the DC magnet in ELEGANT\textemdash that would give an additional kick\textemdash was removed.    
\par
\begin{figure*}[hbt]
  \centering
    \subfigure[$x$-$y$ distribution.]{\label{fig:xy}\includegraphics[scale=0.21]{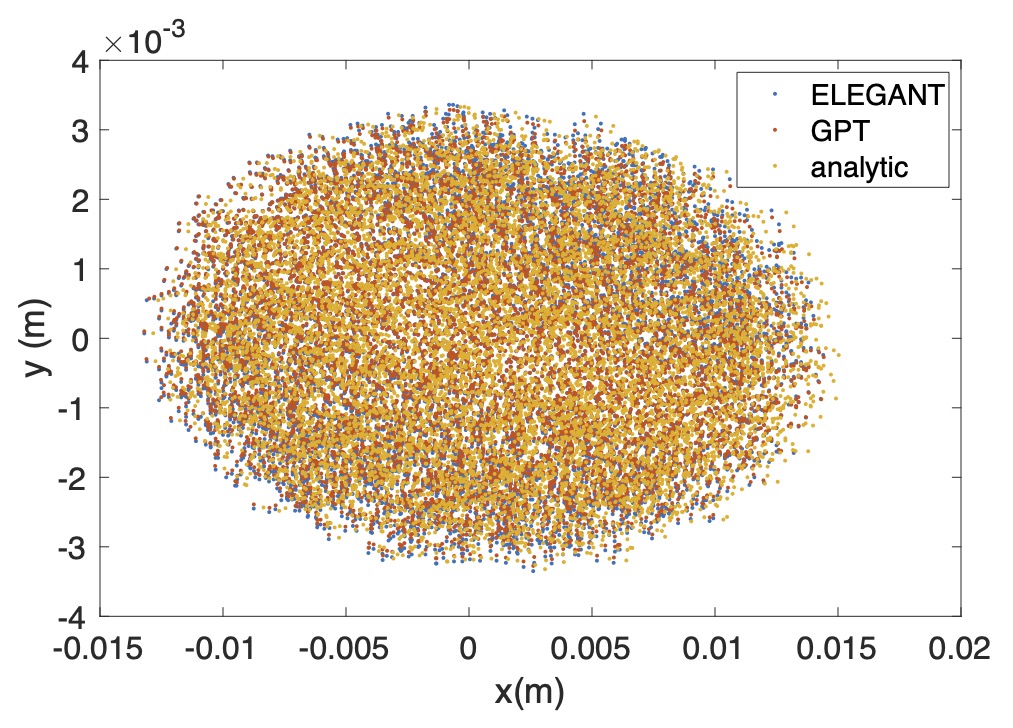}}%
    ~
    \subfigure[$x$-$x'$ distribution.]{\label{fig:xpx}\includegraphics[scale=0.18]{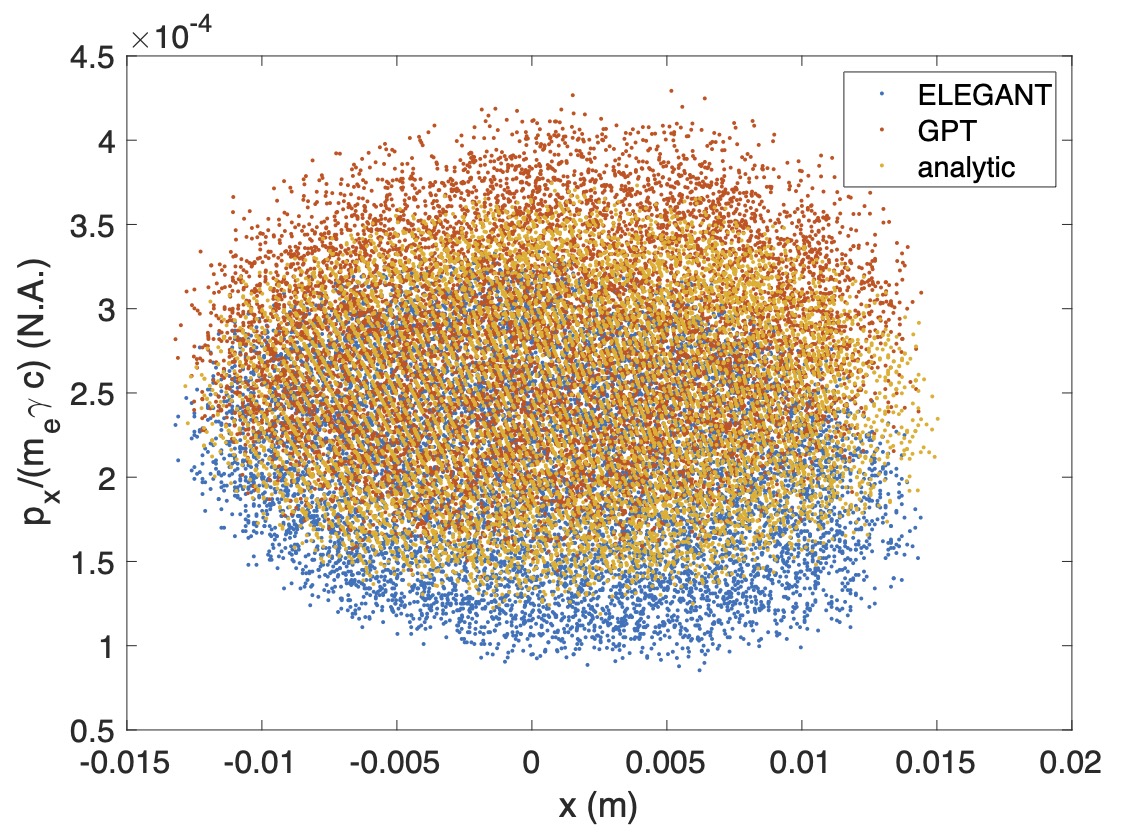}}\\
    \subfigure[$x$-$y'$ distribution.]{\label{fig:xpy}\includegraphics[scale=0.22]{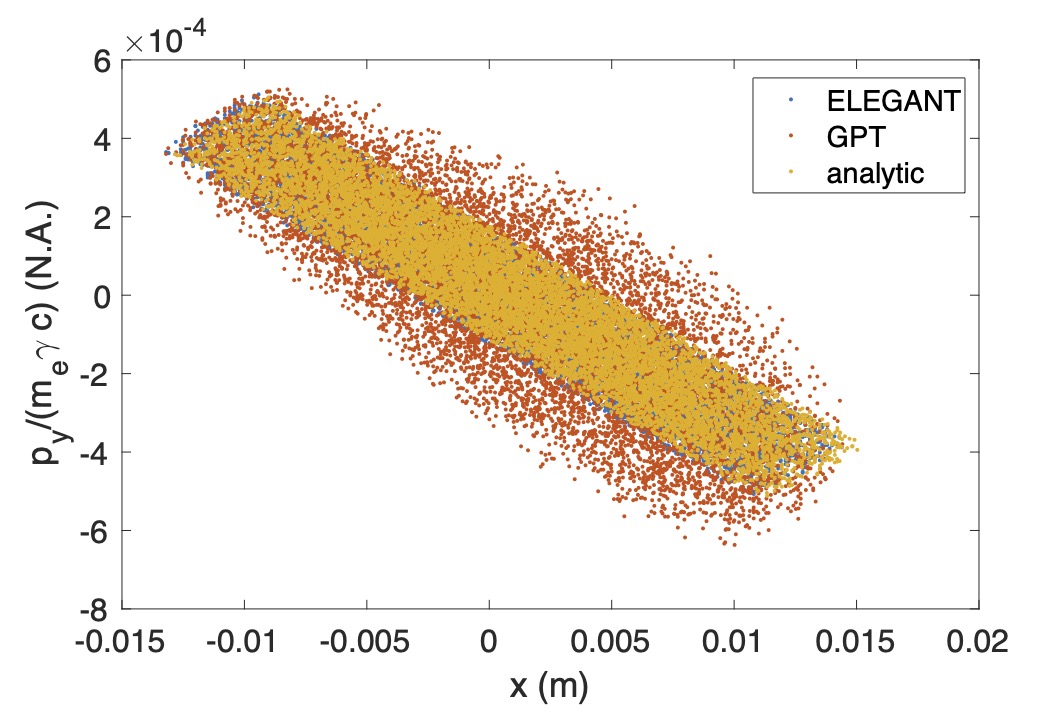}}%
    ~
    \subfigure[$y$-$x'$ distribution.]{\label{fig:ypx}\includegraphics[scale=0.2]{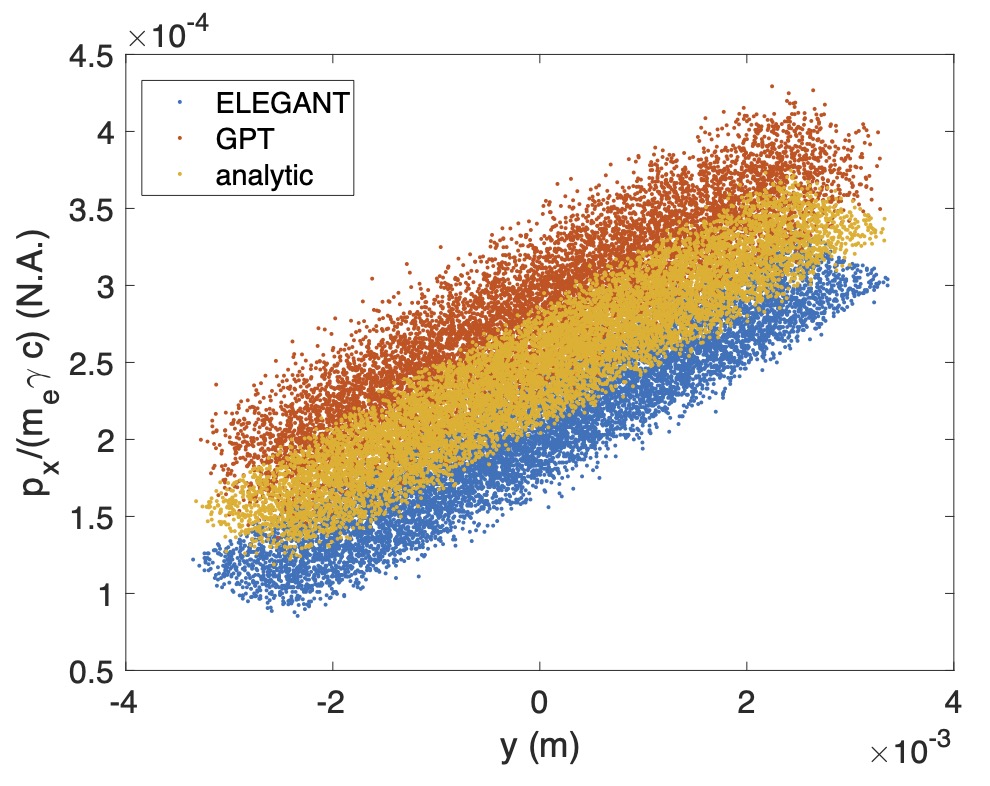}}
  \caption{Benchmark of ELEGANT code and analytic computations against GPT.}
  \label{fig:bm_plot}
\end{figure*}

 The benchmark results of ELEGANT and analytic computations against GPT is shown in Fig.\,\ref{fig:bm_plot}, where various beam distributions are compared. In particular, the $x$-$y$ distributions in Fig.\,\ref{fig:xy} are indistinguishable, confirming the extended kick model closely approximates the actual trajectory as realized by the GPT simulation. However, the slope changes (common to $x$-$x'$ in Fig.\,\ref{fig:xpx} and $y$-$x'$ in Fig.\,\ref{fig:ypx}) show a discrepancy on the order of a few tens of $\mu$rad's. This comes from the inaccuracy of evaluating the multipole moments, discretization of the kick into $N=27$ impulsive kicks with drift spaces. The slope change in Fig.\,\ref{fig:xpy} shows a much smaller change but shows a blurred distribution from GPT. This also can be attributed to the inaccurate evaluation of the multipole coefficients (including skew multipoles). The benchmark in terms of  all the elements of the beam matrix is listed in Table\,\ref{table:bm_Sigma}. Overall, the numbers are close enough to one other, suggesting that the multipole representation in an extended kick model in the ELEGANT is reasonably good approximation to the 3D field maps\textemdash although whether these small errors will stabilize or amplify after the multiple passes remains to be investigated. 


\begin{table}[hbt]
   \centering
   \caption{Benchmark in terms of beam matrix elements at the exit of the kicker.}
   \vspace{5pt}
   \begin{tabular}{llccc}
       \toprule
           Beam & Unit &ELEGANT & GPT & Analytic \\ 
           distributions & \\
           \hline
           $\langle x^2 \rangle$ & $10^{-4}$m$^2$ & 0.3816 & 0.3816 & 0.3882\\
           $\langle xx'\rangle$ & $10^{-7}$m & -0.5617 & 0.2772 & -0.0109\\
           $\langle xy \rangle$ & $10^{-6}$m$^2$ & -0.4755 & -0.5811 & -0.4882\\
           $\langle xy' \rangle$ & $10^{-5}$m & -0.1334 & -0.1325 & -0.1344\\
           $\langle x'^{\,2} \rangle$ & $10^{-8}$ & 0.3154 & 0.3294 & 0.3091\\
           $\langle x'y \rangle$ & $10^{-7}$m & 0.7953 & 0.7641 & 0.7817\\
           $\langle x'y' \rangle$ & $10^{-8}$ & 0.2064 & -0.7258 & 0.0078\\
           $\langle y^2 \rangle$ & $10^{-5}$m$^2$ & 0.2268 & 0.2167 & 0.2230\\
           $\langle yy' \rangle$ & $10^{-6}$m & 0.0224 & -0.1531 & 0.0210 \\
           $\langle y'^{\,2} \rangle$ & $10^{-7}$ & 0.5215 & 0.6626 & 0.5195\\
           $\langle x \rangle$ & $10^{-3}$m & 0.6 & 0.6 & 1.0\\
           $\langle x' \rangle$ & $10^{-3}$ & 0.2087 & 0.2825 & 0.2457\\
           $\langle y \rangle$ & $10^{-5}$m & -0.6085 & -0.7765 & -0.6690\\
           $\langle y' \rangle$ & $10^{-5}$ & 0.1964& -0.4126& -0.4867\\
        \botrule
   \end{tabular}
   \label{table:bm_Sigma}
\end{table}
\end{appendices}

\end{document}